\documentclass[12pt]{article}
\usepackage{jheppub}
\usepackage{mathtools,amssymb,amsthm}
\usepackage[utf8]{inputenc}
\usepackage{simpler-wick}
\usepackage{enumitem}
\usepackage{amsmath}
\usepackage{amssymb}
\usepackage[all]{xy}
\usepackage{ragged2e}
\usepackage{xcolor}
\usepackage{tikz}
\usepackage[export]{adjustbox}

\usepackage{babel}
\begin{document}
\global\long\def\aad{(a\tilde{a}+a^{\dagger}\tilde{a}^{\dagger})}%

\newcommand{\mh}{\mathcal{H}}
\newcommand{\normord}[1]{:\mathrel{#1}:}
\newcommand{\tgvw}{\tilde{G}^{\Delta_V\Delta_W}_4}
\newcommand{\gvw}{G^{\Delta_V\Delta_W}_4}
\newcommand{\gv}{G^{\Delta_V}_2}
\newcommand{\gw}{G^{\Delta_W}_2}

\global\long\def\ad{{\rm ad}}%

\global\long\def\bij{\langle ij\rangle}%

\global\long\def\df{\coloneqq}%

\global\long\def\bs{b_{\alpha}^{*}}%

\def\JX#1{{\color{purple}{[#1]}}}
\def\jx#1{{\color{purple}{[#1]}}}

\newcommand{\ot}{\otimes}
\newcommand{\op}{\oplus}

\global\long\def\bra{\langle}%

\global\long\def\dd{{\rm d}}%

\global\long\def\dg{{\rm {\rm \dot{\gamma}}}}%

\global\long\def\ddt{\frac{{\rm d^{2}}}{{\rm d}t^{2}}}%

\global\long\def\ddg{\nabla_{\dot{\gamma}}}%

\global\long\def\del{\mathcal{\delta}}%

\global\long\def\Del{\Delta}%

\global\long\def\dtau{\frac{\dd^{2}}{\dd\tau^{2}}}%

\global\long\def\ul{U(\Lambda)}%

\global\long\def\udl{U^{\dagger}(\Lambda)}%

\global\long\def\dl{D(\Lambda)}%

\global\long\def\da{\dagger}%

\global\long\def\id{{\rm id}}%

\global\long\def\ml{\mathcal{L}}%

\global\long\def\mm{\mathcal{\mathcal{M}}}%

\global\long\def\mf{\mathcal{\mathcal{F}}}%

\global\long\def\ket{\rangle}%

\global\long\def\kpp{k^{\prime}}%

\global\long\def\lr{\leftrightarrow}%

\global\long\def\lf{\leftrightarrow}%

\global\long\def\ma{\mathcal{A}}%

\global\long\def\mb{\mathcal{B}}%

\global\long\def\md{\mathcal{D}}%

\global\long\def\mbr{\mathbb{R}}%

\global\long\def\mbz{\mathbb{Z}}%

\global\long\def\mh{\mathcal{\mathcal{H}}}%

\global\long\def\mi{\mathcal{\mathcal{I}}}%

\global\long\def\ms{\mathcal{\mathcal{\mathcal{S}}}}%

\global\long\def\mg{\mathcal{\mathcal{G}}}%

\global\long\def\mfa{\mathcal{\mathfrak{a}}}%

\global\long\def\mfb{\mathcal{\mathfrak{b}}}%

\global\long\def\mfb{\mathcal{\mathfrak{b}}}%

\global\long\def\mfg{\mathcal{\mathfrak{g}}}%

\global\long\def\mj{\mathcal{\mathcal{J}}}%

\global\long\def\mk{\mathcal{K}}%

\global\long\def\mmp{\mathcal{\mathcal{P}}}%

\global\long\def\mn{\mathcal{\mathcal{\mathcal{N}}}}%

\global\long\def\mq{\mathcal{\mathcal{Q}}}%

\global\long\def\mo{\mathcal{O}}%

\global\long\def\qq{\mathcal{\mathcal{\mathcal{\quad}}}}%

\global\long\def\ww{\wedge}%

\global\long\def\ka{\kappa}%

\global\long\def\nn{\nabla}%

\global\long\def\nb{\overline{\nabla}}%

\global\long\def\pathint{\langle x_{f},t_{f}|x_{i},t_{i}\rangle}%

\global\long\def\ppp{p^{\prime}}%

\global\long\def\qpp{q^{\prime}}%

\global\long\def\we{\wedge}%

\global\long\def\pp{\prime}%

\global\long\def\sq{\square}%

\global\long\def\vp{\varphi}%

\global\long\def\ti{\widetilde{}}%

\global\long\def\wg{\widetilde{g}}%

\global\long\def\te{\theta}%

\global\long\def\tr{{\rm Tr}}%

\global\long\def\ta{{\rm \widetilde{\alpha}}}%

\global\long\def\sh{{\rm {\rm sh}}}%

\global\long\def\ch{{\rm ch}}%

\global\long\def\Si{{\rm {\rm \Sigma}}}%

\global\long\def\si{{\rm {\rm \sigma}}}%

\global\long\def\sch{{\rm {\rm Sch}}}%

\global\long\def\vol{{\rm {\rm {\rm Vol}}}}%

\global\long\def\reg{{\rm {\rm reg}}}%

\global\long\def\zb{{\rm {\rm |0(\beta)\ket}}}%

\newcommand{\be}{\begin{equation}}
\newcommand{\ee}{\end{equation}}
\newcommand{\tx}{\text}
\newcommand{\tb}{\textbf}
\newcommand{\me}{\mathcal{E}}
\title{On Chord Dynamics and Complexity Growth in Double-Scaled SYK}
\author{Jiuci Xu}

\affiliation{Department of Physics, University of California, Santa Barbara, CA 93106, USA}
\emailAdd{Jiuci\_Xu@ucsb.edu}

\justify
\abstract{
We study the time evolution governed by the two-sided chord Hamiltonian in the double-scaled SYK model, which induces a probability distribution over operators in the double-scaled algebra. Through the bulk-to-boundary map, this distribution translates into dynamical profiles of bulk states within the chord Hilbert space. We derive analytic expressions for such profiles, valid across a broad parameter range and all time scales. Additionally, we demonstrate how distinct semi-classical behaviors emerge by localizing within specific energy regions in the semi-classical limit. We revisit the doubled Hilbert space formalism as an isometric map between the one-particle sector of the chord Hilbert space and the doubled zero-particle sector. Utilizing this map, we obtain analytic results for correlation functions and investigate the dynamical evolution for chord operators. Specifically, we  establish an equivalence between the chord number generating function in presence of matter chords and the crossed four-point correlation function, the latter is closely related to the $6j$-symbol of $U_{\sqrt{q}}(\mathfrak{su}(1,1))$. We also explore finite-temperature effects, showing that operator spreading slows as temperature decreases. In the semi-classical limit, we perform a saddle point analysis and incorporate the one-loop determinant to derive the normalized time-ordered four-point correlation function at infinite temperature. The leading correction reproduces the \(1/N\) connected contribution observed in the large-\(p\) SYK model. Finally, we examine the time evolution of total chord number in presence of matter in the triple-scaled regime, linking it to the renormalized two-sided length in JT gravity with matter.
}

\maketitle

\section{Introduction}
Understanding the nature of quantum chaos has long presented a significant challenge in theoretical physics. In recent years, numerous approaches have been proposed to probe quantum chaos, including operator spreading~\cite{Roberts:2014isa,Roberts:2018mnp,Qi_2019,Lensky:2020ubw,Jian_2021,Schuster_2022}, out-of-time-order correlators (OTOCs)~\cite{Roberts_2015,shenker2015stringyeffectsscrambling,Roberts_2015_butterfly,Hashimoto_2017,Gu_2022}, and various formulations of quantum complexity~\cite{Parker_2019,Balasubramanian_2022,Nielsen2005AGA,PhysRevD.103.026015.QueryC}. Within the framework of the AdS/CFT correspondence~\cite{Susskind:1994vu,Maldacena:1997re,Witten:1998zw,Witten:1998qj,Susskind:1998dq}, these probes serve as useful tools for exploring the microscopic structure underlying the duality. Notably, it has been shown that the time-evolution profile of quantum complexity in the boundary system corresponds to geometric structures in the bulk~\cite{Parker_2019,Haferkamp_2022,Jian_2021,Rabinovici:2023yex,Aguilar-Gutierrez:2024nau,Sergio-C:2024rka,Li:2024kfm}. This correspondence supports the conjecture that certain notion of quantum complexity on the boundary is dual to diffeomorphism-invariant quantities~\cite{Sekino_2008,Lashkari_2013,Susskind:2014rva,Stanford:2014jda,Brown:2015bva,Couch_2017}.

Given these insights, it is desirable to formulate a concrete measure of complexity that is intrinsic to the quantum system and does not rely on external inputs. Identifying such a quantity in the bulk would provide a bridge between the time evolution of boundary complexity and that of diffeomorphism-invariant observables in the bulk, thereby deepening our understanding of the duality.

One promising candidate for this intrinsic measure is Krylov complexity, first introduced in \cite{Parker_2019}. Krylov complexity has been shown to capture the dynamical behavior of quantum complexity over all time scales without requiring external resources \cite{Barb_n_2019,Rabinovici_2021}. To illustrate, let $M$ be a generic operator acting on a Hilbert space $\mathcal{H}$, with $H$ as the system’s Hamiltonian. The time evolution of $M$, governed by the Liouvillian operator, is given by 
\begin{equation}
M(t) = e^{\mathcal{L}t}M \equiv e^{iHt}M e^{-iHt}.
\end{equation}
This evolution can be expanded in terms of a basis of operators \(\{\mathcal{O}_n\}, n=0,1,\dots\) that spans the Krylov subspace of $\mathcal{B}(\mathcal{H})$:
\begin{equation} \label{eq:M-expand}
M(t) = \sum_{n=0}^\infty \phi_n(t) \mathcal{O}_n.
\end{equation}
While the specific behavior of the overlap functions \(\phi_n(t)\) depends on the microscopic details of the theory, Krylov complexity \(\mathcal{K}(t)\), defined as
\be
\mathcal{K}(t) \equiv \sum_{n=0}^\infty n |\phi_n(t)|^2,
\ee
exhibits the universal features expected of quantum complexity.

The Double-Scaled SYK (DSSYK) model, on the other hand, offers a valuable framework in which many of the aforementioned ideas can be explicitly tested and further developed. The model was introduced in \cite{Cotler:2016fpe} by taking the limits \( N \to \infty \) and \( p \to \infty \), while keeping the ratio \(\lambda = 2 p^2 / N\) fixed. The functional dependence of \(\lambda\) for generic values of \( q = e^{-\lambda} \in [0,1) \) was later determined in \cite{Berkooz:2018qkz,Berkooz:2018jqr}, where the limiting behavior of commutation relations between large fermionic string operators, whose sizes scale with \(p\), was mapped onto combinatorial problems involving chord diagrams.

An auxiliary Hilbert space \(\mathcal{H}\) was introduced to effectively describe the action of these limiting operators, or chord operators. The inner product within this Hilbert space organizes intermediate chord configurations in a manner that simplifies the evalu-ation of correlation functions. In \cite{Lin_2023}, this auxiliary Hilbert space was further associated with a bulk interpretation, wherein states with a given number of open chords correspond to a two-sided wormhole with a specified length. For any finite \( q \in (0,1) \), the radial direction of the corresponding wormhole geometry is discrete, with a minimal distance cutoff defined by \(\lambda = -\ln q\). In the limit \(\lambda \to 0\), one expects a continuum description of the bulk geometry to emerge, and several recent proposals aim to connect this limit to lower-dimensional quantum gravity \cite{Susskind:2021esx,Susskind:2022dfz,Lin:2022nss,Susskind:2022bia,Berkooz:2022mfk,Goel:2023svz,Narovlansky:2023lfz,Susskind:2023hnj,Rahman:2024vyg,Rahman:2024iiu,Almheiri:2024ayc,Almheiri:2024xtw,Milekhin:2024vbb}.

However, different types of semi-classical physics emerge as one focuses on different regions of the DSSYK energy spectrum in the \(\lambda \to 0\) limit. These distinct semi-classical regimes suggest that a deeper understanding is needed of how various classical limits arise from the analytic expressions governing the system at generic parameter values. This may also help clarify the emergence of different semi-classical phenomena and their connections to bulk gravitational dynamics in DSSYK.

Moreover, a concrete notion of bulk-to-boundary correspondence exists for general values of \( q \) in DSSYK. It was observed in \cite{Lin:2022rbf} that the orthogonalization of states prepared by acting chord operators on the empty state \(\Omega\) closely parallels the Lanczos algorithm used to construct the Krylov operator basis in \eqref{eq:M-expand}. This insight bridges boundary states, prepared by insertions of one-sided chord operators at the boundary, and bulk two-sided wormhole states, with the wormhole length corresponding to the number of chords. This connection was further extended in \cite{Xu:2024hoc}, establishing an operator-state correspondence between the one-sided chord algebra and the chord Hilbert space.

In addition, \cite{Xu:2024hoc} reveals the underlying reason for the existence of this bulk-to-boundary correspondence for generic \( q \) by showing that the one-sided chord algebra (termed the double-scaled algebra in \cite{Lin:2022rbf}) forms a Type II$_1$ von Neumann factor, with the empty state \(\Omega\) being cyclic and separating. Specifically, the bulk-to-boundary map is realized through a normal-ordering prescription among chord operators, where the Krylov operator basis is represented by normal-ordered chord operators. Applying this prescription to the chord evolution operator \( e^{-iHt} \), one arrives at:
\begin{equation} \label{eq:opbasis}
e^{-iHt} = \sum_{n=0}^{\infty} \frac{\left(-it\right)^{n}}{n!}H^{n} = \sum_{n=0}^{\infty} \frac{\phi_n(t)}{(1-q)^{n/2}[n]_q!} \normord{H^n},
\end{equation}
where \( \sum_{n=0}^{\infty} n |\phi_n(t)|^2 \) defines the Krylov complexity of the state \( e^{-iHt}|\Omega\rangle \).

Similarly, for a matter chord operator \(\hat{M}(t) = e^{i(H_L - H_R)t} M e^{-i(H_L - H_R)t}\), evolved by the two-sided chord Hamiltonian, the normal-ordering prescription leads to:
\begin{equation}
\hat{M}(t) = \sum_{n=0}^{\infty} \phi_n^{\Delta}(t) \hat{\Psi}_n,
\end{equation}
where \(\hat{\Psi}_n\)s are operators corresponding to the Krylov basis, and \( \sum_{n=0}^{\infty} n |\phi_n^{\Delta}(t)|^2 \) gives the operator Krylov complexity of \(\hat{M}(t)\).

In this work, we examine the time evolution generated by the chord Hamiltonian and present analytic results for various quantities that apply for general \( q \), as well as in various semi-classical limits.

\subsection{Organization} 
The remainder of the paper is organized as follows. In Section~\ref{sec:warmup}, we briefly review the construction of the chord Hilbert space and the operator-state correspondence. We also derive analytic expressions for the dynamical profile \(\{\phi_n(t)\}\) in \eqref{eq:opbasis}, valid for \(q \in [0,1)\), and demonstrate the late-time behavior \(\sim t^{-3/2}\) for \(t \gg \sqrt{1-q}/\lambda \).\footnote{We present a more detailed exploration on this point in \cite{Milekhin:2024ToAppear}.}  We then show how, in the \(\lambda \to 0\) limit, our results reduce to the high-temperature limit considered in \cite{Almheiri:2024xtw} by localizing around the center of the energy spectrum, and to the low-temperature Schwarzian regime considered in \cite{Berkooz:2018jqr,Lin_2023,mukhametzhanov2023largepsykchord} by localizing around the edge of the spectrum. In particular, we explicitly show that the two regimes exhibit fundamentally different dynamical properties.

In Section~\ref{sec:complexity}, we analyze the operator growth of chords evolved under the two-sided chord Hamiltonian. We present analytic results for various quantities, including the chord number generating function, defined as the expectation value of \(e^{-\mu \hat{N}}\) in states with given chord configuration. We also formulate the doubled Hilbert space formalism of \cite{Okuyama:2024gsn,Okuyama:2024yya} as an isometric factorization of the one-particle Hilbert space and adopt the one-particle chord wavefunction from \cite{Xu:2024hoc} to facilitate our calculations. Notably, this isometric factorization reveals the equivalence between the chord number generating function and the crossed four-point correlation function by organizing the chord combinatorics in terms of a \(R\)-matrix closely related to the $6j$ symbol of \(U_{\sqrt{q}}(\mathfrak{su}(1,1))\).

In Section~\ref{sec:manifestation}, we revisit the semi-classical analysis of the \(\lambda \to 0\) limit for the uncrossed four-point function and chord number growth. By localizing around the center of the energy spectrum and incorporating one-loop corrections to energy fluctuations, we find a connected contribution to the corresponding correlation function, reminiscent of results from the large \(p\) SYK model at infinite temperature \cite{Choi:2019bmd,Streicher_2020}. We extend the result in \cite{Rabinovici:2023yex,Sergio2024towards} to include matter, showing that the limiting behavior of the operator Krylov complexity of \(\hat{M}(t)\) corresponds to the two-sided length in the presence of bulk matter fields studied in \cite{Harlow:2021dfp}.

In Section~\ref{sec:finite-T}, we explore finite-temperature effects and provide analytic expressions for the overlap of Hartle-Hawking wavefunctions and states with given chord number in the presence of matter. We demonstrate that the speed of operator spreading reaches its maximum at infinite temperature and decreases continuously as the temperature lowers.

Finally, in Section~\ref{sec:Future}, we summarize our results and discuss future research directions. Additional technical details are provided in the appendices.

\section{Warm up: Chord Dynamics in the Zero-Particle Hilbert Space} \label{sec:warmup}
In this section, we examine the chord dynamics in the absence of the matter chord. We first review the construction of the chord Hilbert space and the dynamics introduced by the chord Hamiltonian. We then demonstrate that the bulk-to-boundary map described in \cite{Lin:2022rbf} can be understood as a mapping between the moments of the chord Hamiltonian and the Krylov operator basis introduced in \cite{Bhattacharjee:2022ave,Tang:2023ocr} , and later represented as normal-ordered chord operators in \cite{Xu:2024hoc}. This time evolution leads to a probability distribution among the Krylov operator basis. We will provide analytical solutions for this distribution and investigate their behavior under various limits.

\subsection{The Bulk-to-Boundary Map and Chord Spreading}
The chord Hilbert space was initially introduced in \cite{Berkooz:2018qkz, Berkooz:2018jqr} as an auxiliary linear space designed to accommodate various chord configurations appearing in the moment calculations within the double-scaled limit of fermionic strings in the SYK model. Subsequently, this space was given a bulk interpretation in \cite{Lin:2022rbf}. In the double-scaled limit, the SYK Hamiltonian transitions into the chord Hamiltonian, denoted as \( H \), which operates on the chord Hilbert space. We will specify the precise definition of \( H \) below.

In this section, we investigate the dynamics of time evolution generated by \( H \) within the chord Hilbert space in absence of matter chords, denoted as \( \mathcal{H}_0 \). This subspace \( \mathcal{H}_0 \) comprises states \( |n\rangle \) characterized by a given number of Hamiltonian chords (\( H \)-chords). For example, consider a state with three open chords, which we can depict as follows: 
\be
|3\ket = \begin{tikzpicture}[baseline={([yshift=-0.1cm]current bounding box.center)},scale=0.35]
    \draw[thick] (0,-0.5) arc (0:-180: 2);
    \draw (-1,-0.5) -- (-1,-2.23);
    \draw (-2,-0.5) -- (-2,-2.5);
    \draw (-3,-0.5) -- (-3,-2.23);  
    \node at (-1,-2.23) [circle,fill,black,inner sep=1.2pt]{}; 
    \node at (-2,-2.5) [circle,fill,black,inner sep=1.2pt]{};
    \node at (-3,-2.23) [circle,fill,black,inner sep=1.2pt]{};
\end{tikzpicture} , 
\ee
The chord Hamiltonian is defined by operations that either add or delete an open chord from a state with a fixed number of chords. We introduce creation and annihilation operators to perform these addition and deletion operations as follows:
\be \label{eq:add}
a^\dagger \begin{tikzpicture}[baseline={([yshift=-0.1cm]current bounding box.center)},scale=0.35]
    \draw[thick] (0,-0.5) arc (0:-180: 2);
    \draw (-1,-0.5) -- (-1,-2.23);
    \draw (-2,-0.5) -- (-2,-2.5);
    \draw (-3,-0.5) -- (-3,-2.23);  
    \node at (-1,-2.23) [circle,fill,black,inner sep=1.2pt]{}; 
    \node at (-2,-2.5) [circle,fill,black,inner sep=1.2pt]{};
    \node at (-3,-2.23) [circle,fill,black,inner sep=1.2pt]{};
\end{tikzpicture}    = \begin{tikzpicture}[baseline={([yshift=-0.1cm]current bounding box.center)},scale=0.35]
    \draw[thick] (0,-0.5) arc (0:-180: 2);
    \draw (-1,-0.5) -- (-1,-2.23);
    \draw (-2,-0.5) -- (-2,-2.5);
    \draw (-3,-0.5) -- (-3,-2.23);  
    \draw (-3.5,-0.5) -- (-3.5,-1.8);
    \node at (-1,-2.23) [circle,fill,black,inner sep=1.2pt]{};  \node at (-3.5,-1.8) [circle,fill,black,inner sep=1.2pt]{}; 
    \node at (-2,-2.5) [circle,fill,black,inner sep=1.2pt]{};
    \node at (-3,-2.23) [circle,fill,black,inner sep=1.2pt]{};
\end{tikzpicture}  ,
\ee
\be \label{eq:delete}
a \begin{tikzpicture}[baseline={([yshift=-0.1cm]current bounding box.center)},scale=0.35]
    \draw[thick] (0,-0.5) arc (0:-180: 2);
    \draw (-1,-0.5) -- (-1,-2.23);
    \draw (-2,-0.5) -- (-2,-2.5);
    \draw (-3,-0.5) -- (-3,-2.23);  
    \node at (-1,-2.23) [circle,fill,black,inner sep=1.pt]{}; 
    \node at (-2,-2.5) [circle,fill,black,inner sep=1.pt]{};
    \node at (-3,-2.23) [circle,fill,black,inner sep=1.pt]{};
\end{tikzpicture}    = \begin{tikzpicture}[baseline={([yshift=-0.1cm]current bounding box.center)},scale=0.35]
    \draw[thick] (0,-0.5) arc (0:-180: 2);
    \draw (-1,-0.5) -- (-1,-2.23);
    \draw (-2,-0.5) -- (-2,-2.5);
    \draw (-3.5,-1.8) .. controls (-3,-1.3) .. (-3,-2.23);
    \node at (-1,-2.23) [circle,fill,black,inner sep=1.pt]{};  \node at (-3.5,-1.8) [circle,fill,black,inner sep=1.pt]{}; 
    \node at (-2,-2.5) [circle,fill,black,inner sep=1.pt]{};
    \node at (-3,-2.23) [circle,fill,black,inner sep=1.pt]{};
\end{tikzpicture}  +
\begin{tikzpicture}[baseline={([yshift=-0.1cm]current bounding box.center)},scale=0.35]
    \draw[thick] (0,-0.5) arc (0:-180: 2);
    \draw (-1,-0.5) -- (-1,-2.23);
    \draw (-3,-0.5) -- (-3,-2.23);  
    \draw (-2,-2.5) .. controls (-2.5,-1.3) and (-3,-1.3) .. (-3.5,-1.8);
    \node at (-1,-2.23) [circle,fill,black,inner sep=1.pt]{};  \node at (-3.5,-1.8) [circle,fill,black,inner sep=1.pt]{}; 
    \node at (-2,-2.5) [circle,fill,black,inner sep=1.pt]{};
    \node at (-3,-2.23) [circle,fill,black,inner sep=1.pt]{};
\end{tikzpicture}  +
\begin{tikzpicture}[baseline={([yshift=-0.1cm]current bounding box.center)},scale=0.35]
    \draw[thick] (0,-0.5) arc (0:-180: 2);
    \draw (-2,-0.5) -- (-2,-2.5);
    \draw (-3,-0.5) -- (-3,-2.23);  
    \draw (-1,-2.23) .. controls (-2,-1.3) and (-3,-1.3)  .. (-3.5,-1.8);
    \node at (-1,-2.23) [circle,fill,black,inner sep=1.pt]{};  \node at (-3.5,-1.8) [circle,fill,black,inner sep=1.pt]{}; 
    \node at (-2,-2.5) [circle,fill,black,inner sep=1.pt]{};
    \node at (-3,-2.23) [circle,fill,black,inner sep=1.pt]{};
\end{tikzpicture}  .
\ee
In this setup, we have defined all operators to act on the left side of the boundary. Similarly, one can introduce operators acting on the right side. These left- and right-acting operators are equivalent when only one type of chord is considered. However, they diverge when an additional type of chord---referred to as a ``matter chord''---is introduced. This distinction will be explored in the next section. For each crossing between \( H \)-chords, a penalty factor of \( q \) is incurred. Accordingly, the above equations can be succinctly summarized as follows:
\be
a^\dagger|3\ket = |4\ket , \quad a|3\ket = (1+q+q^2)|2\ket = [3]_q|2\ket,
\ee
where we introduced $q$-deformed integers $[n]_q=1+q+\cdots+q^{n-1}$.  It's easy to write down the action of $a/a^\dagger$ on arbitrary chord states:
\be
a^{\da}|n\ket=|n+1\ket,\quad a|n\ket=[n]_{q}|n-1\ket,\quad\bra n|m\ket=[n]_{q}!\del_{nm},
\ee
and verify the $q$-commutation relation $[a,a^\dagger]_q=aa^\dagger-q a^\dagger a =1$. We also denote the zero-chord state $|0\ket$ as $|\Omega\ket$ in the following discussion. 
A chord Hamiltonian can thus be defined as \footnote{We choose the normalization of $H$ to be unit for convenience. It's easy to restore the dimensionful normalization factor which is usually $J/\sqrt{\lambda}$ by a redefinition of time. }
\be\label{eq:H-chord}
H=\frac{1}{2C}(a+a^\dagger).
\ee
The overall normalization $(2C)^{-1}$ is determined from the normalization of the quadratic moment $\langle H^2 \rangle$. In Section~\ref{sec:manifestation}, we set $2C=\sqrt{\lambda}$ to match the standard SYK results when taking the semi-classical limit, while in other sections, we set it to $1$ for simplicity. While the action of \( H \) clearly involves adding a new mark to the left of the boundary, its full operation on the chord Hilbert space is more involved. It includes both chord addition and deletion components, which can contract among each other. The action of multiple \( H \) operators on the empty state \( |\Omega\rangle \) takes the form:
\be
H^{n}|\Omega\ket=|n\ket+\text{states with chords less than }n.
\ee
In \cite{Xu:2024hoc}, it is proposed that a normal-ordered operator basis, denoted \( \normord{H^n} \), can be defined by subtracting all contractions among the \( H \) operators. This basis can be recursively defined as follows:
\be \label{eq:Nord}
\normord{H^{n+1}}\equiv H\normord{H^{n}}-\wick{\c H  \normord{ \c H^{n}}}=H \normord{H^{n}}-[n]_{q} H^{n-1},\quad \normord{H}=H,
\ee
It then follows by definition that it corresponds to a chord state with fixed chord number as:
\be \label{eq:os-corres}
\normord{H^{n}}|\Omega\ket=|n\ket.
\ee
The two operator basis are related by the following relation:
\be \label{eq:bb-map}
H^{n}=\sum_{k=0}^{[\frac{n}{2}]}\frac{c_{n,k}}{\left(1-q\right)^{k}}:H^{n-2k}:,
\ee 
where the coefficients are defined as:
\begin{equation} \label{eq:clk-def}
c_{l, k} = \sum_{m=0}^{l} (-1)^m q^{\binom{m+1}{2}} \frac{k-2l+2m+1}{k+1} \binom{k+1}{l-m} \binom{k-2l+m}{m}_q.
\end{equation}
This construction corresponds to the bulk-to-boundary map described in \cite{Lin:2022rbf}. The operator on the left-hand side of \eqref{eq:bb-map} can be interpreted as a boundary operator that inserts an endpoint of an \( H \)-chord at the boundary, analogous to local operators in the boundary theory of the standard AdS/CFT dictionary. Meanwhile, the normal-ordered operators on the right-hand side each generate bulk states from the vacuum with a definite wormhole length.

\begin{figure}
    \centering
    \includegraphics[width=0.8\linewidth]{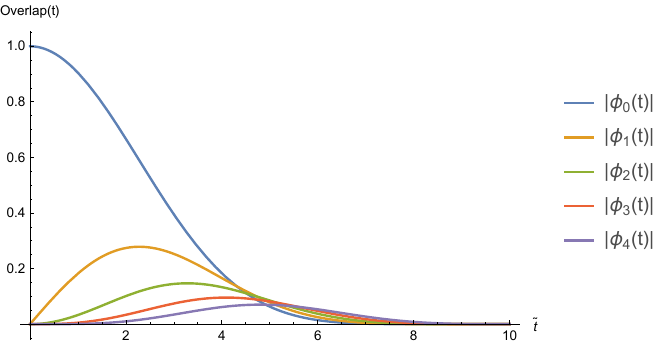}
    \caption{A plot of the solutions \(\phi_0(t), \phi_1(t), \dots, \phi_4(t)\) from Eq.~\eqref{eq:overlap-00} is shown, where \(q = 0.8\), and only the first 15 terms in the series are included. The plot displays the functional dependence on the rescaled time \(\tilde{t}\). The discrepancy between this finite series approximation and the numerical results obtained through direct integration \eqref{eq:phi-def} is illustrated in Fig.~\ref{fig:error}.}
    \label{fig:phi-plot}
\end{figure}

\begin{figure}
    \centering
    \includegraphics[width=0.8\linewidth]{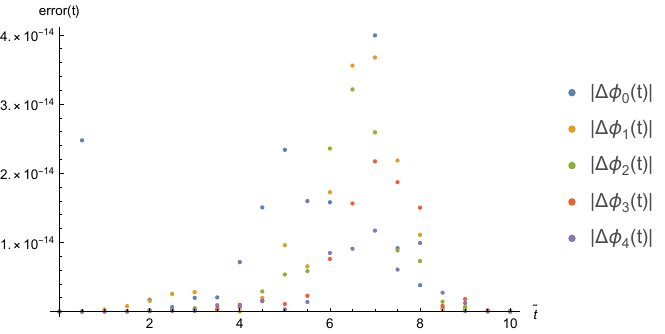}
    \caption{A plot of the discrepancy between the solutions \(\phi_0(t), \dots, \phi_4(t)\) with $q=0.8$, using only the first 15 terms in the series from Eq.~\eqref{eq:overlap-00}, and the numerical integral of Eq.~\eqref{eq:phi-def}, is shown as a function of the rescaled time \(\tilde{t}\). The values of \(\tilde{t}\) are sampled at \(0, 0.5, 1, \dots, 9.5, 10\), and the error is less than \(10^{-13}\). }
    \label{fig:error}
\end{figure}

The simplest approach to derive \eqref{eq:bb-map} is to utilize the energy eigenbasis \( |\te\ket \), which consists of eigenstates of the chord Hamiltonian \( H \):
\be
H|\te\ket=E\left(\te\right)|\te\ket,\quad E\left(\te\right)=\frac{2\cos\te}{\sqrt{1-q}}.
\ee
Through out this paper, we adopt the following normalization of $|\te\ket$:
\be
\bra\te_{1}|\te_{2}\ket=\frac{1}{\mu\left(\te_{1}\right)}\del\left(\te_{1}-\te_{2}\right),\quad\dd\mu\left(\te\right)=\frac{\left(e^{\pm2i\te},q;q\right)_{\infty}}{2\pi}\dd\te,\quad\mathbf{1}=\int_{0}^{\pi}\dd\mu\left(\te\right)|\te\ket\bra\te|.
\ee
We can then sandwich the two sides of \eqref{eq:bb-map} with $\bra\te|\cdot|\Omega\ket$, and use the following equations:
\be \label{eq:melements}
\begin{split}
   & \langle\theta| H^n|\Omega\rangle=\left(\frac{2 \cos \theta}{\sqrt{1-q}}\right)^n,\\
&\bra\te|\normord{H^{k}}|\Omega\ket=\bra\te|k\ket=\frac{H_{k}\left(\cos\te|q\right)}{\left(1-q\right)^{k/2}},
\end{split}
\ee
where in the first line we used the fact that $\bra\te|\Omega\ket=1$. It's then straightforward to verify \eqref{eq:bb-map} combined with the following identity ~\cite{Berkooz:2018jqr} that expresses $\cos(\theta)$ in terms of $q$-Hermite polynomial:
\be \label{eq:cos-expr}
\left(2\cos\te\right)^{n}=\sum_{k=0}^{[\frac{n}{2}]}c_{n,k}H_{n-2k}\left(\cos\te|q\right).
\ee
The map \eqref{eq:bb-map} has been shown to be equivalent to the Schmidt diagonalization procedure of the Krylov basis \cite{Rabinovici:2023yex}, and the time evolution \( e^{-iHt} \) thus describes the spreading within the Krylov (sub-)space. In this context, we can apply the operator-state correspondence \footnote{There is a reason for the existence of such an operator-state correspondence for chord operators. See \cite{Xu:2024hoc} for a perspective from Von Neumann algebra.} \eqref{eq:os-corres} to map this spreading onto a spreading among the normal-ordered operator basis: 
\be \label{eq:opbasis-0}
e^{-iHt}=\sum_{n=0}^{\infty}\frac{\left(-it\right)^{n}}{n!}H^{n}=\sum_{n=0}^{\infty}\frac{\phi_{n}\left(t\right)}{(1-q)^{n/2}[n]_{q}!}\normord{H^{n}},
\ee
where the normalization is chosen such that the spreading function $\phi_n (t)$ is defined via:
\be \label{eq:phi-def}
\phi_{n}\left(t\right)\equiv\int_{0}^{\pi}\dd\mu\left(\te\right)e^{-iE\left(\te\right)t}H_{n}\left(\cos\te|q\right).
\ee
In terms of the evolution of the chord state \( |\Omega(t)\rangle = e^{-iHt}|\Omega\rangle \), \( \phi_n(t) \) represents the projection of \( |\Omega(t)\rangle \) onto the fixed-length state:
\be
\phi_{n}\left(t\right)=(1-q)^{n/2}\bra n|\Omega\left(t\right)\ket.
\ee
We present a detailed derivation for an analytic expression of $\phi_n(t)$s valid for all parameter regime $q$ and $t$ in \ref{app:Overlap}. The results can be expressed as:
\be \label{eq:overlap-00}
\begin{aligned}
& \phi_0(t)=\sum_{k \in \mathbb{Z}} q^{\binom{k}{2}} J_{2 k}(2 \tilde{t}), \\
& \phi_n(t)=(-i)^n \sum_{k=0}^{\infty} q^{\binom{k}{2}} \frac{(q ; q)_{n+k}}{(q ; q)_k} \frac{1-q^{n+2 k}}{1-q^{n+k}} J_{n+2 k}(2 \tilde{t}).
\end{aligned}
\ee
Here, the argument in the Bessel functions is a rescaled time, \( \tilde{t} \equiv t / \sqrt{1 - q} \). For general \( q = e^{-\lambda} \in [0, 1) \), the series converges rapidly, as it decays like a Gaussian distribution for large \( k \). This rapid decay allows for efficient numerical exploration of relevant quantities at general values of \( q \in [0, 1) \), as high accuracy can be achieved by retaining only a few terms in the series expansion \eqref{eq:overlap-00}. Figure \ref{fig:phi-plot} shows a plot of \( \phi_0(t), \ldots, \phi_4(t) \) for \( q = 0.8 \) with the first 15 terms in the series \eqref{eq:overlap-00} included, and Figure \ref{fig:error} displays the corresponding error sampled over a discrete set of time points.

\begin{figure}
    \centering
    \includegraphics[width=0.8\linewidth]{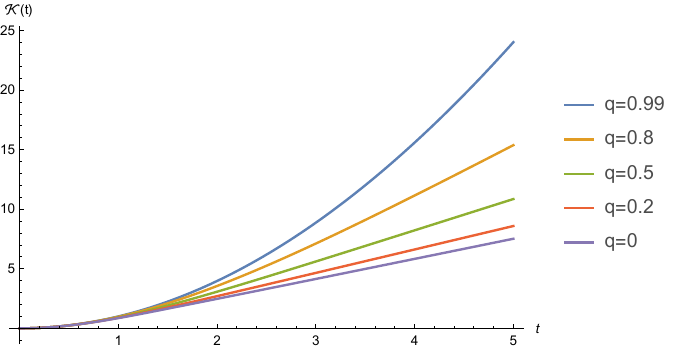}
    \caption{The plot illustrates the Krylov complexity, $\mathcal{K}(t)$ as a function of time $t$ for various values of $q$. We truncate the sum in Eq.~\eqref{eq:K-def} to the first 50 terms, with \(q\) ranging over \(\{0, 0.2, 0.5, 0.8, 0.99\}\). The results show that the growth of complexity becomes faster as $q$ approaches 1.
}
    \label{fig:K-growth}
\end{figure}

The result indeed shows that the distribution begins as localized on the empty chord state \( |\Omega\rangle \) and subsequently spreads to states with larger chord numbers. For a given \( n > 0 \), there exists a characteristic time \( t_n > 0 \) at which \( |\phi_n(t_n)| \) reaches its maximum. Beyond this point, the wavefunction starts to decay, indicating that the wave packet continues to move toward states with a greater number of chords. \cite{Rabinovici:2023yex} identified this chord spreading with the spreading among the Krylov basis. If we introduce a normalized basis \( \{|n)\} \):
\be
|n)=([n]_{q}!)^{-1/2}|n\ket,
\ee
One can show that the action of \eqref{eq:H-chord} can be put into the standard form in Krylov basis:
\be
\begin{aligned}
H|n) & \left.\left.=b_{n+1}  |n+1\right)+b_n |n-1\right), \\
b_n & =\sqrt{\frac{1-q^n}{1-q}}.
\end{aligned}
\ee
Therefore, the expectation value of chord number $\hat{n}$ in state $|\Omega(t)\ket$ becomes the value of Krylov complexity:
\be
\mathcal{K}\left(t\right)	=\bra\Omega\left(t\right)|\hat{n}|\Omega\left(t\right)\ket.
\ee
It is straightforward to relate the above result to the overlap functions \( \{\phi_n(t)\} \):
\be
(n|\Omega\left(t\right))	=\int_{0}^{\pi}\dd\mu\left(\te\right)(n|\te)(\te|e^{-iHt}|\Omega)=\frac{1}{\sqrt{\left(q;q\right)_{n}}}\phi_{n}\left(t\right),
\ee
where we have identified the energy eigenstates $|\te)\equiv|\te\ket$. It's easy to derive that:
\be \label{eq:K-def}
\mk\left(t\right)=\sum_{n=0}^{\infty}\frac{n}{\left(q;q\right)_{n}}|\phi_{n}\left(t\right)|^{2}.
\ee
We compare the growth of \( \mathcal{K}(t) \) for different values of \( q \) and present illustrative results in Figure \ref{fig:K-growth}. It is shown that as \( q \) approaches \( 1 \), the growth becomes more rapid.

The representation \eqref{eq:overlap-00} is applicable for any finite \( q \in [0,1) \). Specifically, for \( q = 0 \), it truncates to
\be
\phi_{n}(t) = (-i)^{n} \left( J_{n}(2t) + J_{n+2}(2t) \right) = (-i)^{n} \frac{(n+1)J_{n+1}(2t)}{t},
\ee
which aligns with the random matrix limit of this model as studied in \cite{Tang:2024xgg}. Note that the time dependence is consistently associated with the rescaling \( \tilde{t} = t/\sqrt{1-q} \), which prevents the expression from describing the \( q \to 1 \) limit. In this limit, intriguing physics may emerge, as one might expect the Krylov complexity \eqref{eq:K-def} to grow more rapidly compared to any other finite \( q \) cases.

Another interesting question concerns how the wavefunctions \eqref{eq:overlap-00} decay for very late times. We will first address this for general \( q \in [0,1) \), demonstrating a universal late-time polynomial decay of \( \sim t^{-3/2} \), which is independent of \( q \). We then investigate the detailed behavior of chord dynamics across various relevant regimes in the energy spectrum as \( q \to 1 \).

\subsection{Late time behavior for all $q\in [0,1)$}
While the early-time behavior of chord spreading can be determined by expanding the evolution operator and evaluating the contributions at each order in $t$ as moments of the Hamiltonian $H$, the late-time behavior becomes more intriguing due to the $q$-deformed nature of the solutions \eqref{eq:overlap-00}. In this section, we demonstrate that for a general $q \in [0,1)$ and for $t \gg \sqrt{1-q}\lambda^{-1}$, the wavefunctions decay polynomially over time as $\sim t^{-3/2}$.

We start with the evolution of empty chord state, where
\be \label{eq:overlap-00-late}
\phi_{0}\left(t\right)=\sum_{n\in\mbz} q^{\binom{n}{2}}J_{2n}\left(2\tilde{t}\right),
\ee
for $\tilde{t}\gg n^2$\footnote{This is evident in the higher-order terms of the asymptotic expansion of \(J_n(\tilde{t})\).}, we would like to use the asymptotic behavior of Bessel functions:
\be \label{eq:J-asymptotic}
J_{2n}\left(2\tilde{t}\right)=\left(-1\right)^{n}\sqrt{\frac{1}{\pi\tilde{t}}}\cos\left(2\tilde{t}-\frac{\pi}{4}\right)+\left(-1\right)^{n-1}(n^{2}-\frac{1}{16})\sqrt{\frac{1}{\pi\tilde{t}^{3}}}\sin\left(2\tilde{t}-\frac{\pi}{4}\right)+O\left(t^{-5/2}\right),
\ee
Plugging back into \eqref{eq:overlap-00-late}, we find that the coeffcient of the leading order term at $O(\tilde{t}^{-1/2})$ vanishes:
\be \label{eq:vanishing-1}
\sum_{n=-\infty}^{\infty}\left(-1\right)^{n}q^{\binom{n}{2}} =0,
\ee
which holds for all $q\in [0,1)$. Introducing 
\be
\vartheta=\sum_{n\in\mbz}\left(-1\right)^{n-1}q^{\binom{n}{2}}n^{2},
\ee
we thus declare that in the regime where  \eqref{eq:J-asymptotic} holds, we have
\be \label{eq:phi0-approx}
\phi_{0}\left(t\right)=\frac{\vartheta}{\sqrt{\pi}}\tilde{t}^{-3/2}+O\left(\tilde{t}^{-5/2}\right).
\ee
Let’s now examine condition \eqref{eq:J-asymptotic} more closely. For the asymptotic expansion to be valid, it is required that the time parameter $\tilde{t}$ be significantly larger than the index, i.e., $\tilde{t} \gg n$. In applying the vanishing condition \eqref{eq:vanishing-1}, we assumed this asymptotic relationship holds for all $n$. Consequently, errors will arise from terms where $n$ is comparable to or larger than $|\tilde{t}|^{1/2}$. However, due to the factor $q^{\binom{n}{2}} = \exp(-\lambda n(n-1)/2)$, these large $n$ terms are exponentially suppressed as long as $n \gg \lambda^{-1/2}$. Combining these conditions, we find that for the approximation \eqref{eq:phi0-approx} to be valid, it is necessary to have $\tilde{t} \gg \lambda^{-1}$, or equivalently $t \gg \sqrt{1-q}\lambda^{-1}$.

One can move on to examine the late time behavior of $\phi_n (t)$ in the same regime as we discussed above. Due to the following identity:
\be \label{eq:vanishing-2}
\sum_{k=0}^{\infty}\left(-1\right)^{k}q^{\binom{k}{2}}\frac{\left(q;q\right)_{n+k}}{\left(q;q\right)_{k}}\frac{1-q^{n+2k}}{1-q^{n+k}}=0,\quad q\in[0,1).
\ee
one can directly verify that all the wavefunctions $|\phi_n (t)|$ decay as $\simeq t^{-3/2}$ for $t\gg \sqrt{1-q}/\lambda$, for $q\in [0,1)$. 

In the above discussion, we did not include the case where $q \to 1$ and $\lambda \to 0$, as in this scenario, the corresponding time regime extends infinitely late, rendering the identities \eqref{eq:vanishing-1} and \eqref{eq:vanishing-2}—which we use to show the decay behavior—ambiguous. In fact, by zooming in to different portions of the spectrum as $\lambda \to 0$, the functional time dependence of the wavefunctions varies significantly. Below, we present two examples of such limits: the triple-scaled high-temperature regime studied in \cite{Almheiri:2024xtw} and the triple-scaled low-temperature regime examined in \cite{Lin_2023}. We then state the late-time behavior of the limiting wavefunctions as valid within each corresponding regime.

\subsection{Triple-Scaled High Temperature Regime around the Spectrum Center}
We investigate the analytic properties of the spreading wavefunctions $\{\phi_n(t)\}$ within the triple-scaled high-temperature limit analyzed in \cite{Almheiri:2024xtw}. This limit is characterized by a rescaling of the inverse temperature, $\beta \approx \lambda \tilde{\beta}$ with $\tilde{\beta}$ fixed, while maintaining $\lambda n \approx \lambda \tilde{\beta}^2$ as $\lambda \to 0$. In this regime, the model simplifies significantly, as the theory becomes dominated by the center of the energy spectrum, which follows a Gaussian distribution. Consequently, the associated wavefunctions exhibit Gaussian decay over time. In this section, we delve into this aspect in reverse order: we focus on the near-center region of the energy spectrum while taking the $\lambda \to 0$ limit. We explicitly demonstrate that the asymptotic behavior of our analytic solutions, given in \eqref{eq:overlap-00}, reproduces the Gaussian decay observed in \cite{Almheiri:2024xtw}.

Note that as $\lambda \to 0$, the $q$-Hermite polynomial simplifies as follows:
\be
H_{n}\left(\cos\theta \,|\, q\right) = \left(2\cos\theta\right)^{n} + O\left(\lambda\right).
\ee
Therefore, from the definition of $\phi_{n}(t)$, we deduce:
\be \label{eq:phin-in-phi0}
\lim_{\lambda \to 0} \phi_{n}\left(t\right) = \lim_{\lambda \to 0} \left(i\partial_{\tilde{t}}\right)^{n} \phi_{0}\left(t\right),
\ee
which indicates that we essentially only need to examine the functional dependence of $\phi_0(t)$ in $t$. To manifest this time dependence, we first rewrite the Bessel function $J_{2n}(2\tilde{t})$ as:
\be
\begin{aligned}
J_{2 n}(2 \tilde{t}) &= \frac{(-1)^n}{\pi} \int_0^\pi e^{2 i \tilde{t} \cos \theta} \cos(2 n \theta) \, \mathrm{d} \theta \\
&= (-1)^n \int_0^\pi \frac{\mathrm{d} \theta}{2 \pi} e^{2 i \tilde{t} \cos \theta} \left(e^{2 i n \theta} + e^{-2 i n \theta}\right).
\end{aligned}
\ee
Commuting the infinite sum with the integral, and introducing the $q$-Theta function:
\be
\sum_{n\in\mathbb{Z}} (-1)^{n} q^{\frac{n^{2}}{2}-\frac{n}{2}} e^{2in\theta} = \sum_{n\in\mathbb{Z}} q^{\frac{n^{2}}{2}} x^{n} = \Theta_{q}\left(x\right), \quad \ln x = 2i\theta - i\pi + \frac{\lambda}{2},
\ee
where $\theta \in [0, \pi]$ and the principal branch of the logarithmic function are used in defining $x$. We can now express $\phi_0(t)$ as:
\be
\phi_{0}\left(t\right) = \int_{0}^{\pi} \frac{\mathrm{d}\theta}{2\pi} e^{2i\tilde{t}\cos\theta} \Theta_{q}\left(x\right).
\ee
To analyze the $\lambda \to 0$ behavior, we apply the following asymptotic expansion of  $\Theta_q(x)$ at small $\lambda$ (or $q\to 1^-$) as:
\be \label{eq:phi0-expanded}
\Theta_{q}\left(x\right) \stackrel{\lambda \to 0}{\simeq} \sqrt{\frac{\pi}{\lambda}} \exp\left(\frac{\left(2i\theta - i\pi + \lambda/2\right)^{2}}{4\lambda}\right),
\ee
where we have neglected $O(1)$ contributions in the exponent. Substituting this into \eqref{eq:phi0-expanded}, we find:
\be
\begin{aligned}
\phi_0 (t) & \stackrel{\lambda \rightarrow 0}{\simeq} \int_0^\pi \frac{\mathrm{d} \theta}{\sqrt{4\pi \lambda}} e^{2 i \tilde{t} \cos \theta} e^{(2 i \theta - i \pi + \lambda / 2)^2 / 4 \lambda} \\
&= \int_{-\frac{\pi}{2 \sqrt{\lambda}}}^{\frac{\pi}{2 \sqrt{\lambda}}} \frac{\mathrm{d} s}{\sqrt{4\pi}} \exp \left(-2 i \tilde{t} \sin \sqrt{\lambda} s - s^2\right),
\end{aligned}
\ee
where we have used the parameterization $\theta = \frac{\pi}{2} + \sqrt{\lambda} s$ to capture the energy fluctuations near the center of the spectrum. Thus, in the $\lambda \to 0$ limit, the asymptotic behavior of $\phi_0(t)$ is:
\be
\phi_{0}(t) \stackrel{\lambda \to 0}{\simeq} \int_{-\infty}^{\infty} \frac{\mathrm{d} s}{\sqrt{4\pi}} \exp\left(-2its - s^2\right) = \frac{1}{2} \exp\left(-t^{2}\right).
\ee
where we have approximated $\sin\sqrt{\lambda}t\simeq \sqrt{\lambda}t $ and $\tilde{t}\sqrt{\lambda}\simeq t$.  Applying \eqref{eq:phin-in-phi0}, we similarly find:
\be \label{eq:phi-Gaussian}
\phi_{n}(t) \stackrel{\lambda \to 0}{\simeq} \frac{\left(2it\right)^{n} }{2} \exp\left(-t^{2}\right), \quad n = 0, 1, 2, \dots
\ee
The final result can be viewed as the evolution of coherent states associated with the Heisenberg-Weyl algebra studied in \cite{caputa2021geometrykrylovcomplexity}, and aligns with the results of \cite{Rabinovici:2023yex} up to a normalization factor of the Hamiltonian.

\subsection{Triple-Scaled Low Temperature Regime Near the Spectrum Edge}

As shown in \cite{Lin_2023}, zooming in on the edge of the energy spectrum while taking the $\lambda \to 0$ limit reveals that the theory approaches the low-energy, low-temperature limit of conventional SYK. In the context of the chord Hilbert space, this involves keeping
\be
l = \lambda n - 2 \log \lambda,
\ee
fixed as $\lambda \to 0$ and $n \to \infty$, while considering $O(\lambda)$ energy fluctuations above the ground state.\footnote{Here, we define the Hamiltonian as $-H$, and zooming in on the left edge $\theta = 0$ of the spectrum. Since the theory is invariant under $H \to -H$, we could equivalently define $H$ as the Hamiltonian and consider the right edge $\theta = \pi$.} This implies:
\be
\theta = \lambda s, \quad E(s) = -\frac{2 \cos \lambda s}{\sqrt{1 - q}} = E_{0} + \lambda s^{2} + O(\lambda^{3}),
\ee
suggesting that at very late times, contributions to the overlap functions $\{\phi_n(t)\}$ in this regime are dominated by low-energy states. The polynomial decay behavior \eqref{eq:phi0-approx} at late times thus remains valid as $\lambda \to 0$ \cite{Mertens:2022irh}. In this section, we confirm this by taking the appropriate limit of our analytic solutions.

Following \cite{Lin:2022rbf}, we adopt the limiting behavior:
\be \label{eq:t-replace}
\begin{aligned}
\mathrm{d} \mu(\theta) & \rightarrow \rho(s) \mathrm{d} s = s \sinh 2\pi s \, \mathrm{d} s, \\
H_n(\cos \theta \,|\, q) & \rightarrow K_{2 i s}\left(2 e^{-l / 2}\right), \\
e^{-i E t} & \rightarrow e^{-i E_0 t} \, e^{-i \lambda s^2 t / 2},
\end{aligned}
\ee
and reformulate $\phi_n(t)$ as:
\be
\begin{aligned}
\phi_n(t) &= \int_0^\pi \mathrm{d} \mu(\theta) \, H_n(\cos \theta \,|\, q) \, e^{-i E(\theta) t} \\
& \stackrel{\lambda \rightarrow 0}{\simeq} \int_0^{\infty} \rho(s) \, \mathrm{d} s \, \left(2 K_{2 i s}\left(e^{-l / 2}\right)\right) \, e^{-i s^2 \lambda t / 2},
\end{aligned}
\ee
where, in the last step, we implemented the triple-scaling limit by substituting \eqref{eq:t-replace}, while neglecting the divergent phase factor $e^{-i E_0 t}$ and other $\lambda$-dependent overall coefficients. 

Using the integral representation of the Bessel function \cite{Saad:2019pqd}:
\be
K_{2is}\left(e^{-l/2}\right) = \frac{1}{2} \int_{-\infty}^{\infty} \mathrm{d}x \, e^{-e^{-l/2}\cosh x} \, e^{-2isx},
\ee
we can interchange the order of integration. Integrating over $s$ first, we obtain:
\be\label{eq:t-decay-2}
\int_{0}^{\infty} \rho(s) \, e^{-is^{2}\lambda t / 2 - 2isx} \, \mathrm{d} s \simeq \frac{e^{-i (2\pi - ix)^2 / 4\lambda t}}{t^{3/2}}.
\ee
The remaining integral over $x$ captures the functional dependence in $l$, which depends on the precise structure of the limiting theory. In deriving \eqref{eq:t-decay-2}, we have assumed $t \gg \lambda^{-1}$ so that contributions are dominated by $s$ near $0$. This allows us to approximate $\rho(s) \sim s^3$, resulting in a $t^{-3/2}$ decay. The result differs significantly from \eqref{eq:phi-Gaussian}, suggesting that exploring different regions of the energy spectrum in the $\lambda \to 0$ limit may reveal completely distinct semi-classical theories.

\section{Complexity Growth of Chords and Isometric Factorization} \label{sec:complexity}
In this section, we extend the discussion of chord dynamics from the previous section to include the case with a matter chord \( M \). An \( M \)-chord is a new type of chord that does not commute with \( H \), and incurs a penalty factor \( q^\Delta \) for crossing an \( H \)-chord, where \( \Delta \geq 0 \). We keep \( \Delta \) general for most of the following discussion.

We begin by mapping the Liouvillian time evolution of a single matter chord $M$ to the corresponding time evolution of a one-particle state governed by the two-sided chord Hamiltonian \( H_L - H_R \). This is done using an extended version of the bulk-to-boundary map \eqref{eq:bb-map} that includes a matter chord. Next, we develop an isometric factorization map \cite{Okuyama:2024yya, Xu:2024hoc}, which provides an equivalent description of the one-particle chord Hilbert space in terms of a doubled zero-particle chord Hilbert space. Using this map, we then present various analytic results for relevant quantities. In particular, we evaluate the chord-number generating function in the spirit of \cite{Qi_2019} and establish its equivalence to the crossed four-point function studied in \cite{Berkooz:2018jqr}.

\subsection{Liouvillian Operator and Krylov Operator Basis}
 Let $\mo$ be an operator acting on a Hilbert space $\mh_0$. One can associate $\mo$ with a corresponding state $|\mo\rangle$ in the doubled Hilbert space $\mh_0 \otimes \mh_0$, defined as
\begin{equation} \label{eq:operator-state}
|\mo\rangle = \sum_{n,m} \langle n|\mo|m \rangle \, |n\rangle \otimes |m\rangle,
\end{equation}
where $|n\rangle$ and $|m\rangle$ form a complete orthonormal basis of $\mh_0$. The Heisenberg time evolution of $\mo$ can be formulated in terms of the Liouvillian operator $\ml$, which acts on $\mo$ via
\begin{equation} \label{eq:Heisenberg}
\ml(\mo) \equiv i [H, \mo],
\end{equation}
with $H$ denoting the Hamiltonian of the system. Accordingly, the time-evolved state $|\mo(t)\rangle$ in the doubled Hilbert space, corresponding to the evolved operator $\mo(t) = e^{i H t} \mo \, e^{-i H t}$, can be expressed as
\begin{equation}
|\mo(t)\rangle = e^{\ml t} |\mo\rangle = \sum_{n=0}^{\infty} \frac{(\ml t)^n}{n!} |\mo\rangle.
\end{equation}
In the double-scaled SYK model with both $H$-chord and $M$-chord operators, it is unnecessary to duplicate the chord Hilbert space $\mh$ \footnote{The space is defined as the span of states with arbitrary configurations of various types of chords. In the case where the theory includes only $H$-chords and $M$-chords, this space can be described as:
\begin{equation}
\mh = \text{span}\{|n_0, \dots, n_k\rangle\},
\end{equation}
where the inner product is determined according to the relevant chord statistics. For details, see \cite{Xu:2024hoc}.
}when examining operator growth. This is because the one-sided double-scaled algebra $\mathcal{A}_{L/R}$, generated by two types of chord operators acting from either the left or right, naturally differentiates between the two sides of the system. Consequently, a state in $\mathcal{H}$ can be regarded as an infinitely entangled state connecting both sides. This setup prevents the chord Hilbert space $\mh$ from factorizing into boundary Hilbert spaces $\mh_L \otimes \mh_R$ such that $\mathcal{A}_{L/R} = \mathcal{B}(\mh_{L/R})$. Instead, the notions of ``left'' and ``right'' emerge from the respective double-scaled algebras $\mathcal{A}_L$ and $\mathcal{A}_R$, which consist of chord operators that act from the left and right, respectively, and mutually commute with each other.

In this framework, the analogue of \eqref{eq:operator-state} is an extended version of operator-state correspondence in \eqref{eq:bb-map}, which establishes a one-to-one mapping between a chord operator in $\mathcal{A}_L$ or $\mathcal{A}_R$ and a state in $\mh$, owing to the cyclic property of the empty chord state $|\Omega\rangle$ for both $\mathcal{A}_L$ and $\mathcal{A}_R$. Specifically, a chord operator $\hat{M}_L \in \mathcal{A}_L$, which either adds or removes a matter chord with weight $\Delta$ from the left, corresponds to the one-particle state $|\Delta; 0, 0\rangle$ \cite{Lin_2023}. Namely, we have:
\be
\hat{M}_L|\Omega\ket = |\Delta;0,0\ket. 
\ee
A general normal ordering prescription for a string of chord operators has been established in \cite{Xu:2024hoc}. For states with a one-particle insertion, this prescription yields:
\begin{equation} \label{eq:bb-map-1}
\normord{H^{n_L} \hat{M} H^{n_R}} |\Omega\ket = |\Delta; n_L, n_R\rangle.
\end{equation}
Here, \( |\Delta; n_L, n_R \rangle \) denotes a state with one \( M \)-chord inserted, along with \( n_L \) \( H \)-chords to its left and \( n_R \) \( H \)-chords to its right respectively. We then introduce the Liouvillian operator, following the spirit of the Heisenberg evolution in \eqref{eq:Heisenberg}. This operator acts on both sides and can be expressed in terms of the left and right chord Hamiltonians as:
\begin{equation} \label{eq:Liouvillian}
\ml = i (H_L - H_R).
\end{equation}
The time evolution of $\hat{M}_L$ generated by $\ml$ generally produces a two-sided operator:
\begin{equation} \label{eq:opbasis-1}
\hat{M}_L(t) = e^{\ml t} \hat{M}_L = \sum_{n=0}^{\infty} \frac{t^n}{n!} \ml^n \hat{M}_L = \sum_{n=0}^{\infty} \frac{i^n}{n!} \sum_{k=0}^{n} (-1)^k \binom{n}{k} H_L^k \hat{M}_L H_R^{n-k}.
\end{equation}
The time evolution of the state obtained by acting $\hat{M}_L(t)$ on the empty state can be expressed as
\begin{equation} \label{eq:time-evolution-1}
|\Delta(t)\rangle \equiv e^{\ml t} \hat{M}_L |\Omega\rangle = e^{i H_L t} \hat{M}_L e^{-i H_L t} |\Omega\rangle,
\end{equation}
where in the second equality we have used the fact that $(H_L - H_R) |\Omega\rangle = 0$. This shows that the action of $e^{\ml t} \hat{M}_L$ on the empty state is equivalent to the standard Heisenberg evolution of $\hat{M}_L$ under the one-sided Hamiltonian $H_L$.

The time evolution of \( \hat{M} \) spreads within the Krylov basis, which spans a subspace of the one-particle Hilbert space \( \mathcal{H}_1 \), since the number of matter chords is conserved during evolution. The Krylov basis can be constructed through the standard procedure: starting with \( |O_0\rangle \equiv |\Delta;0,0\rangle \) and \( |O_1\rangle = \frac{1}{\sqrt{2 + 2q^\Delta}} \left( |\Delta;1,0\rangle - |\Delta;0,1\rangle \right) \), we recursively define \( |O_n\rangle \) for \( n = 2,3, \dots \) using the Lanczos algorithm:
\be
\begin{aligned}
\mathcal{L}|O_{n}\rangle &= |A_{n+1}\rangle + b_{n}|O_{n-1}\rangle, \\
b_{n+1} &= \langle A_{n+1} | A_{n+1} \rangle, \quad b_1 = \sqrt{2 + 2q^\Delta}, \\
|O_{n+1}\rangle &= b_{n+1}^{-1} |A_{n+1}\rangle.
\end{aligned}
\ee
The algorithm yields a set of orthonormal basis operators \( \{ |O_n \ket \} \) that satisfy \( \bra O_n | O_m\ket = \delta_{nm} \), along with a sequence of real, positive coefficients \( \{ b_n \} \), known as the Lanczos coefficients.  The large \( n \) behavior of \( b_n \) has recently been utilized to detect chaotic behavior in various systems, see \cite{Rabinovici:2023yex, Tang:2023ocr, Hashimoto_2023, Gorsky_2020} for example. The functional dependence of $n$ is determined by the moments \( \langle \Delta; 0,0 | \mathcal{L}^n | \Delta; 0,0 \rangle \) of the Liouvillian operator, as outlined in the algorithm from \cite{Parker_2019}. In Appendix \ref{app:momentsL}, we present our results for these moment calculations, employing techniques developed in the following discussion.

The statement of operator-state correspondence \eqref{eq:bb-map-1} then says that for each $|O_n\ket\in\mh_1$, there exists a unique operator $\hat{\Psi}_n$, such that
\be 
|O_{n}\ket= \hat{\Psi}_{n}|\Omega\ket,
\ee
Therefore, similar to \eqref{eq:opbasis-0}, the two-sided time evolution of matter chord operator results in a probability distribution in the Krylov operator basis, specifically, we have:
\be\label{eq:opbasis-2}
\hat{M}\left(t\right)=e^{\ml t}\hat{M}=\sum_{n=0}^{\infty}\phi_{n}^{\Del}\left(t\right)\hat{\Psi}_{n},
\ee
and the operator Krylov complexity of $\hat{M}(t)$ is defined analogous to \eqref{eq:K-def} as:
\be
\mathcal{K}_{\Del}(t)=\sum_{n=0}^{\infty}n|\phi_{n}^{\Del}(t)|^{2}.
\ee
One might be tempted to identify the Krylov basis index \( n \) with the chord number, defined as
\be \label{eq:N-def}
\hat{N}|\Delta; n_L, n_R\rangle = (n_L + n_R) |\Delta; n_L, n_R\rangle,
\ee
as in the matter-free case. However, this identification does not hold in general. For arbitrary values of \( q \) and \( \Delta \), the Krylov basis elements \( \hat{\Psi}_n \) are not simply the normal-ordered versions of the operator basis defined in \eqref{eq:opbasis-1}. Instead, they include additional contributions arising from contractions involving the \( H \)-chords.

This distinction stems from the structure of the Hilbert space. In the zero-particle sector \( \mathcal{H}_0 \), the subspace at fixed chord number is one-dimensional and is therefore fully spanned by the normal-ordered monomial \( \normord{H^n} \). In contrast, when matter is present, this is no longer true. For instance, the normal-ordered version of the polynomial operator
\be
\hat{\Phi}_n = \sum_{k=0}^n (-1)^k \binom{n}{k} \normord{H^k \hat{M} H^{n-k}}
\ee
does not remain within the span of the original operator basis \( \{\Phi_i\} \).\footnote{For a more detailed analysis of the Krylov basis in the one-particle sector, we refer the reader to the recent discussion in \cite{2025KComplexity}.}

Nevertheless, following the spirit of \eqref{eq:K-def}, one can examine the time evolution of chord number operator defined as:
\be \label{eq:K-def-2}
N_\Delta(t) \equiv \langle \Delta(t) | \hat{N} | \Delta(t) \rangle,
\ee
and we will show in the next section that this quantity matches the two-sided wormhole length in JT gravity with conformal matter of weight \( \Delta \), in the semi-classical limit.

In the remainder of this section, we consider a slightly generalized version of \eqref{eq:opbasis-1}, incorporating general two-sided time evolution generated independently by \( H_L \) and \( H_R \):
\be \label{eq:time-evolve-r1}
|\Delta(t_L, t_R)\rangle = e^{i(H_L t_L + H_R t_R)} |\Delta; 0, 0\rangle.
\ee
We will study the time evolution of this state for arbitrary values of \( q \) and \( \Delta \) in the one-particle sector \( \mathcal{H}_1 \). The special case \( t_L = -t_R = t \) corresponds to Liouvillian evolution.  We also introduce the chord-number generating function, analogous to \cite{Qi_2019}, defined by
\be \label{eq:I-def0}
I_\Delta^\mu(t_L, t_R) = \langle \Delta(t_L, t_R) | e^{-\mu \hat{N}} | \Delta(t_L, t_R) \rangle.
\ee
We will explore the functional dependence of \( |\Delta(t_L, t_R)\rangle \) and \( I_\Delta^\mu \) on \( t_L \) and \( t_R \), and demonstrate how this is related to the crossed four-point function. In particular, we show how this is related to the \(6j\)-symbol of the quantum group \( U_{\sqrt{q}}(\mathfrak{su}(1,1)) \).

\subsection{Isometric Factorization of the One-particle Chord Space} \label{sec:factorize}
The evaluation of \eqref{eq:time-evolve-r1} and \eqref{eq:I-def0} is intrinsically challenging, as the inner product in \( \mathcal{H}_1 \) is defined to reflect the chord statistics. Consequently, the states \( |\Delta; k, n - k \rangle \), for \( k = 0, 1, \dots, n \), are not orthogonal, and their overlaps are generally given by complicated expressions involving \( q \) and \( q^\Delta \).

To facilitate this calculation, we revisit an equivalent description of the one-particle Hilbert space \( \mathcal{H}_1 \) by introducing a factorization map \( \mathcal{F}: \mathcal{H}_1 \to \mathcal{H}_0 \otimes \mathcal{H}_0 \), following \cite{Okuyama:2024yya}. We demonstrate that \( \mathcal{F} \) can be made into an isometry by appropriately modifying the Fock space inner product in \( \mathcal{H}_0 \otimes \mathcal{H}_0 \) with a suitable matter density.

In \cite{Xu:2024hoc} it was found that given a one-particle chord state $|\Delta;m,n\ket$ with given number left and right Hamiltonian, one can find a corresponding state in the doubled $0$-particle Hilbert space $\mh_0\otimes\mh_0$ as:
\be \label{eq:F-def1}
\mf|\Del;m,n\ket=\sum_{k=0}^{{\rm min}\left(m,n\right)}\frac{\left(-1\right)^{k}q^{\binom{k}{2}}q^{\Del k}}{\left(q;q\right)_{k}}\sqrt{\frac{\left(q;q\right)_{m}\left(q;q\right)_{n}}{\left(q;q\right)_{m-k}\left(q;q\right)_{n-k}}}|m-k\ket\otimes|n-k\ket,
\ee
We can thus define $\mf$ as a linear map between the two Hilbert spaces by linearly extending it on any linear superposition of the basis states.\footnote{The linear extension of $\mf$ is valid only if the states $\{|\Delta; m, n \rangle\}$ in $\mathcal{H}_1$ are linearly independent. This condition holds for general parameters $0 \leq q < 1$ and $\Delta > 0$, but fails when $\Delta = 0$. In the latter case, $\mf$ cannot be defined as a linear map in this way, as it maps linearly dependent states to linearly independent ones.  For more details, see relevant discussion in \cite{Lin_2023}.} We use the following simplified notation :
\be
|m,n)\equiv \frac{1}{\sqrt{[m]_q! [n]_q!}}|m\ket\otimes|n\ket,\quad(m_{L},m_{R}|n_{L},n_{R})=\del_{m_{L}n_{L}}\del_{m_{R}n_{R}},
\ee
which uses round bracket for states and the Fock space inner product in $\mh_0\otimes\mh_0$. We also introduce the left and right $q$-ladder operators as:
\be \label{eq:ladder}
\begin{aligned}
&a_{L}^{\da}|n_{L},n_{R})	=\sqrt{1-q^{n_{L}+1}}|n_{L}+1,n_{R}), \\
&a_{L}|n_{L},n_{R})	=\sqrt{1-q^{n_{L}}}|n_{L}-1,n_{R}),\\
&a_{R}^{\da}|n_{L},n_{R})	=\sqrt{1-q^{n_{R}+1}}|n_{L},n_{R}+1),\\
&a_{R}|n_{L},n_{R})	=\sqrt{1-q^{n_{R}}}|n_{L},n_{R}-1).
\end{aligned}
\ee
They satisfy the  $q$-commutation relation $[a_L,a^{\dagger}_L]_q=[a_L,a^{\dagger}_L]_q=1-q$. In particular, one can show
\be
a_{L}^{k}a_{R}^{l}|m,n)=\sqrt{\frac{\left(q;q\right)_{m}\left(q;q\right)_{n}}{\left(q;q\right)_{m-k}\left(q;q\right)_{n-l}}}|m-k,n-l).
\ee
Therefore, \eqref{eq:F-def1} can be repackaged as \cite{Okuyama:2024yya}:
\be
\mf|\Del;m,n\ket=\sum_{k=0}^{{\rm min}\left(m,n\right)}\frac{\left(-1\right)^{k}q^{\binom{k}{2}}q^{\Del k}}{\left(q;q\right)_{k}}a_{L}^{k}a_{R}^{k}|m,n)=\me^{-1}_{\Del}|m,n),
\ee
where the two-sided linear operator $\mathcal{E}_\Delta$ on $\mh_0 \otimes\mh_0$ is defined via:
\be
\begin{aligned}
& \mathcal{E}_{\Delta}^{-1}=\left(q^{\Delta} a_L a_R ; q\right)_{\infty}=\sum_{k=0}^{\infty} \frac{(-1)^k q^{\binom{k}{2}} q^{\Delta k}}{(q ; q)_k} a_L^k a_R^k,\\
& \mathcal{E}_{\Delta}=\frac{1}{\left(q^{\Delta} a_L a_R ; q\right)_{\infty}}=\sum_{k=0}^{\infty} \frac{q^{\Delta k}}{(q ; q)_k} a_L^k a_R^k,
\end{aligned}
\ee
The operator $\me_\Delta$ was introduced in the context of doubled-Hilbert space formalism in \cite{Okuyama:2024yya}, characterized by its correspondence to the state $|q^{\Delta \hat{n}})$ defined via \eqref{eq:operator-state}. Concretely, its conjugate creates the operator state from the doubled empty state:
\be
|q^{\Delta\hat{n}})\equiv\sum_{n=0}^{\infty}q^{\Del n}|n,n)=\sum_{n=0}^{\infty}\frac{q^{\Del n}}{\left(q;q\right)_{n}}a_{L}^{\da n}a_{R}^{\da n}|0,0)=\me_{\Del}^{\da}|0,0).
\ee
In our current context, it arranges the matrix component of $\mf$ between the Lin-Stanford states in $\mathcal{H}_1$ and the Fock space states in $\mathcal{H}_0 \otimes \mathcal{H}_0$ based on its own matrix component in terms of states in the doubled Hilbert space. We also introduce its conjugate that connects the bra states following similar spirit:
\be
\bra\Del;m,n|\mf^{\da}=(m,n|\me_{\Del}^{-1\da}.
\ee
We now introduces the energy basis $|\te_L,\te_R) \in \mh_0\ot\mh_0 $ which plays an important role in bridging the inner product of the two sides. The left and right Hamiltonian acting on $\mh_0\ot\mh_0$ are given by:
\be
h_{L}=\frac{a_{L}+a_{L}^{\da}}{\sqrt{1-q}},\quad h_{R}=\frac{a_{R}+a_{R}^{\da}}{\sqrt{1-q}}.
\ee 
Let $H_{L}$ and $H_R$ be the corresponding left and right chord Hamiltonian, it can be shown that they are connected by the factorization map as
\be \label{eq:Hh-relation}
\mf\circ H_{L/R}\circ\mf^{-1}=h_{L/R},
\ee
where $\circ$ means the composite of maps.  Therefore, they share the same eigenvalues and we denote the corresponding eigenstates for $h_L,h_R$ as $|\te_L,\te_R)$, which satisfy:
\be
h_{L/R}|\te_{L},\te_{R})=E_{L/R}\left(\te_{L/R}\right)|\te_{L},\te_{R}),\quad E_{L/R}\left(\te_{L/R}\right)=\frac{2\cos\te_{L/R}}{\sqrt{1-q}}.
\ee
We can now restate the relevant results from \cite{Xu:2024hoc} (see also \cite{Okuyama:2024gsn}) by noting that $\mf$ becomes an isometry when $\mathcal{H}_0 \otimes \mathcal{H}_0$ is endowed with the inner product defined by the measure $\mathrm{d} \mu_\Delta(\theta_1, \theta_2)$:
\be \label{eq:reformulation-1}
\bra\psi_{1}|\psi_{2}\ket=\int_{[0,\pi]^{2}}\dd\mu_{\Del}\left(\te_{L},\te_{R}\right)\bra\psi_{1}|\mf^{\da}|\te_{L},\te_{R})(\te_{L},\te_{R}|\mf|\psi_{2}\ket,\quad\forall\psi_{1},\psi_{2}\in\mh_{1},
\ee
where the two-sided measure incorporates the matter density of state:
\be \label{eq:measure-matter}
\dd\mu_{\Del}\left(\te_{L},\te_{R}\right)=\dd\mu\left(\te_{L}\right)\dd\mu\left(\te_{R}\right)\bra\te_{L}|q^{\Del\hat{n}}|\te_{R}\ket,
\ee
with the inner product involved in $\bra\te_L|q^{\hat{n}}|\te_R\ket$ defined in $\mh_0$, and can be expressed compactly as:
\be\label{eq:n-def}
\bra\te_{L}|q^{\Delta\hat{n}}|\te_{R}\ket=\frac{\left(q^{2\Delta};q\right)_{\infty}}{\left(q^{\Delta}e^{\pm i\te_{L}\pm i\te_{R}};q\right)_{\infty}}=N(q,\Delta)\frac{\Gamma_q (\Delta\pm i\te_L/\lambda \pm i\te_R/\lambda)}{\Gamma_q (2\Delta)},
\ee
In the second equality, we employ an expression reminiscent of the matter density in JT gravity as discussed in \cite{Mertens:2022irh}. Here, \( N(q,\Delta) \) is a normalization factor depending solely on \( q \) and \( \Delta \). The functions \( \Gamma_q \) represent \( q \)-deformed versions of the ordinary Gamma function, which reduce to the usual Gamma function in the \( q \to 1 \) limit. Therefore, in the doubled Hilbert space \( \mathcal{H}_0 \otimes \mathcal{H}_0 \), the matter chord functions as an entangler, creating correlations between the energy spectra of the two sides. This perspective has been previously discussed in \cite{Okuyama:2024yya, Okuyama:2024gsn}, along with various efforts to derive \eqref{eq:n-def} from various putative dual theories of DSSYK. For a detailed discussion on the correspondence with spectral density and matter correlation functions in such theories, we refer the reader to \cite{Verlinde:2024zrh, mukhametzhanov2023largepsykchord}.
  
A particular interesting case of \eqref{eq:reformulation-1} is the overlap between the states \( \{|\Delta; n_L, n_R\rangle\} \) with fixed $H$-chord number. With the help of \eqref{eq:measure-matter}, they can be evaluated as:
\begin{equation} \label{eq:reformulate-2}
\langle \Delta; m_{L}, m_{R} | \Delta; n_{L}, n_{R} \rangle = \int_{[0, \pi]^2} d\mu_{\Delta}(\theta_{L}, \theta_{R}) \, (m_{L}, m_{R} | \mathcal{E}_{\Delta}^{-1\dagger} | \theta_{L}, \theta_{R}) \, (\theta_{L}, \theta_{R} | \mathcal{E}^{-1}_{\Delta} | n_{L}, n_{R}),
\end{equation}
where the integral is performed over the range \((\te_L,\te_R)\in [0, \pi]^2\).

The essence of \eqref{eq:reformulate-2} lies in that it reformulates the Lin-Stanford inner product, originally defined through \( q \)-combinatorics as a weighted sum over chord configurations,  now as the inner product between states in the factorized Hilbert space \( \mathcal{H}_0 \otimes \mathcal{H}_0 \), with the matter insertion creating energy correlations between the two sides. We adopt this framework in evaluation of various quantities in the following discussion.

\subsection{Operator Spreading of Chords}
\begin{figure}
    \centering
    \includegraphics[width=0.8\linewidth]{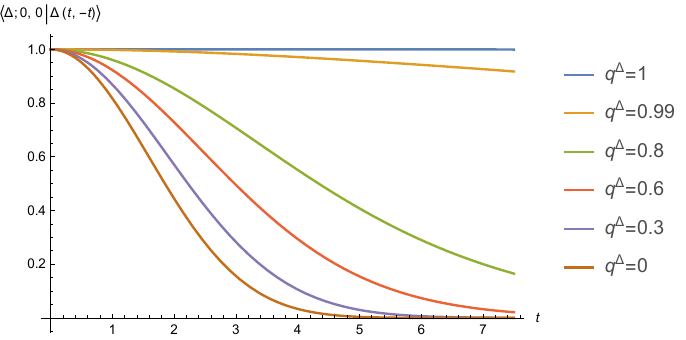}
    \caption{The figure shows the time dependence of $\bra\Delta;0,0|\Delta(t,-t)\ket$, with $q=0.8$ and different values of $\Delta$ as a signature of operator spreading. The decay becomes faster with the increase of matter weight $\Del$. In particular, in the no-particle case with $\Delta=0$ the result stays a constant for all time. The case with $\Delta=\infty$ shows fastest decaying. }
    \label{fig:Liouvillian-1}
\end{figure}

We examine the spreading of the one-particle state $|\Delta;0,0\ket$. As a quantification, we shall calculate the overlap $\bra\Delta;m_L,m_R|\Delta(t_L,t_R)\ket$ where the two-sided time evolution is defined in \eqref{eq:time-evolve-r1}.   We adopt the factorization formulation introduced in the last section, and start with the overlap $\bra\Delta;0,0|\Delta(t_L,t_R)\ket$ as:
\be \label{eq:overlap-r1-1}
\bra\Del;0,0|\Del\left(t_{L},t_{R}\right)\ket=\int_{[0,\pi]^{2}}\dd\mu_{\Del}\left(\te_{L},\te_{R}\right)\bra\Del;0,0|\mf^{\da}|\te_{L},\te_{R})(\te_{L},\te_{R}|\mf|\Del\left(t_{L},t_{R}\right)\ket,
\ee 
With the help of \eqref{eq:Hh-relation}, $\mathcal{F}|\Delta(t_L,t_R)\ket$ can be simplified as:
\be
\mf|\Del\left(t_{L},t_{R}\right)\ket=\mf e^{i(H_{L}t_{L}+H_{R}t_{R})}|\Del;0,0\ket=e^{i(h_{L}t_{L}+h_{R}t_{R})}|0,0),
\ee 
where we have used the fact that $\me^{-1}_\Delta |0,0)=|0,0)$. We conclude that the overlap between $|\te_1,\te_2)$ and $\mf |\Delta(t_L,t_R)\ket$ evaluates to a phase factor:
\be \label{eq:phase}
(\te_{L},\te_{R}|\mf|\Del\left(t_{L},t_{R}\right)\ket=e^{i(E_{L}t_{L}+E_{R}t_{R})}.
\ee
Plugging back into \eqref{eq:overlap-r1-1}, we obtain
\be\label{eq:overlap-r1-2}
\bra\Del;0,0|\Del\left(t_{L},t_{R}\right)\ket=\int_{0}^{\pi}\dd\mu\left(\te_{L}\right)\dd\mu\left(\te_{R}\right)\bra\te_{L}|q^{\Delta\hat{n}}|\te_{R}\ket\exp\left(iE_{L}t_{L}+iE_{R}t_{R}\right),
\ee 
We evaluate the integral above by first expressing the matter density in terms of $0$-particle energy wavefunction as:
\be
\bra\te_{L}|q^{\Delta\hat{n}}|\te_{R}\ket=\sum_{k=0}^{\infty}q^{\Del k}\bra\te_{L}|k\ket\bra k|\te_{R}\ket,
\ee
Combined with the fact that:
\be
\bra\Omega|\te_{L}\ket=\bra\te_{R}|\Omega\ket=1,
\ee 
The integrand of \eqref{eq:overlap-r1-2} can be re-written as:
\be
\begin{aligned}
\left\langle\Delta ; 0,0 | \Delta\left(t_L, t_R\right)\right\rangle & =\sum_{k=0}^{\infty} q^{\Delta k}\left[\left(\int_0^\pi \mathrm{d} \mu\left(\theta_L\right)\left\langle\Omega | \theta_L\right\rangle\left\langle\theta_L | k\right\rangle e^{i E_L t_L}\right)\right. \\
& \left.\times\left(\int_0^\pi \mathrm{d} \mu\left(\theta_R\right)\left\langle k | \theta_R\right\rangle\left\langle\theta_R | \Omega\right\rangle e^{i E_R t_R}\right)\right].
\end{aligned}
\ee
The integral over energy parameter $\te_{L/R}$ can be implemented as completeness relation, for example, we have:
\be
\int_{0}^{\pi}\dd\mu\left(\te_{R}\right)\bra k|\te_{R}\ket\bra\te_{R}|\Omega\ket e^{iE_{R}t_{R}}=\int_{0}^{\pi}\dd\mu\left(\te_{R}\right)\bra k|\te_{R}\ket\bra\te_{R}|e^{iHt_{R}}|\Omega\ket=\bra k|\Omega\left(-t_{R}\right)\ket,
\ee
where the overlap function involved has been computed in \eqref{eq:overlap-00}. Therefore, we conclude that
\be \label{eq:overlap-r1-result}
\bra\Del;0,0|\Del\left(t_{L},t_{R}\right)\ket=\sum_{k=0}^{\infty}q^{\Del k}\bra\Omega\left(t_{L}\right)|k\ket\bra k|\Omega\left(-t_{R}\right)\ket=\sum_{k=0}^{\infty} \frac{q^{\Delta k}}{(q;q)_k} \phi^{*}_k(t_L)\phi^{*}(t_R).
\ee
As a cross-check, we expect the result to reduce to the zero-particle case when we send $\Delta\to0$. This is indeed the case:
\be
\lim_{\Del\to0}\bra\Del;0,0|\Del\left(t_{L},t_{R}\right)\ket=\sum_{k=0}^{\infty}\bra\Omega\left(t_{L}\right)|k\ket\bra k|\Omega\left(-t_{R}\right)\ket=\bra\Omega|\Omega\left(t_{L}+t_{R}\right)\ket,
\ee
where we used completeness relation in the first equality, and used the reality of the inner product to deduce that $\bra\Omega|\Omega(-t)\ket=\bra\Omega(t)|\Omega\ket=\bra\Omega|\Omega(t)\ket$ in the second equality. This can also be seen directly from the $\Delta\to0$ limit of \eqref{eq:time-evolve-r1}, where the distinction of left and right Hamiltonian vanishes we simply obtain:
\be
\lim_{\Del\to0}e^{i(H_{L}t_{L}+H_{R}t_{R})}|\Del;0,0\ket=e^{iH(t_{L}+t_{R})}|\Omega\ket.
\ee
In the opposite limit, $\Delta \rightarrow \infty$, any intersection between $H$-chords and the $M$-chord is forbidden. Consequently, the one-particle subspace $\mathcal{H}_1$ factorizes into left and right subsystems, with dynamics occurring only within each subsystem. This fact is appreciated by \eqref{eq:overlap-r1-result}, where:
\be
\lim_{\Del\to\infty}\bra\Del;0,0|\Del\left(t_{L},t_{R}\right)\ket=\bra\Omega|\Omega\left(t_{L}\right)\ket\bra\Omega|\Omega\left(t_{R}\right)\ket.
\ee
In either case we find the result decays as a function of each time variable. Below we present our results in~Fig \ref{fig:Liouvillian-1} where the time evolution is generated by the Liouvillian operator \eqref{eq:Liouvillian} for different matter weight $\Delta$. We find that the initial decay generally becomes faster with the increase of matter weight $\Delta$. In extremal case where $\Delta=0$, the results stays constant for all time, suggesting a lack of operator spreading. From the perspective of the double-scaled algebra, the $\Delta \rightarrow 0$ limit of $|\Delta, 0,0\rangle$ is equivalent to the empty chord state $|\Omega\rangle$. This state serves as the unique tracial state of the algebra and corresponds to the identity operator via the operator-state correspondence, and exhibits trivial Liouvillian time evolution.

We can employ a similar strategy to evaluate the projection of \( |\Delta(t_L, t_R)\rangle \) onto a state with a fixed chord number, given by
\begin{equation}
\langle \Delta; m_{L}, m_{R} | \Delta(t_{L}, t_{R}) \rangle = \int d\mu_{\Delta}(\theta_{L}, \theta_{R}) \, e^{iE_{L}t_{L} + iE_{R}t_{R}} (\theta_{L}, \theta_{R} | \mathcal{E}_{\Delta}^{-1} | m_{L}, m_{R}).
\end{equation}
The result is expressed in terms of the overlap functions \( \{ \phi_n(t) \} \) as follows:
\begin{equation}
\begin{aligned}
& \frac{\langle \Delta; m_{L}, m_{R} | \Delta(t_{L}, t_{R}) \rangle}{\sqrt{(q;q)_{m_{L}}(q;q)_{m_{R}}}} = \sum_{n=0}^{\infty} \sum_{k=0}^{\min (m_L, m_R)} \sum_{l_{L/R}=0}^{\min (n, m_{L/R} - k)} \frac{(-1)^k \, q^{\binom{k}{2}} \, q^{\Delta(n+k)}}{(q; q)_n \, (q; q)_k \, (q; q)_{m_L - k} \, (q; q)_{m_R - k}} \\
& \quad \times \binom{n, m_L - k}{l_L}_q \, \binom{n, m_R - k}{l_R}_q \, \phi_{n + m_L - k - 2 l_L}^*(t_L) \, \phi_{n + m_R - k - 2 l_R}^*(t_R).
\end{aligned}
\end{equation}
A detailed derivation is provided in Appendix \ref{app:Overlap-2}. With the overlap functions at hand, it becomes straightforward, though tedious, to determine the operator spreading coefficients \( \{ \phi^{\Delta}_n(t) \} \) as in \eqref{eq:opbasis-2}.

\begin{figure}
    \centering
    \includegraphics[width=0.8\linewidth]{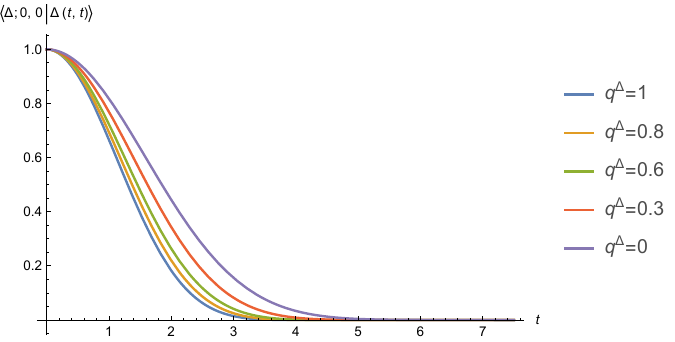}
    \caption{A plot of $\bra\Delta; 0,0|\Delta(t,t)\ket$ with increasing matter weight $q^\Del=1, 0.8,0.5, 0.2$ and $0$ respectively and $q=0.8$. The initial decay of the overlapping function slows down as one increases the matter weight. }
    \label{fig:r1}
\end{figure}

\paragraph{Time Evolution Generated by $H_L+H_R$} In addition, one might be interested in comparing the decay rate of the 1-particle state $|\Delta ; 0,0\ket$ with that of the vacuum state $|\Omega\ket$. Note that the time evolution described by equation  \eqref{eq:overlap-00} is not governed by the Liouvillian operator. Instead, using the condition that the left and right Hamiltonian act identically on the vacuum state, we can express equation \eqref{eq:overlap-00} as follows:
\be
\bra\Omega|\Omega(t)\ket = \bra\Omega|\exp(-iHt)|\Omega\ket= \bra \Omega|\exp(-i(H_L+H_R)t/2)|\Omega\ket
\ee
This corresponds to the choice $t_L = t_R = t/2$ in equation \eqref{eq:time-evolve-r1}. We then present the result for \( \langle \Delta;0,0 \,|\, \Delta(t/2,t/2) \rangle \) with different matter weights \( \Delta \) in Figure \ref{fig:r1}. The results indicate that the overlap generally decays with time \( t \), and the initial decay rate increases as the matter weight decreases. Thus, we conclude that the empty chord state (identified with the \( \Delta \to 0 \) limit of \( | \Delta;0,0 \rangle \)) exhibits the fastest initial decay under time evolution generated by $H_L+H_R$ compared to all one-particle states with positive matter weight.

\subsection{Chord Number Generating Function}
\begin{figure}
    \centering
    \includegraphics[width=0.8\linewidth]{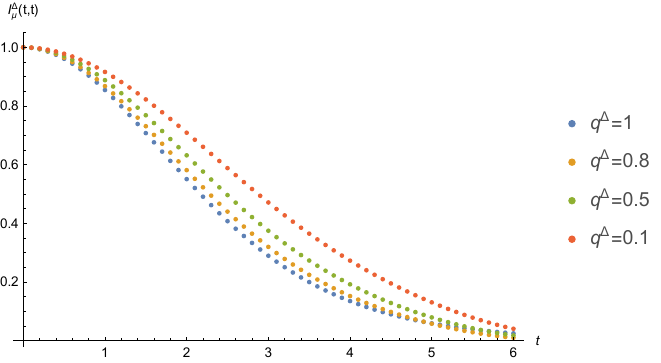}
    \caption{A plot of $I^{\Delta}_\mu(t,t)$ with $e^{-\mu}=q=0.8$, with $q^{\Delta}$ varying from $0.1$ to $1$. We find that the decay of $I^{\Delta}_\mu (t,t)$ slows down with increasing $\Delta$ (decreasing $q^\Delta$).  }
    \label{fig:partitionr1}
\end{figure}

We examine the growth of total chord number  defined in \eqref{eq:K-def-2}:
\be  \label{eq:chord-number-r1}
N_{\Delta}(t_{L},t_{R})=\bra\Del\left(t_{L},t_{R}\right)|\hat{N}|\Del\left(t_{L},t_{R}\right)\ket.
\ee
Similar to \cite{Qi_2019}, we introduce the chord number generating function as:
\be\label{eq:I-def}
I_{\Del}^{\mu}(t_{L},t_{R})=\bra\Del\left(t_{L},t_{R}\right)|e^{-\mu\hat{N}}|\Del\left(t_{L},t_{R}\right)\ket.
\ee
It's clear that the functional dependence of $I^{\mu}_\Delta$ in $t_{L/R}$ contains all information about moments of $\hat{N}$ in $|\Delta(t_L,t_R)\ket$, which manifests as:
\be 
\bra\Del\left(t_{L},t_{R}\right)|\hat{N}^{k}|\Del\left(t_{L},t_{R}\right)\ket=\frac{(-1)^{k}}{k!}\frac{\partial^{k}I_{\mu}^{\Delta}}{\partial\mu^{k}}|_{\mu=0}.
\ee
According to \eqref{eq:reformulation-1}, \eqref{eq:I-def} can be formulated  as
\be
I_{\Del}^{\mu}\left(t_{L},t_{R}\right)	=\int\dd\mu_{\Del}\left(\te_{1},\te_{2}\right)\bra\Del\left(t_{L},t_{R}\right)|\mf^{\da}|\te_{1},\te_{2})(\te_{1},\te_{2}|\mf e^{-\mu\hat{N}}|\Del\left(t_{L},t_{R}\right)\ket.
\ee
Inserting an identity $\mathbf{1}=\mf^{-1}\mf$ to the right of $e^{-\mu \hat{N}}$, and combined with \eqref{eq:phase} we end up with the following integral: \footnote{We don't get matter density in terms of $\te_3$ and $\te_4$ since we are only using the completeness relation in $\mh_0\otimes\mh_0$. The apparent asymmetry in $\te_{1,2}$ and $\te_{3,4}$ origins from a specific choice of slicing from the bulk Hilbert space perspective. One can show that $I^{
\mu}_\Delta$ is invariant under the exchange $(\te_1,\te_2,q^{\Delta})\leftrightarrow (\te_3,\te_4,e^{-\mu})$ of the integrand \cite{Berkooz:2018jqr,Okuyama:2024yya}. }
\be
\begin{aligned}
I_{\Del}^{\mu}\left(t_{L},t_{R}\right)	&=\int\prod_{i=1}^{4}\dd\mu\left(\te_{i}\right)e^{-i(E_{1}-E_{3})t_{L}-i(E_{2}-E_{4})t_{R}} \\
	&\times\bra\te_{1}|q^{\Del\hat{n}}|\te_{2}\ket(\te_{1},\te_{2}|\mf e^{-\mu\hat{N}}\mf^{-1}|\te_{3},\te_{4}).
\end{aligned}
\ee
The component $(\te_{1},\te_{2}|\mf e^{-\mu\hat{N}}\mf^{-1}|\te_{3},\te_{4})$ can be expressed with in terms of operators that act on $\mh_0\otimes\mh_0$. We introduce the left and right number operators as
\be
\hat{n}_{L/R}|n_{L},n_{R})=n_{L/R}|n_{L},n_{R}),
\ee
combined with the definition of $\hat{N}$ in \eqref{eq:N-def}, one can verify that
\be
\mf\hat{N}\mf^{-1}=\me_{\Del}^{-1}(\hat{n}_{L}+\hat{n}_{R})\me_{\Del}.
\ee
Therefore, we map the integrand to expressions in the doubled-Hilbert space and organize the chord number generating function $I^{\mu}_\Delta$ as: 
\be \label{eq:Idef-3}
\begin{aligned}
& I_{\Delta}^{\mu}\left(t_L, t_R\right)= \\
& \int_0^\pi \prod_{i=1}^4 \mathrm{~d} \mu\left(\theta_i\right)\bra\theta_1|q^{\Delta \hat{n}}| \theta_2\ket  (\theta_1, \theta_2|\mathcal{E}_{\Delta}^{-1} e^{-\mu\left(\hat{n}_L+ \hat{n}_R\right)} \mathcal{E}_{\Delta}| \theta_3, \theta_4) e^{-i\left(E_1-E_3\right) t_L-i\left(E_2-E_4\right) t_R}.
\end{aligned}
\ee 
Setting \( e^{-\mu} = q^{-\Delta_2} \), the matrix element can be shown to be equivalent to the crossed 4-point function derived in \cite{Berkooz:2018jqr}, which is expressed as follows:
\begin{equation}\label{eq:R-matrix-0}
\begin{aligned}
\langle \theta_{1} | q^{\Delta_{1} \hat{n}} | \theta_{2} \rangle (\theta_{1}, \theta_{2} | \mathcal{E}_{\Delta_{1}}^{-1} q^{\Delta_{2}(\hat{n}_{L} + \hat{n}_{R})} \mathcal{E}_{\Delta_{1}} | \theta_{3}, \theta_{4}) &= \gamma_{\Delta_{1}}^{(q)}(\theta_{1}, \theta_{2}) \, \gamma_{\Delta_{1}}^{(q)}(\theta_{3}, \theta_{4}) \, \gamma_{\Delta_{2}}^{(q)}(\theta_{1}, \theta_{3}) \, \gamma_{\Delta_{2}}^{(q)}(\theta_{2}, \theta_{4}) \\
&\quad \times R_{\theta_{2} \theta_{4}}^{(q)} 
\begin{bmatrix}
\theta_{1} & \Delta_{1} \\
\theta_{3} & \Delta_{2} 
\end{bmatrix},
\end{aligned}
\end{equation}
where \( \gamma^{(q)}_\Delta (\theta_1, \theta_2) = \sqrt{\langle \theta_1 | q^{\Delta \hat{n}} | \theta_2 \rangle} \), and the $R$-matrix is closely related to the $6j$-symbol of $U_{\sqrt{q}} (\mathfrak{su}(1,1))$. The underlying crossing symmetry of this integrand has been explored in \cite{Okuyama:2024yya}; see also \cite{Gaiotto:2024kze}. For those interested in the derivation, details are provided in Appendix \ref{app:gen-I}. With some effort, we are able to perform the integral over the \( \theta_i \)'s, yielding the following results:
\be
\begin{aligned}
I_{\Delta}^{\mu}\left(t_L, t_R\right) & =\sum_{n_1, m_1, m_2, l=0}^{\infty} \frac{\left(e^{-2 \mu} ; q\right)_l e^{-\mu\left(m_1+m_2\right)} q^{\Delta\left(n_1+l\right)}}{(q ; q)_{m_1}(q ; q)_{m_2}(q ; q)_{n_1}(q ; q)_l} \\
& \times \sum_{k_1=0}^{\min \left(n_1, m_1\right)} \sum_{k_2=0}^{\min \left(n_1, m_2\right)}\binom{n_1, m_1}{k_1}_q\binom{n_1, m_2}{k_2}_q \\
& \times \phi_{n_1+m_1-2 k_1}\left(t_L\right) \phi_{m_1+l}^*\left(t_L\right) \phi_{n_1+m_2-2 k_2}\left(t_R\right) \phi_{m_2+l}^*\left(t_R\right).
\end{aligned}
\ee
Here, we have expressed our results in terms of the overlap functions \( \{ \phi_n(t) \} \), with the \( q \)-combinatoric coefficients defined as:
\begin{equation}
\binom{n_{1},n_{2}}{k}_{q} = \frac{(q;q)_{n_{1}} (q;q)_{n_{2}}}{(q;q)_{n_{1}-k} (q;q)_{n_{2}-k} (q;q)_{k}}.
\end{equation}
As a consistency check, we consider the no-particle limit where \( \Delta \to 0 \) while keeping \( q \) finite. In this case, the matter density becomes diagonal in the energy eigenbasis:
\begin{equation}
\langle \theta_{1} | q^{\Delta \hat{n}} | \theta_{2} \rangle \xrightarrow{\Delta \to 0} \mu^{-1}(\theta_{2}) \, \delta(\theta_{1} - \theta_{2}),
\end{equation}
and furthermore, we show in Appendix \ref{app:energy-basis} that
\begin{equation}
(\theta_{1}, \theta_{1} | \mathcal{E}_{0}^{-1} e^{-\mu(\hat{n}_{L} + \hat{n}_{R})} \mathcal{E}_{0} | \theta_{3}, \theta_{4}) \xrightarrow{\Delta \to 0} \mu^{-1}(\theta_{4}) \, \delta(\theta_{3} - \theta_{4})\bra \te_1|e^{-\mu \hat{n}}|\te_3\ket .
\end{equation}
Thus, we find that the generating function becomes 
\begin{equation}
I_{0}^{\mu}(t_{L}, t_{R}) = \int d\mu(\theta_{1}) \, d\mu(\theta_{2}) \, \langle \theta_{1} | e^{-\mu \hat{n}} | \theta_{2} \rangle \, e^{-i(E_{1} - E_{2})(t_{L} + t_{R})}.
\end{equation}
In particular, for Liouvillian time evolution with \( t_L + t_R = 0 \), we find that \( I_{0}^{\mu} \) remains constant, indicating the absence of evolution in \( N_{\Delta}(t) \). Our numerical results, plotted for \( q = e^{-\mu} = 0.8 \) with \( q^{\Delta} \) varying from \( 0.1 \) to \( 1 \), confirm this point. This is consistent with our expectation: as \( \Delta \to 0 \) represents the absence of the chord operator, it corresponds to an empty chord state with trivial evolution. We also plot the results for time evolution generated by \( H_L + H_R \) in Figure~\ref{fig:partitionr1}, corresponding to the choice \( t_L = t_R = t \). In this case, we observe that \( I^{\Delta}_\mu(t, t) \) exhibits the fastest initial decay for the empty chord state \( \Omega \), which is obtained as the \( \Delta \to 0 \) limit of \( |\Delta; 0, 0\rangle \).

\begin{figure}
    \centering
    \includegraphics[width=0.8\linewidth]{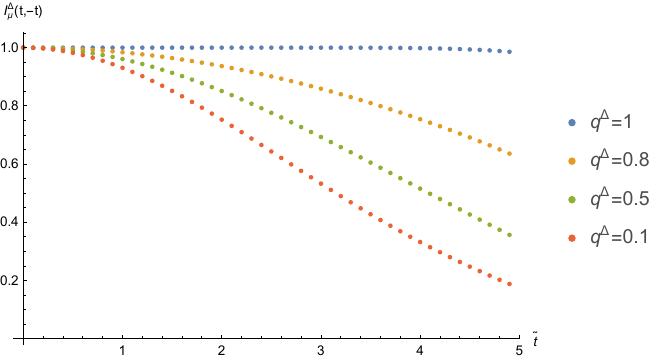}
    \caption{A plot of \( I^{\Delta}_\mu(t, -t) \) as function of $\tilde{t}$ is shown with \( e^{-\mu} = q = 0.8 \), and \( q^{\Delta} \) varying from \( 0.1 \) to \( 1 \). We observe that \( \langle \hat{N} \rangle = 0 \) when \( \Delta = 0 \) and \( q^\Delta = 1 \), indicating the absence of the chord operator. As \( \Delta \) increases for a fixed \( q \), \( I^{\Delta}_\mu \) decays, signaling the growth of \( \mathcal{K}^{\Delta} \).}
    \label{fig:OTOC-L}
\end{figure}

\subsection{Two Point Function in Presence of Matter}\label{ssec:4pt}
It has been shown that for the Brownian double-scaled SYK, the decay of matter chord two-point functions slows down in the presence of matter \cite{Milekhin:2023bjv}. In this section, we verify this statement for the double-scaled SYK model across a general parameter regime for the infinite temperature state \( |\Omega\rangle \).

The relevant quantity we consider is the (unnormalized) time-ordered 4-point correlation function:
\begin{equation} \label{eq:G4-def}
G_{4}^{\Delta_{W}\Delta_{V}} = \langle \Omega | W(T_{f}) V(T_{2}) V(T_{1}) W(0) | \Omega \rangle, \quad T_{f} \geq T_{2} \geq T_{1} \geq 0,
\end{equation}
where \( W \) and \( V \) are matter chord operators with weights \( \Delta_W \) and \( \Delta_V \), respectively. For the time configurations given in \eqref{eq:G4-def}, this quantity can be interpreted as the two-point correlation function of the \( W \)-particle in the presence of the \( V \)-particle. Reference \cite{Berkooz:2018jqr} represents \eqref{eq:G4-def} as the following energy integral:
\begin{equation} \label{eq:integral-4pt}
\begin{aligned}
G_4^{\Delta_W \Delta_V} &= \int \prod_{i=1}^3 \, d\mu(\theta_i) \, e^{-i E_1(T_f - (T_2 - T_1))} \, e^{-i E_2(T_2 - T_1)} \, e^{i E_3 T_f} \\
& \quad \times \langle \theta_1 | q^{\Delta_V \hat{n}} | \theta_2 \rangle \, \langle \theta_1 | q^{\Delta_W \hat{n}} | \theta_3 \rangle,
\end{aligned}
\end{equation}
which can also be derived from the overlap between states in the chord Hilbert space. The derivation is detailed in Appendix \ref{app:bulk-dev}. It is evident from the above expression that \( G_4^{\Delta_V \Delta_W} \) depends on \( T_2 - T_1 \) and \( T_f \). Following the approach introduced in the previous section, we can integrate this expression and write the result in terms of the overlap functions \( \{ \phi_n(t) \} \) as:
\begin{equation} \label{eq:4pt-un}
\begin{aligned}
G_4^{\Delta_W \Delta_V}(T_f, T_2 - T_1) &= \sum_{n_1, n_2, n_3 = 0}^{\infty} q^{\Delta_V(n_1 + n_3) + \Delta_W(n_2 + n_3)} \\
& \quad \times \frac{\phi_{n_1 + n_2}(T_f - (T_2 - T_1)) \, \phi_{n_1 + n_3}(T_2 - T_1) \, \phi_{n_2 + n_3}(-T_f)}{(q; q)_{n_1} (q; q)_{n_2} (q; q)_{n_3}}.
\end{aligned}
\end{equation}
For general \( (T_f, \Delta_W) \) and \( (T_2 - T_1, \Delta_V) \), the result does not factorize. In the case where the duration of the \( V \)-particle is zero (\( T_2 - T_1 = 0 \)), we find:
\begin{equation}
G_{4}^{\Delta_{W}\Delta_{V}}(T_{f}, 0) = \sum_{n_{2} = 0}^{\infty} \frac{q^{\Delta_{W} n_{2}}}{(q; q)_{n_{2}}} \, \phi_{n_{2}}(T_{f}) \, \phi_{n_{2}}(-T_{f}) = G_{2}^{\Delta_{W}}(T_{f}),
\end{equation}
where we used \( \phi_n(0) = \delta_{n0} \), reducing the expression to the two-point function of the \( W \)-particle, \( G^{\Delta_W}_{2}(T_f) = \langle \Omega | W(T_f) W(0) | \Omega \rangle \).

\begin{figure}
    \centering
    \includegraphics[width=0.8\linewidth]{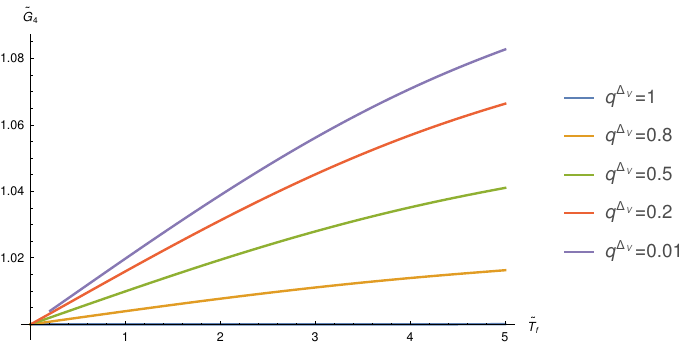}
    \caption{A plot of $\tilde{G}_4$ as a function of $\tilde{T}_f$ for fixed $q=0.8,q^{\Delta_W}=0.8$ and $|T_{21}|=0.5$. We find that $\tilde{G}_4$ increases as one increase the weight $\Delta_V$ of intermediate particle, suggesting that the decay rate of $W$-particle becomes slower in presence of other particles. }
    \label{fig:decay-weight}
\end{figure}

We then analyze the functional dependence of the normalized 4-point function on both the duration \( T_2 - T_1 \) of the intermediate particle and its weight \( \Delta_V \). We find that, for a given particle duration \( |T_{12}| = |T_2 - T_1| \), increasing the weight of the intermediate particle increases the normalized four-point function \( \tilde{G}_4 \) at a fixed time \( T_f \) (see Figure \ref{fig:decay-weight} for an illustration). In other words, the decay of the four-point function \( G^{\Delta_V \Delta_W}_4 \) becomes slower compared to its factorized counterpart \( G^{\Delta_V}_2(T_{21}) G^{\Delta_W}(T_f) \) as \( \Delta_V \) increases.

\begin{figure}
    \centering    \includegraphics[width=0.8\linewidth]{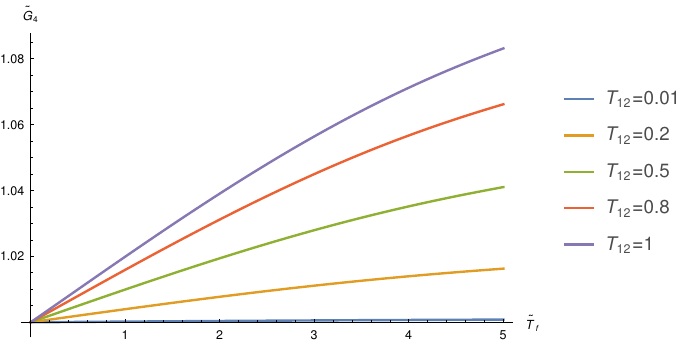}
    \caption{A plot of $\tilde{G}_4$ as a function of $\tilde{T}_f$ for fixed $q=0.8, q^{\Delta_V}=0.5,q^{\Delta_W}=1$. We find that $\tilde{G}_4$ increases as one increase the duration $|T_{12}|$ of the intermediate particle.}
    \label{fig:decay-t12}
\end{figure}

Moreover, for a given particle weight \( \Delta_V \), we demonstrate that increasing the duration \( |T_{12}| \) of the intermediate particle also leads to an increase of \( \tilde{G}_4 \) at a fixed time \( T_f \) (see Figure \ref{fig:decay-t12} for an illustration).

\paragraph{Initial Growth Rate of \(\tilde{G}_4^{\Delta_V\Delta_W}\)} To gain quantitative insight into the initial spreading rate's sensitivity to matter insertion, we examine the first derivative of \(\tilde{G}_4^{\Delta_V\Delta_W}\) with respect to \(T_{21}\). A detailed calculation (see \ref{app:4pt-growth})shows
\be
\partial_{T_{21}} \tilde{G}_4^{\Delta_V \Delta_W} \big|_{T_{21} = 0} = \frac{(1 - q^{\Delta_V})(1 - q^{\Delta_W})}{\sqrt{1 - q}} \sum_{n=0}^{\infty} \frac{q^{\Delta_W n}}{(q; q)_n} \left[(-i) \phi_n(T_f) \phi_{n+1}(-T_f)\right],
\ee
where we have arranged the summand to ensure that the term in brackets is real. Assuming \(0\leq T_{21} \ll T_f \ll \sqrt{\lambda}\), we further expand in \(\tilde{T}_f = T_f / \sqrt{\lambda}\), yielding
\be
\tilde{G}_4^{\Delta_V \Delta_W} (T_f,T_{21})= 1 + (1 - q^{\Delta_V})(1 - q^{\Delta_W}) T_f T_{21} + O(\tilde{T}_f^2).
\ee
This confirms the initial increase in \(\tilde{G}_4^{\Delta_V \Delta_W}\), or equivalently, the moderated decay of the two-point function in the presence of intermediate particles.

The dependence of the growth rate on the matter weight and interaction time aligns with similar observations in \cite{Milekhin:2023bjv}. In the following section, we lift the small-time assumption and explore the semi-classical limit \(\lambda \to 0\), confirming this functional dependence in that regime.

\section{Bulk Manifestation in Semi-Classical Regimes} \label{sec:manifestation}
The results from the previous section may seem unfamiliar due to the $q$-deformed nature of the chord statistics. In this section, we explore the semi-classical limit by taking \(\lambda = -\log(q) \to 0\) and focusing on specific regions of the energy spectrum when computing the relevant quantities. This approach provides greater control over the system's behavior in this limit, enabling us to extract analytic results from general expressions discussed in previous sections. 

We start with the calculation of the time-ordered four-point correlator \eqref{eq:integral-4pt} by evaluating both the saddle-point value around center of energy spectrum and one-loop corrections of fluctuations around it. The latter result is reminiscent of the leading \(1/N\) correction to the connected part of the correlation function in large $p$ SYK at infinite temperature, providing an explicit demonstration of the slowdown in decay noted in last section. 

Next, we consider the growth of total chord number $N_{\Delta}(t)$ in presence of a single chord operator in the semi-classical limit. Our findings indicate that this behavior aligns with the two-sided renormalized length in JT gravity with $SL(2,\mathbb{R})$ matter fields \cite{Harlow:2021dfp}. At late times, it exhibits linear growth, consistent with previous results in \cite{Rabinovici:2023yex}. In the early-time regime, $t < (\lambda J)^{-1}$, however, we observe exponential growth, deviating from the quadratic time-dependence seen in the initial growth of state complexity of $|\Omega(t)\rangle$, as discovered in \cite{Rabinovici:2023yex}.

\subsection{The Semi-classical Limit of the Uncrossed Four-Point Function}
In \ref{ssec:4pt} we declared numerically that inserting matter chord generally slows down the decay of two point function at infinite temperature. In other words, we expect the normalized time-ordered four-point function to receive a positive contribution from its connected part. While this has been made explicit in the Brownian model \cite{Milekhin:2023bjv}, it's not easy to verify directly from \eqref{eq:4pt-un} due to the complexity of the expression. Here we turn to the semi-classical regime and declare the statement. In this section, we stick to the definition $q=e^{-\lambda}$, and mainly work with the asymptotic analysis for small $\lambda$.  

We consider the following integral analogues to \eqref{eq:integral-4pt} with Euclidean time:
\be \label{eq:IE4}
G_{E}^{\Delta_V\Delta_W}(\tau_1,\tau_2,\tau_3)=\int_{0}^{\pi}\prod_{i=1}^{3}\dd\mu\left(\te_{i}\right)e^{-E_{1}\tau_{1}-E_{2}\tau_{2}-E_{3}\tau_{3}}(\te_{1}|q^{\Del_{V}\hat{n}}|\te_{2})(\te_{1}|q^{\Del_{W}\hat{n}}|\te_{3}),
\ee
At infinite temperature, the correlation satisfies \(\sum_{i=1}^3 \tau_i = 0\). Henceforth, we simplify the notation by referring to the Euclidean four-point function at infinite temperature as \(G_4\). We consider the integral under the limit $\lambda\to0$, $\lambda_{V/W}\equiv\lambda\Delta_{V/W}$ fixed as independent parameter, it can be written as:
\be \label{eq:IE4-2}
G_{4} = \int_0^\pi \prod_{i=1}^3 \frac{\mathrm{~d} \theta_i}{2\pi} \exp \left(-\frac{I_0}{\lambda}+I_{c}\right),
\ee
where $\simeq$ here means we only keep the leading order in $\lambda$ contribution to the argument in the exponential, and the term $I_c=I_c(\lambda,\lambda_{V/W})$ takes into account the normalization constant:
\be \label{eq:def-f0}
\begin{aligned}
e^{I_c} & \equiv(q ; q)_{\infty}^3\left(q^{2 \Delta_V}, q^{2 \Delta_W} ; q\right)_{\infty} \\
& \stackrel{\lambda \rightarrow 0}{=} \exp \left(-\frac{5 \pi^2}{6 \lambda}-\frac{2 \lambda_V\left(1-\log \lambda_V\right)+2 \lambda_W\left(1-\log \lambda_W\right)}{\lambda}+O(1)\right).
\end{aligned}
\ee
The $I_0$ in \eqref{eq:IE4-2} is defined by picking up the $\theta$ dependent part of the integrand:
\be \label{eq:f-def}
I_0=\sum_{i=1}^{3}{\rm Li}_{2}\left(e^{\pm2i\te_{i}}\right)-\sum_{i\in\mathcal{I}}{\rm Li}_{2}\left(e^{-\lambda_i\pm i\te_{1}\pm i\te_{i}}\right)+\sum_{i=1}^{3}2\tau_{i}\cos\te_{i},
\ee
where the notation $\text{Li}_2 (e^{\pm 2i\te})\equiv \text{Li}_2 (e^{2i\te})+\text{Li}_2 (e^{-2i\te})$, is defined to sum over all sign configurations. The index set \(\mathcal{I}\) is defined such that 
\be
(\lambda_i, \Delta_i, \tau_i, \theta_i) \in \{(\lambda_V, \Delta_V, \tau_2, \theta_2), (\lambda_W, \Delta_W, \tau_3, \theta_3)\}, \quad i \in \mathcal{I},
\ee
and will be used frequently in the following discussion. For example, we have 
\be
\sum_{i \in \mathcal{I}} f(\theta_i, \Delta_i, \tau_i) \equiv f(\lambda_V, \Delta_V, \tau_2) + f(\lambda_W, \Delta_W, \tau_3),
\ee
according to this notation, for some function \(f\).

We emphasize that \(I_0 = I_0(\theta_i, \lambda_{V/W})\) represents the leading-order energy dependent part of the integrand in the perturbative expansion of \(\lambda\), where the parameters \(\lambda_{V/W}\) are treated independently of \(\lambda\). \(I_0\) is obtained by isolating the \(\exp(O(\lambda^{-1}))\) contribution from \eqref{eq:IE4}, using the asymptotic behavior:
\be
\left(x;q\right)_{\infty} \stackrel{\lambda \to 0}{=} \exp\left(-\frac{1}{\lambda}{\rm Li}_{2}(x) + O(1)\right),
\ee
and depends only on the energy parameters \(\theta_i\) and \(\lambda_{V/W}\).

\paragraph{The Saddle-Point Value}
We now evaluate the saddle point value of the integrand \eqref{eq:IE4-2}. The calculation closely follows the approach in \cite{Goel:2023svz}. We keep track of the derivation here for later reference, as it will be useful when addressing the one-loop contribution, which is largely based on expanding around the saddle, retaining terms quadratic in $\lambda_{V}$ and $\lambda_W$. 

For $\te_i$s within the range $[0,\pi]$, we have 
\be
\begin{aligned}
\mathrm{Li}_2\left(e^{ \pm 2 i \theta}\right) & =2 \sum_{k=1}^{\infty} \frac{\cos (2 k \theta)}{k^2}=2 \theta^2-2 \pi \theta+\frac{\pi^2}{3} ,\\
\partial_\theta \mathrm{Li}_2\left(e^{ \pm 2 i \theta}\right) & =-4 \sum_{k=1}^{\infty} \frac{\sin 2 k \theta}{k}=4 \theta-2 \pi,
\end{aligned}
\ee 
and simiarly for the interacting term:
\be
\begin{aligned}
\partial_{\theta_1} \mathrm{Li}_2\left(q^{\Delta_V} e^{i\left( \pm \theta_1 \pm \theta_2\right)}\right) & =-2 \sum_{k=1}^{\infty} e^{-\lambda_V}\left(\frac{\sin k\left(\theta_1-\theta_2\right)}{k}+\frac{\sin k\left(\theta_1+\theta_2\right)}{k}\right) \\
& =2 \arctan \left(\frac{\sin \left(\theta_1 \pm \theta_2\right)}{e^{\lambda_V}-\cos \left(\theta_1 \pm \theta_2\right)}\right).
\end{aligned}
\ee
Therefore, the extremal condition $\partial_{\te_i}I_0=0$ can be organized into the following:
\be \label{eq:extremal}
\begin{aligned}
& 2 \theta_1-\pi+\arctan \left(\frac{\sin \left(\theta_1 \pm \theta_2\right)}{e^{\lambda_V}-\cos \left(\theta_1 \pm \theta_2\right)}\right)+\arctan \left(\frac{\sin \left(\theta_1 \pm \theta_3\right)}{e^{\lambda_W}-\cos \left(\theta_1 \pm \theta_3\right)}\right)-\tau_1 \sin \theta_1=0, \\
& 2 \theta_2-\pi+\arctan \left(\frac{\sin \left(\theta_1+\theta_2\right)}{e^{\lambda_V}-\cos \left(\theta_1+\theta_2\right)}\right)-\arctan \left(\frac{\sin \left(\theta_1-\theta_2\right)}{e^{\lambda_W}-\cos \left(\theta_1-\theta_2\right)}\right)-\tau_2 \sin \theta_2=0, \\
& 2 \theta_2-\pi+\arctan \left(\frac{\sin \left(\theta_1+\theta_3\right)}{e^{\lambda_V}-\cos \left(\theta_1+\theta_3\right)}\right)-\arctan \left(\frac{\sin \left(\theta_1-\theta_3\right)}{e^{\lambda_W}-\cos \left(\theta_1-\theta_3\right)}\right)-\tau_3 \sin \theta_3=0.
\end{aligned}
\ee
It's easy to see that the saddle point localizes around $\te_1=\te_2=\te_3=\pi/2$ without the $\tau_i$ terms. The $O(\lambda_{V/W})$ deviations from $\pi/2$ enters the interacting term and get enhanced to a term that starts at $O(\lambda_{V/W}^0)$. Therefore, by expanding the above equation in $\lambda_V$ and $\lambda_W$ and solve for $\te$s up to linear order in $\lambda_V$ and $\lambda_W$, we find the saddle point values are:
\be \label{eq:saddles-0}
\begin{aligned}
& \theta_1^*=\frac{\pi}{2}+\frac{1}{3} \lambda_V \tan \tau_2+\frac{1}{3} \lambda_W \tan \tau_3, \\
& \theta_2^*=\frac{\pi}{2}-\frac{2}{3} \lambda_V \tan \tau_2+\frac{1}{3} \lambda_W \tan \tau_3, \\
& \theta_3^*=\frac{\pi}{2}+\frac{1}{3} \lambda_V \tan \tau_2-\frac{2}{3} \lambda_W \tan \tau_3,
\end{aligned}
\ee
where we have imposed the infinite temperature condition  $\tau_1+\tau_2+\tau_3=0$, and denote the on-shell value of $\te$s as $\te^*$.  Plugging back into \eqref{eq:f-def}, we find:
\be
\begin{aligned}
-\left.\frac{I_0}{\lambda}\right|_{\theta_i=\theta_i^*} & =\frac{5 \pi^2}{6 \lambda}-\frac{2 \lambda_V\left(1-\log \lambda_V\right)+2 \lambda_W\left(1-\log \lambda_W\right)}{\lambda} \\
& -\log \cos ^{2 \lambda_V / \lambda} \tau_2-\log \cos ^{2 \lambda_V / \lambda} \tau_3+O\left(\lambda_V, \lambda_W\right).
\end{aligned}
\ee
The first line precisely cancels with the divergent terms in $\exp(f_0)$ \eqref{eq:def-f0}. Therefore, if we introduce matter weights as $\lambda_{V/W}=\lambda \Delta_{V/W}$\footnote{Note that the saddle point location \(\theta^*\) is determined by treating \(\lambda_{V/W}\) as independent from \(\lambda\). This rescaling is applied only when evaluating the action at the saddle point. } in the limit $\lambda\to 0$, and take the real time configuration:
\be
\tau_2\to i(T_2-T_1),\quad \tau_3\to (i T_f),
\ee
we find that
\be
\begin{split}
\lim_{\lambda\to0}G_{4}^{\Del_{W}\Del_{V}}	&=\lim_{\lambda\to0}\Big[\int\prod_{i=1}^{3}\dd\mu\left(\te_{i}\right)e^{-iE_{1}(T_{f}-(T_{2}-T_{1}))}e^{-iE_{2}(T_{2}-T_{1})}e^{iE_{3}T_{f}} \\
	&\times(\te_{1}|q^{\Del_{V}\hat{n}}|\te_{2})(\te_{1}|q^{\Del_{W}\hat{n}}|\te_{3}) \Big]\\
	&=\frac{1}{\cosh^{2\Del_{V}}(T_{2}-T_{1})\cosh^{2\Del_{W}}T_{f}}=G^{\Delta_V}_2 (T_2-T_1)G^{\Delta_W}_2 (T_f).
\end{split}
\ee
This reproduces the factorized result \(\langle V(t_2)V(t_1)\rangle \langle W(T)W(0)\rangle\) at infinite temperature. In the final step, we take \(\lambda_{V/W} = \lambda \Delta_{V/W} \to 0\) in the limit \(\lambda \to 0\), corresponding to the strong semi-classical limit described in \cite{Rahman:2024vyg}, where all quantum fluctuations vanish. Conversely, a weak semi-classical limit is defined by keeping \(\lambda_{V/W}\) finite, allowing energy fluctuations—characterized by deviations from the center of the spectrum—to remain quantum and governed by these parameters. In the following, we evaluate the leading-order corrections from such fluctuations by calculating the corresponding one-loop determinant.

\paragraph{The One-loop Correction}
We evaluate the one-loop corrections to the integral \eqref{eq:IE4}. Such corrections have two origins: the $O(\lambda^0)$ corrections to the integrand and the one-loop determinants from integrating out the fluctuations around the saddle.  We calculate them separately below.

We now incorporates the $O(\lambda^0)$ contribution to the integrand:
\be
G_4=\int_{0}^{\pi}\prod_{i=1}^{3}\frac{\dd\te_{i}}{2\pi}\exp\left(-\frac{I_0}{\lambda}+I_1+I_{c}+O\left(\lambda\right)\right),
\ee
where the term $I_1=I_1(\lambda_{V/W},\te_i)$ in the exponential is obtained by picking up the $O(\lambda^0)$ contribution from the integrand \eqref{eq:IE4}
\be
I_1	=\frac{1}{2}\sum_{i=1}^{3}\left(\log\left(1-e^{\pm2i\te_{i}}\right)-\tau_{i}\cos\te_{i}\right)-\frac{1}{2}\sum_{i\in\mi}\log\left(1-e^{-\lambda_{i}\pm i\te_{1}\pm i\te_{i}}\right).
\ee
Note that in the $\lambda\to0$ limit  the saddles are still determined by the extremality conditions $\partial_{\te_i} I_0 \mid_{\te_i=\te^{*}_i}=0$ as we treat $\lambda_{V/W}$ as independent parameters. Around the saddle we expand $f$ to quadratic order:
\be \label{eq:f-expansion}
I_0=I_0\mid_{\te_{i}=\te_{i}^{*}}+\frac{1}{2}\sum_{i,j=1}^{3}\frac{\partial^{2}I_0}{\partial\te_{i}\partial\te_{j}}|_{\te_{i}=\te_{i}^{*}}\left(\te_{i}-\te_{i}^{*}\right)\left(\te_{j}-\te_{j}^{*}\right)+O\left(\lambda_{V/W}^{3}\right),
\ee
and perform the corresponding Gaussian integral, which leads to:
\be
G_{4}\simeq {\rm det}\left[\frac{\partial^{2}I_0}{\partial\te_{i}\partial\te_{j}}\mid_{\te_{i}=\te_{i}^{*}}\right]^{-1/2}\exp\left(-\frac{I_{0}|_{\te_{i}=\te_{i}^{*}}}{\lambda}+I_1|_{\te_{i}=\te_{i}^{*}}+I_{c}\right),
\ee
where $\simeq$ here means we neglect all the numerical coefficients. Note that for \eqref{eq:f-expansion} to be valid, we need the extremality conditions  \eqref{eq:extremal} to be satisfied up to $O(\lambda_{V/W})$ \footnote{From the solutions \eqref{eq:saddles-0}, it is clear that the deviations $\theta - \theta^*$ are controlled by $\lambda_{V/W}$, not by $\lambda$. While we focus on loop corrections from $\lambda$, the higher-order corrections in $O(\lambda_{V/W})$ to the saddle point position must be properly calculated to obtain the correct Hessian determinant of $f$.
}, whereas the solutions \eqref{eq:saddles-0} derived above only satisfy the leading order extremality conditions.  Therefore, we calculate the saddle point values using the following anstaz \cite{Goel:2023svz}:
\be \label{eq:saddle-2}
\begin{aligned}
\theta_1^* & =\frac{\pi}{2}+\frac{1}{3} \lambda_V \tan \tau_2+\frac{1}{3} \lambda_W \tan \tau_3+\eta_1+\gamma_1 ,\\
\theta_2^* & =\frac{\pi}{2}-\frac{2}{3} \lambda_V \tan \tau_2+\frac{1}{3} \lambda_W \tan \tau_3+\eta_2+\gamma_2 ,\\
\theta_3^* & =\frac{\pi}{2}+\frac{1}{3} \lambda_V \tan \tau_2-\frac{2}{3} \lambda_W \tan \tau_3+\eta_3+\gamma_3,
\end{aligned}
\ee
where $\eta_i\in O(\lambda_{V/W})$ and $\gamma_i \in O(\lambda^{2}_{V/W})$. We find that the extremality condition at $O(\lambda_{V/W})$ requires $\eta_1=\eta_2=\eta_3=\eta$, and only depends on the two of the differences $\gamma_1-\gamma_2,\gamma_1-\gamma_3$ among the coefficients at quadratic order. Therefore, we choose $\gamma_1=0$ and the rest of parameters are solved as:
\be
\begin{aligned}
& \eta=-\frac{1}{6}\left(\lambda_V \tan \tau_2+\lambda_W \tan \tau_3\right), \\
& \gamma_2=\frac{\lambda_V}{2} \frac{-\lambda_V \tan \tau_2+\lambda_W \tan \tau_3}{\cos ^2 \tau_2}, \\
& \gamma_3=\frac{\lambda_W}{2} \frac{\lambda_V \tan \tau_2-\lambda_W \tan \tau_3}{\cos ^2 \tau_3}.
\end{aligned}
\ee 
Now let's evaluate the saddle point value of $I_1$, denoted as $I^{4\text{pt}}_{1*} (\lambda_V,\lambda_W)\equiv I_{1}\mid_{\te_i=\te^{*}_i}$, we find:
\be\label{eq:gstar}
\begin{aligned}
I^{\text{4pt}}_{1*} & =\frac{1}{2} \lambda_V\left(1+\sec ^2 \tau_2+\tau_2 \tan \tau_2\right)+\frac{1}{2} \lambda_W\left(1+\sec ^2 \tau_3+\tau_3 \tan \tau_3\right) \\
& -\frac{1}{2}\left(\lambda_V+\lambda_W\right) \tan \tau_2 \tan \tau_3+O\left(\lambda_{V / W}^2\right) \\
& =I_{1*}^{2 \mathrm{pt}}\left(\tau_2, \lambda_V\right)+I_{1*}^{2 \mathrm{pt}}\left(\tau_3, \lambda_W\right)-\frac{1}{2}\left(\lambda_V+\lambda_W\right) \tan \tau_2 \tan \tau_3+O\left(\lambda_{V / W}^2\right),
\end{aligned}
\ee
where we have denoted $I_{1*}^{{\rm 2pt}}\left(\tau_{2},\lambda_{V}\right)=\frac{1}{2}\lambda_{V}\left(1+\sec^{2}\tau_{2}+\tau_{2}\tan\tau_{2}\right)$ to be the corresponding saddle point value in the two-point function at infinite temperature. The result can be obtained by taking the $\beta\to0$ limit of equation (3.10) in \cite{Goel:2023svz}. We then move on to evaluate the determinant:
\be
{\rm Det}^{VW}_{4}\equiv{\rm det}\left[\frac{\partial^{2}I_0}{\partial\te_{i}\partial\te_{j}}\mid^{\te_{i}=\te_{i}^{*}}_{i=1,2,3}\right]=\frac{16\cos^{2}\tau_{2}\cos^{2}\tau_{3}}{\lambda_{V}\lambda_{W}}\left[1+\ml^{{\rm 4pt}}+O\left(\lambda_{V/W}^{2}\right)\right],
\ee 
where $\ml^{4\text{pt}}\in O(\lambda_{V/W})$ is explicitly calculated as:
\be
\begin{aligned}
\ml^{{\rm 4pt}}	&=\frac{1}{2}\lambda_{V}\left(2+3\tan^{2}\tau_{2}+\tau_{2}\tan\tau_{2}\right)+\frac{1}{2}\lambda_{W}\left(2+3\tan^{2}\tau_{3}+\tau_{3}\sin2\tau_{3}\right)\\
	&-(\lambda_{V}+\lambda_{W})\tan\tau_{2}\tan\tau_{3}\\
	&=:\ml^{{\rm 2pt}}\left(\tau_{2},\lambda_{V}\right)+\ml^{2{\rm pt}}\left(\tau_{3},\lambda_{W}\right)+\ml^{{\rm conn}}.
\end{aligned}
\ee
As in the previous case, the integrands decompose into a disconnected contribution, $\ml^{\text{2pt}}$, arising from individual two-point functions, and a connected contribution, $\ml^{\text{conn}}$:
\be
\begin{aligned}
    &\ml^{{\rm 2pt}}\left(\tau_{2},\lambda_{V}\right)	=2\lambda_{V}\left(2+3\tan^{2}\tau_{2}+\tau_{2}\tan\tau_{2}\right),\\
&\ml^{{\rm conn}}	=-(\lambda_{V}+\lambda_{W})\tan\tau_{2}\tan\tau_{3}.
\end{aligned}
\ee
$\ml^{\text{2pt}}$ can be obtained by taking the infinite temperature limit of equation (3.11) in \cite{Goel:2023svz}. For later convenience, we present the result of one-loop determinant in two-point correlator below:
\be
{\rm Det}_{2}^{V}\equiv{\rm det}\left[\frac{\partial^{2}I_{0}^{{\rm 2pt}}}{\partial\te_{i}\partial\te_{j}}\mid_{i=1,2}^{\te_{i}=\te_{i}^{*}}\right]=\frac{8\cos^{2}\tau_{2}}{\lambda_{V}}\left[1+\ml^{{\rm 2pt}}\left(\tau_{2},\lambda_{V}\right)+O\left(\lambda_{V}^{2}\right)\right]
\ee
Therefore, for the normalized four point function, the ratio between the corresponding determinants can be evaluated as:
\be
\begin{aligned}
\left[\frac{{\rm Det}_{4}^{VW}}{{\rm Det}_{2}^{V}{\rm Det}_{2}^{W}}\right]^{-1/2}	&=2\left[1-\frac{1}{2}\ml^{{\rm conn}}+O\left(\lambda_{V/W}^{2}\right)\right]\\
	&=2\left[1+\frac{1}{2}(\lambda_{V}+\lambda_{W})\tan\tau_{2}\tan\tau_{3}+O\left(\lambda_{V/W}^{2}\right)\right]
\end{aligned}
\ee
Notably, the corrections at order $O(\lambda_{V/W})$ cancel out with that from $I_{1*}^{4\text{pt}}$ in \eqref{eq:gstar}, suggesting that the correction from their combination to the normalized four-point correlator starts at $O(\lambda^{2}_{V/W})$, therefore, we deduce that:
\be
\left[\frac{{\rm Det}_{4}^{VW}}{{\rm Det}_{2}^{V}{\rm Det}_{2}^{W}}\right]^{-1/2}	 \exp\left(I^{4\text{pt}}_{1*}-\sum_{i\in\mi}I_{1*}^{{\rm 2pt}}\left(\lambda_{i},\tau_{i}\right)\right)=2\left[1+O\left(\lambda_{V/W}^{2}\right)\right].
\ee
Finally, we consider the corrections to the saddle point value of $f$, which can be explicitly calculated as:
\be
-\frac{I_{0*}}{\lambda}+I_{c}=-\log\cos^{2\Del_{V}}\tau_{2}\cos^{2\Del_{W}}\tau_{3}+\lambda(\Del_{V}^{2}+\Del_{W}^{2}+\frac{1}{2}\left(\Del_{V}\tan\tau_{2}-\Del_{W}\tan\tau_{2}\right)^{2}+O\left(\lambda^{2}\right),
\ee
where we have set $\lambda_{V/W} =\lambda \Delta_{V/W}$ in the final step and then conduct the $\lambda$ expansion. Similarly, the leading order logarithmic terms come from the two-point calculation, for the normalized case, we have:
\be
\exp\left[\left(-\frac{I_{0*}}{\lambda}+I_{c}\right)-\sum_{i\in\mi}\left(-\frac{I_{0*}^{{\rm 2pt}}}{\lambda}+I_{c}^{{\rm 2pt}}\right)\right]\simeq\left(1-\lambda\Del_{V}\Del_{W}\tan\tau_{2}\tan\tau_{3}+O\left(\lambda_{V/W}^{2}\right)\right).
\ee
Finally, converting $\tau_2 \to iJ(T_2-T_1)$, $\tau_3 \to i J T_f$, and taking the strong semi-classical limit, we recover the leading order connected contribution in Large $p$ SYK at infinite temperature:
\be
\frac{\bra V\left(T_f\right)W\left(T_{2}\right)W\left(T_{1}\right)V\left(0\right)\ket}{\bra V\left(T_f\right)V\left(0\right)\ket\bra W\left(T_{2}\right)W\left(T_{1}\right)\ket}\stackrel{\lambda\to0}{\simeq}1+\lambda\Del_{V}\Del_{W}\tanh J \left(T_{2}-T_{1}\right)\tanh J  T_f+O\left(\lambda_{V/W}^{2}\right).
\ee
This is consistent with the observation that the initial decay of the two-point function slows down in the presence of matter chord. A similar phenomenon was previously discovered in the Brownian version of double-scaled SYK \cite{Milekhin:2023bjv}, and also the high temperature triple scaled limit of DSSYK in \cite{Almheiri:2024xtw}, where the theory becomes tractable by the Heisenberg algebra. 

Notably, the one-loop correction from energy fluctuations around the saddle exhibits the same time dependence as the leading \(1/N\) correction in the infinite temperature limit of the large \(p\) SYK model, as discussed in \cite{Choi:2019bmd}. Here, with \(\lambda \equiv 2p^2/N\), \(\Delta_V = p_V/p\), and \(\Delta_W = p_W/p\), taking the limit \(\Delta_V \to 0, \Delta_W \to 0\) implies that the size of the matter insertion are small compared to the size of the SYK Hamiltonian, similar to the large \(p\) limit in the conventional SYK model. In this regime, the result becomes \(\lambda \Delta_V \Delta_W = p_V p_W / N \simeq O(N^{-1})\), which accounts for the \(1/N\)-suppression. Therefore, our results imply that incorporating loop contributions from energy fluctuations provides a systematic way to recovering \(1/N\) corrections in the large-\(p\) SYK model from the double-scaled regime. A detailed investigation of this prescription is left for future work.

\subsection{The Growth of Chord Number and its Bulk Manifestation}
In \cite{Rabinovici:2023yex}, the authors demonstrated that, in the zero-particle sector $\mathcal{H}_0$, the states with a fixed chord number can be identified as the Krylov basis, while the expectation value of the chord number operator quantifies the Krylov complexity of the corresponding state. In the triple-scaling limit, where the edge of the energy spectrum is probed ($\tilde{e} \simeq \lambda s$, $\lambda \to 0$) and $\lambda^2 e^{-\lambda n}$ remains fixed, they showed that the Krylov basis aligns with the eigenbasis of the two-sided (renormalized) length operator. Thus, the geometric interpretation of Krylov complexity naturally corresponds to the (normalized) two-sided geodesic distance in pure JT gravity.

In this section, we extend this framework to JT gravity coupled with bulk matter. Here, we show that the geometric manifestation of the total chord number $N_\Delta (t)$, in presence of the matter chord, can be interpreted as the two-sided length operator in the presence of matter, which has been studied in \cite{Harlow:2021dfp}. We also present the exact time-dependence of this operator when evolved under the two-sided Hamiltonian, $H_L$ and $H_R$.

It has been shown in \cite{Lin:2022rbf} by shifting the ground state energy and picking up the leading-order non-trivial contribution in the triple scaling limit above,   the left and right chord Hamiltonian can be written in continuous varaibles $l_{L/R}$ parametrizing the renormalized left and right length as welll their conjugate momentum $k_{L/R}$ as:
\be \label{eq:two-H}
\begin{aligned}
& H_L=\frac{1}{2 C}\left[k_L^2-i\left(k_R-k_L+i \Delta\right) e^{-l_L}+e^{-l_L-l_R}\right], \\
& H_R=\frac{1}{2 C}\left[k_R^2+i\left(k_R-k_L-i \Delta\right) e^{-l_R}+e^{-l_L-l_R}\right],
\end{aligned}
\ee
where $\Delta$ is the weight of the matter chord and $(2C)^{-1}=\lambda J$. The appearance of imaginary coefficient in the second term of each line is unusual, which origins from the failure of $l_L$ and $l_R$ being Hermitian as operators in the original Hilbert space. Yet for our purpose in this section, these are not relevant for the discussion of the two-sided time-evolution of the total length $L=l_L+l_R$ generated by \eqref{eq:two-H}, As we will demonstrate below, the dynamics of \( L \) are well-described in a real covariant phase space with given boundary conditions to be specified in the following discussion. The time evolution in the aforementioned phase space in terms of conjugate variables \((l_{L/R}, k_{L/R})\) is defined through the canonical equations of motion induced by left/right Hamiltonians as follows:
\begin{align}
\frac{\partial l_{L / R}}{\partial t_L} & = \frac{\partial H_L}{\partial k_{L / R}}, \\
\frac{\partial l_{L / R}}{\partial t_R} & = \frac{\partial H_R}{\partial k_{L / R}}, \\
\frac{\partial k_{L / R}}{\partial t_L} & = -\frac{\partial H_L}{\partial l_{L / R}}, \\
\frac{\partial k_{L / R}}{\partial t_R} & = -\frac{\partial H_R}{\partial l_{L / R}},
\end{align}
where we introduce left and right times \( t_{L/R} \) for generality. We are interested in the functional dependence of \( L \) on both \( t_L \) and \( t_R \). Given the Hamiltonian in \eqref{eq:two-H}, it follows that:
\begin{align} \label{eq:dyn-L}
\frac{\partial L}{\partial t_{L / R}} & = \frac{k_{L / R}}{C}, \\
\frac{\partial k_{L / R}}{\partial t_{L / R}} & = H_{L / R} - \frac{k_{L / R}^2}{2C}, \\
\frac{\partial k_{L / R}}{\partial t_{R / L}} & = \frac{1}{2C} e^{-L}, \\
\frac{\partial H_{L / R}}{\partial t_{L / R}} & = 0, \\
\frac{\partial H_{L / R}}{\partial t_{R / L}} & = 0.
\end{align}
The last two equations in \eqref{eq:dyn-L} indicate that \( H_{L/R} \) are conserved in two-sided time evolution, allowing them to be specified as initial conditions for a given solution. With \( k_{L/R}|_{t_L = t_R = 0} = 0 \) and \( L|_{t_L = t_R = 0} = L_0 \), the general solution is:
\be \label{eq:L-1}
L = 2 \log \left[ e^{L_0 / 2} \cosh\left(\alpha_{L} t_{L}\right) \cosh\left(\alpha_{R} t_{R}\right) + \frac{e^{-L_0 / 2}}{2C \sqrt{E_L E_R}} \sinh\left(\alpha_{L} t_{L}\right) \sinh\left(\alpha_{R} t_{R}\right) \right],
\ee
where \( E_{L/R} = H_{L/R} \) are the conserved left/right energies and \(\alpha_{L/R}\) is defined as:
\be
\alpha_{L/R} = \sqrt{\frac{E_{L/R}}{2C}}.
\ee
The initial conditions \( l_L = l_R = L_0 / 2 \), \( k_L = k_R = 0 \) imply:
\be \label{eq:initial-E}
E_L = E_R = E = \frac{1}{2C} e^{-L_0} (1 + \tilde{\Delta}), \quad \tilde{\Delta} = e^{L_0 / 2} \Delta.
\ee
Given that \( (2C)^{-1} = \lambda J \), energy fluctuations around the ground state are of order \( O(\lambda J) \) in the triple scaling limit, consistent with \cite{Rabinovici:2023yex}. Furthermore, as shown in \cite{Xu:2024hoc}, the addition of a single matter chord does not affect the range of the energy spectrum. Nevertheless, equation \eqref{eq:initial-E} indicates that it does contribute to the energy of a two-sided wormhole with fixed length. These two observations are consistent, as in the $\lambda \to 0$ limit, the energy spectrum range extends to infinity.

The length \( L \) can be expressed in terms of energy fluctuations as:
\be \label{eq:fixed-E}
\begin{aligned}
L & = 2 \log \left( \cosh(t_L + t_R) \sqrt{\lambda J E} + \tilde{\Delta} \cosh(\sqrt{\lambda J E_0} t_L) \cosh(\sqrt{\lambda J E} t_R) \right) \\
& \quad - 2 \log \sqrt{\frac{E (1 + \tilde{\Delta})}{\lambda J}}.
\end{aligned}
\ee 
When \(\Delta \to 0\), this reduces to the no-particle case explored in \cite{Rabinovici:2023yex}. To explore the functional dependence of \( L(t_L, t_R) \) under small perturbations, we set \( t_L = -t_R = t \) and find, for \(\Delta \ll 1\):
\be
\begin{aligned}
L & \approx 2 \log \left( 1 + \Delta \cosh^2(\lambda J t) \right) \\
& = \begin{cases} 2 \Delta e^{2 \lambda J t} & t \lesssim \frac{\log \Delta^{-1}}{2 \lambda J}, \\
2 \lambda J t & t \gg \frac{\log \Delta^{-1}}{2 \lambda J}. \end{cases}
\end{aligned}
\ee 
Thus, the two-sided length initially exhibits exponential growth before transitioning to linear growth at later times.

\section{Finite Temperature Effect}\label{sec:finite-T}
In this section we examine the role of finite temperature effect in double-scaled SYK. We introduce the Hartle-Hawking wavefunction in both the $0$-particle sector and $1$-particle sector, and compare the speed of state spreading at different temperatures. In particular, we show that such effects always slow down the spreading speed at initial times.
\subsection{Hartle-Hawking Wavefunction in $0$-Particle Sector}
In $0$-particle sector the Hartle-Hawking state at inverse temperature $\beta$ can be created by acting on the empty chord state $|\Omega\ket$ 
\be \label{eq:HH-def-r0}
|\Psi\left(\beta\right)\ket\equiv e^{-\beta H/2}|\Omega\ket=\sum_{n=0}^{\infty}\Psi_{\beta}\left(n\right)|n\ket.
\ee 
One might question why \eqref{eq:HH-def-r0} qualifies as the Hartle-Hawking state, as it appears to be a pure state in $\mh_0$. However, as shown in \cite{Xu:2024hoc}, the inclusion of matter chords leads to the emergence of a pair of Type II$_1$ von Neumann algebras—namely, the left and right double-scaled algebras, $\mathcal{A}_{L/R}$—with the empty state serving as the unique tracial state for both algebras. In this framework, the empty chord state can be interpreted as the maximally entangled state between the two sides. Consequently, the action of $e^{-\beta H}$ on this state prepares a thermal field double state of the two sides. Another illustration of the idea is to consider the no-particle limit of the $1$-particle state, in which case one have a clear distinction between the left and right, and we will show that the limit agrees with \eqref{eq:HH-def-r0}.

\begin{figure}
    \centering
\includegraphics[width=0.8\linewidth]{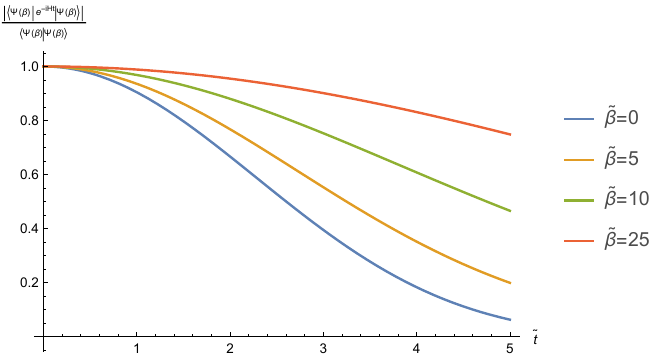}
    \caption{We plot the normalized overlap between \( e^{-iHt}|\Psi(\beta)\rangle \) and \( |\Psi(\beta)\rangle \) as a function of \( \tilde{t} = t/\sqrt{1-q} \). The plot indicates that finite temperature effects generally slow down the initial decay of the overlap, implying that the infinite temperature state exhibits the fastest initial spreading compared to any finite temperature state. }
    \label{fig:decay-HH-r0}
\end{figure}

The overlap function $\Psi_{\beta}(n)$ can be evaluated following similar strategy introduced in previous section:
\be
\Psi_{\beta}\left(n\right)=\bra n|\Psi\left(\beta\right)\ket=\int\dd\mu\left(\te\right)H_{n}\left(\cos\te|q\right)e^{-\tilde{\beta}\cos\te}\left(q;q\right)_{n}^{-1/2}.
\ee 
The energy integral can be evaluated following similar procedure in \eqref{eq:expansion}, with $it$ replaced by $\beta$. The result can be given explicitly as:
\be
\Psi_{\beta}\left(n\right)=\sum_{m=0}^{\infty}(-1)^m q^{\binom{m}{2}}\frac{\left(q;q\right)_{n+m}}{\sqrt{\left(q;q\right)_{n}}\left(q;q\right)_{m}}\frac{1-q^{n+2m}}{1-q^{n+m}}I_{n+2m}\left(\tilde{\beta}\right),
\ee
where $\tilde{\beta}=\beta/\sqrt{1-q}$ and $I_n$ is the $n$-th Bessel $I$ function. The partition function $Z(\beta)$ can then be calculated as the overlap between the Hartle-Hawking state as:
\be
Z\left(\beta\right)=\bra\Psi(\beta)|\Psi\left(\beta\right)\ket=\sum_{m\in\mbz}^{\infty}\left(-1\right)^{m}q^{\binom{m}{2}}I_{2m}(2\tilde{\beta}).
\ee 
We can then examine the spreading of $\Psi_\beta$ under the time-evolution generated by $H$. This probed by considering 
\be \label{eq:HH-decay-r0}
Z^{-1}(\beta)\bra \Psi(\beta)|e^{-iHt} |\Psi(\beta)\ket=Z^{-1}(\beta )\bra \Psi(\beta)|\Psi(\beta+ 2 i t) \ket .
\ee
We find that the initial spreading becomes slower as the temperature is decreased. We present the numeric plot of  \eqref{eq:HH-decay-r0} in figure \ref{fig:decay-HH-r0}.

\subsection{Hartle-Hawking Wavefunction in the 1-Particle Sector}

We now define the Hartle-Hawking wavefunction within the 1-particle Hilbert space, incorporating the actions of both the left and right Hamiltonians: \cite{Lin_2023}
\be
|\Psi_{\beta_{L},\beta_{R}}^{\Del}\rangle \equiv \exp\left(-\frac{\beta_{L}}{4}H_{L} - \frac{\beta_{R}}{4}H_{R}\right) |\Del;0,0\rangle = \sum_{n_{L},n_{R}=0}^{\infty} \Psi_{\beta_{L},\beta_{R}}^{\Del}(n_{L},n_{R}) |\Del;n_{L},n_{R}\rangle.
\ee
However, determining the coefficients precisely poses a challenge since the states \( |\Delta, n_L, n_R \rangle \) are not orthogonal. As a result, projecting \( |\Psi^{\Delta}_{\beta_L, \beta_R} \rangle \) onto a fixed-length state does not directly yield the corresponding coefficient function:
\be
\Psi_{\beta_{L},\beta_{R}}^{\Del}(n_{L},n_{R}) \not= \langle \Del;n_{L},n_{R} | \Psi_{\beta_{L},\beta_{R}}^{\Del} \rangle.
\ee
To compute these coefficients, we utilize the factorization map introduced in section \ref{sec:factorize}. The key is to analyze the wavefunction in the factorized Hilbert space, where an orthogonal basis \( |n_L, n_R) \) exists, based on the natural inner product in \( \mh_0 \otimes \mh_0 \). Thus, we have:
\be
\mf |\Psi_{\beta_{L}, \beta_{R}}^{\Del} \rangle = \sum_{n_{L}, n_{R}=0}^{\infty} \tilde{\Psi}_{\beta_{L}, \beta_{R}}^{\Del}(n_{L}, n_{R}) |n_{L}, n_{R}).
\ee
Here, the coefficients are determined by:
\be
\tilde{\Psi}_{\beta_{L}, \beta_{R}}^{\Del}(n_{L}, n_{R}) = (n_{L}, n_{R} | \mf | \Psi_{\beta_{L}, \beta_{R}}^{\Del} \rangle.
\ee
Using the conjugation relation \eqref{eq:Hh-relation}, we now deduce:
\be \label{eq:image-Psi}
\begin{aligned}
\mathcal{F} \left| \Psi_{\beta_L, \beta_R}^{\Delta} \right\rangle &= \mathcal{F} e^{-\beta_L H_L / 4 - \beta_R H_R / 4} \mathcal{F}^{-1} \mathcal{E}_{\Delta}^{-1} | 0,0 ) \\
&= e^{-\beta_L h_L / 4 - \beta_R h_R / 4} | 0,0 ) = |\Psi_{\beta_L / 2}, \Psi_{\beta_R / 2} ).
\end{aligned}
\ee
This shows that the image of \( |\Psi^{\Delta}_{\beta_L, \beta_R} \rangle \) under the factorization map is precisely a tensor product of 0-particle wavefunctions. Thus we know $\tilde{\Psi}^{\Delta}_{\beta_L,\beta_R}=\Psi_{\beta_L/2}(n_L) \Psi_{\beta_R/2}(n_R)$.  Moreover, we note:
\be
\begin{aligned}
\mathcal{F} \left| \Psi_{\beta_L, \beta_R}^{\Delta} \right\rangle &= \sum_{n_L, n_R=0}^{\infty} \Psi_{\beta_L, \beta_R}^{\Delta}(n_L, n_R) \mathcal{F} \left| \Delta; n_L, n_R \right\rangle \\
&= \sum_{n_L, n_R=0}^{\infty} \Psi_{\beta_L, \beta_R}^{\Delta}(n_L, n_R) \mathcal{E}_{\Delta}^{-1} | n_L, n_R ).
\end{aligned}
\ee
Matching this with the second line of equation \eqref{eq:image-Psi}, we obtain a consistent expression for the coefficients for arbitrary \( n_L, n_R \):
\be
\Psi_{\beta_{L}, \beta_{R}}^{\Del}(n_{L}, n_{R}) = (n_{L}, n_{R} | \mathcal{E}_{\Delta} | \Psi_{\beta_{L}/2}, \Psi_{\beta_{R}/2} ).
\ee
Now, we consider the time evolution governed by the operator \( \exp(-i(H_L t_L + H_R t_R)) \), with the normalization factor defined as:
\begin{equation}
Z_\Delta(\beta_L, \beta_R) = \langle \Psi_{\beta_L, \beta_R}^{\Delta} | \Psi_{\beta_L, \beta_R}^{\Delta} \rangle = \int \mathrm{d} \mu_{\Delta} (\theta_L, \theta_R) \, e^{-\beta_L E_L / 2 - \beta_R E_R / 2}.
\end{equation}
This can be evaluated similar to \eqref{eq:overlap-r1-2}, and it gives $Z_\Delta$ in terms of the $0$-particle Hartle-Hawking state as:
\begin{equation}
Z_{\Delta}(\beta_L,\beta_R)= \sum_{n=0}^{\infty} q^{\Delta n} \Psi_{\beta_L}(n) \Psi_{\beta_R}(n).
\end{equation}
It is evident from this definition that as \(\Delta \to 0\), we have \(\lim_{\Delta \to 0} Z_\Delta(\beta_L, \beta_R) = Z(\beta_L + \beta_R)\).

Upon making the substitution \( \beta_{L/R} \to \beta_{L/R} + 2i t_{L/R} \), we can derive the following expression:
\begin{equation}
\langle \Psi_{\beta_L, \beta_R}^{\Delta} | e^{-iH_L t_L - iH_R t_R} | \Psi_{\beta_L, \beta_R}^{\Delta} \rangle = \sum_{n=0}^{\infty} q^{\Delta n} \Psi_{\beta_L + 2i t_L}(n) \Psi_{\beta_R + 2i t_R}(n).
\end{equation}
As an illustration, we plot the quantity \( Z^{-1}(\beta_L, \beta_R) \langle \Psi_{\beta_L, \beta_R}^{\Delta} | e^{-\mathcal{H} t} | \Psi_{\beta_L, \beta_R}^{\Delta} \rangle \), specifically with \(\beta_L = \beta_R = \beta\), and examine the functional dependence in $t$. We observe that for a fixed time \( t \), the result is minimized at infinite temperature $\beta=0$. As the temperature decreases (i.e., as \(\beta\) increases), the result generally increases, indicating a slower decay rate.

\begin{figure}
    \centering
    \includegraphics[width=0.8\linewidth]{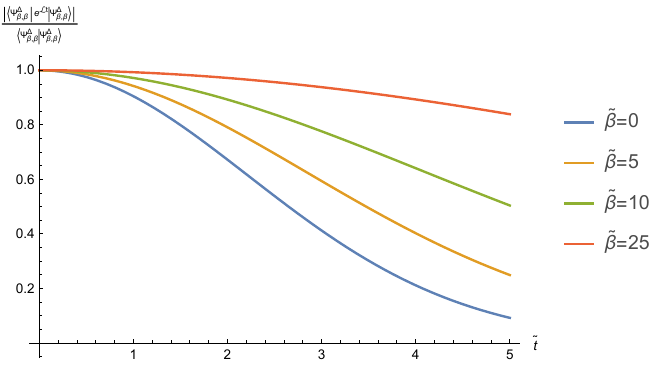}
    \caption{We plot the overlap function $Z^{-1}(\beta_L,\beta_R)\bra \Psi^{\Delta}_{\beta_L,\beta_R}|e^{\ml t}|\Psi^{\Delta}_{\beta_L,\beta_R}\ket$ as a function of $\tilde{t}$ for $\beta_L=\beta_R=\beta$ and fixed $q=0.8, q^\Delta=0.5$. The result shows that the spreading under Liouvillian time evolution slows down as the decrease of temperature (increase of $\beta$).}
    \label{fig:decay-HH-r1}
\end{figure}

The results align with the expectation that the operator spreading speed decreases as the temperature is lowered. This phenomenon has been explicitly confirmed in the large-$q$ SYK model at all temperatures, as demonstrated in \cite{Qi_2019}. The latter model can be realized as the semi-classical limit of the double-scaled SYK model in the current study, as shown in \cite{mukhametzhanov2023largepsykchord,Goel:2023svz}. Our results indicate that a similar behavior extends to the double-scaled limit of SYK, where both $q$ and $N$ are taken to infinity in a specific way. In this regime, chord operators correspond to infinitely massive fermionic strings in the original model, yet they remain local compared to the overall system as $p/N \to 0$. Consequently, non-trivial spreading still occurs in this case, with the chord operator offering an effective description of a regulated version of the spreading. Specifically, it treats an infinitely heavy fermionic string as a single element localized at the boundary.

Notably, our results demonstrate that the spreading speed associated with chord operators remains finite at infinite temperature, with the slowdown persisting as the temperature decreases all the way to low temperatures.

\section{Discussions and Future Prospective}\label{sec:Future}
In this work, we presented analytical results for the real-time evolution generated by the chord Hamiltonian, valid generally for \( q \in [0,1) \) and all temperatures. This evolution leads to a probability distribution over the operator basis, which spans a subspace in the double-scaled chord algebra \(\mathcal{A}_{L/R}\). At infinite temperature, we explicitly show that, at very late times \( t \gg \lambda^{-1/2} \), the projection of this probability onto a fixed operator basis, quantified by the wavefunctions \(\{ \phi_n(t) \}\), decays universally as \(\sim t^{-3/2}\), independent of \( q \) in the valid regime. This result also implies a polynomial decay of the two-point function \(\langle \Omega|V(t)V(0)|\Omega \rangle\) for general matter weight \(\Delta > 0\) at infinite temperature, reminiscent of the late-time behavior of matter correlations in disk JT gravity \cite{Mertens:2022irh}.

We also extended this analysis to the operator growth of a single matter chord under time evolution generated by the Liouvillian operator. We developed an isometric factorization map that provides an analytical tool for evaluating the inner product in the one-particle chord Hilbert space \(\mathcal{H}_1\). With this approach, we examined various quantities and their functional dependencies on both the matter weight \(\Delta\) and time. Interestingly, we established the equivalence between the crossed four-point matter chord correlation function and the expectation value of the chord number-generating function in the one-particle state. Furthermore, we demonstrated that the presence of other particles slows down the decay of two-point functions, a result found previously in the Brownian version of DSSYK \cite{Milekhin:2023bjv} and conjectured to hold for specific correlators in the static patch of de Sitter space \cite{Milekhin:2024vbb}.

Throughout, we consider a general, fixed \(q \in [0,1)\). The limit \(q \to 1^{-}\) requires careful analysis. We showed that by focusing on different regions of the energy spectrum, one obtains semiclassical physics with drastically different behavior. Near the spectrum's edge, where we consider energy fluctuations of \(O(\lambda)\) above the ground state, we recover the Schwarzian description, consistent with the low-temperature limit of the conventional large \(N\) SYK model. This behavior restores the \(\sim t^{-3}\) decay of the two-point function at extremely late times. In contrast, localizing around the center of the spectrum yields Gaussian decay of correlation functions governed by the Heisenberg algebra \cite{Almheiri:2024xtw}. Including one-loop corrections from energy fluctuations around the center, we determine leading-order $1/N$ corrections to the uncrossed four-point function, aligning with the infinite-temperature limit of large \(q\) SYK \cite{Streicher_2020}. This further supports the notion that particle decay slows in the presence of matter in this regime. Additionally, we extended the concept of Krylov complexity \cite{Rabinovici:2023yex} to the one-particle case. In the semiclassical limit of the chord algebra \cite{Lin_2023}, we recover the gravitational algebra by Harlow and Wu \cite{Harlow:2021dfp}, with the total chord number $N_\Delta (t)$ manifesting as the two-sided length operator in JT gravity with bulk matter.

Finally, we examined the effects of finite temperature and evaluated the overlap between the Hartle-Hawking wavefunction and chord number states. Using our analytical expressions, we compared the spreading at different temperatures and found that chord spreading is fastest and finite at infinite temperature, decreasing with lower temperatures. This behavior was signaled by examining the time dependence of overlaps between the Hartle-Hawking state under Liouville time evolution and its initial state. Verifying this by evaluating finite-temperature chord number generating functions, as in \cite{Qi_2019}, would be a promising calculation, made feasible by the analytical tools developed here.

We list several interesting future prospectives of the current work below:

\paragraph{Supersymmetric Extension} The extension of the isometric factorization map to the supersymmetric chord Hilbert space, as developed in \cite{Berkooz:2020xne}, would be desirable, especially for computing the four-point function. This is of particular interest because it provides a way to determine the Lyapunov exponent. In addition, \cite{Boruch:2023bte} observed that the chord Hamiltonian demonstrates enhanced $\mathcal{N}=4$ supersymmetry in the absence of matter chords. This raises the intriguing question of whether such enhancement leaves an imprint on the real-time chord dynamics explored in this paper.

\paragraph{Generalization to Muli-particle states} The isometric factorization map offers an equivalent description that captures the dynamical aspects of the one-particle chord Hilbert space, $\mh_1$. Extending this framework to states with multiple particles would be particularly valuable. Specifically, one should be able to derive the crossed four-point function by directly evaluating the inner product between two-particle states, where a swap of the endpoints of two matter chords induces the crossing. This approach would provide a bulk perspective on the emergence of the quantum $6j$ symbol, reminiscent of the structure uncovered in \cite{Lam:2018pvp}.

\paragraph{Towards Stringy Bulk and Finite $G_N$} A quantized Euclidean disk with discrete radial directions has been examined in earlier studies \cite{qdisk1999-1, qdisk1999-2} and revisited recently in \cite{Almheiri:2024ayc}, where it was shown to support a $q$-deformed version of the holographic dictionary. An open question remains whether the boundary dual of a quantized free field on the quantum disk corresponds to the $q$-Schwarzian theory, previously derived from a bulk formulation\cite{Blommaert:2023wad}. Moreover, applying the boundary operator algebra construction from \cite{Gesteau:2024rpt} to this setup could help elucidate the emergent quantized radial direction, potentially extending the concept of bulk causality to a stringy regime.

Furthermore, DSSYK incorporates a form of noncommutative geometry, analogous to introducing finite gravitational coupling $G_N$ in the bulk (in addition to the $\alpha'$ corrections). The approach in \cite{Berkooz:2022mfk} suggests an alternative formulation of the theory, interpreting it as a quantized particle propagating in a noncommutative bulk. Further exploration of this perspective could yield insights into the nature of quantum gravity at finite $G_N$.

\section{Acknowledgement}
{\justify
I would like to express my gratitude to Xi~Dong, Elliott~Gestaeu, Jeremy van der Heijden, Alexey Milekhin, Vladimir Narovlansky, Xiaoliang Qi, Erik P. Verlinde and Herman L. Verlinde for stimulating discussions and suggestions. I appreciate valuable insights from Henry W. Lin, Douglas Stanford and Leonard Susskind. I thank Ahmed Almheiri, Micha Berkooz, Yiming Chen, Felix M. Haehl, Sergio E. Aguilar-Gutierrez, Henry Maxfield, Kazumi Okuyama, Sirui Shuai, Brian Swingle, Haifeng Tang, Gustavo J. Turiaci, Jinzhao Wang, Wayne W. Weng, Cynthia Yan, Zhenbin Yang, Mengyang Zhang, Ying Zhao and Yuzhen Zhang for helpful comments and correspondence. I gained valuable insights and knowledge at the “Workshop on Quantum Information and Physics”, sponsored by the Institute for Advanced Study, where the majority of this work was completed. I extend my gratitude to the coordinators for organizing such a wonderful event. I acknowledge the support of the U.S. Department of Energy, Office of Science, Office of High Energy Physics, under Award Number DE-SC0011702.
}

\appendix

\section{Useful Formulas} \label{app:Formulas}
We present a self-contained derivation of the following formulas: 
\be \label{eq:integral-H0}
\int_{0}^{\pi}\dd\mu\left(\te\right)e^{2ik\te}=\frac{1}{2}\left(-1\right)^{k}q^{\binom{k}{2}}\left(1+q^{k}\right).
\ee 
\be\label{eq:integral_Hn}
\int_{0}^{\pi}\dd\mu\left(\te\right)H_{n}\left(\cos\te|q\right)	e^{i\left(n+2k\right)\te}=
	\frac{1}{2}\left(-1\right)^{k}q^{\binom{k}{2}}\frac{\left(q;q\right)_{n+k}}{\left(q;q\right)_{k}}\frac{\left(1-q^{n+2k}\right)}{\left(1-q^{n+k}\right)},\qq k=0,1,\cdots.
\ee
For a more comprehensive derivation, we refer the reader to \cite{Askey1983AGO}. We start to prove the first integral \eqref{eq:integral-H0} by introducing the $\beta$-deformed measure:
\be
\dd\mu_{\beta}\left(\te\right)=\frac{\left(e^{\pm2i\te},q;q\right)_{\infty}}{\left(\beta e^{\pm2i\te};q\right)_{\infty}}\frac{\dd\te}{2\pi},\quad\lim_{\beta\to0}\dd\mu_{\beta}\left(\te\right)=\dd\mu\left(\te\right).
\ee
It exhibits a simpler expansion in terms of the angular part $e^{2i\theta}$:
\be
\frac{\left(e^{2i\te};q\right)_{\infty}}{\left(\beta e^{2i\te};q\right)_{\infty}}=\sum_{n=0}^{\infty}\frac{\left(\beta^{-1};q\right)_{n}\beta^{n}}{\left(q;q\right)_{n}}e^{2in\te}.
\ee
Expanding the integrand in terms of $\beta$, we get 
\be
\int_{0}^{\pi}\dd\mu_{\beta}\left(\te\right)e^{2ik\te}	=\sum_{m,n=0}^{\infty}\frac{\left(\beta^{-1};q\right)_{m}\left(\beta^{-1};q\right)_{n}}{\left(q;q\right)_{m}\left(q;q\right)_{n}}\beta^{m+n}\int_{0}^{\pi}\frac{\dd\te}{2\pi}e^{2i\left(k+m-n\right)\te}.
\ee 
The integral over $\theta$ makes the summand supported in  $k+m=n$, therefore, we find
\be \label{eq:intermidiate-1}
\begin{aligned}
\int_0^\pi \mathrm{d} \mu_\beta(\theta) e^{2 i k \theta} & =\frac{1}{2} \sum_{m=0}^{\infty} \frac{\left(\beta^{-1} ; q\right)_m\left(\beta^{-1} ; q\right)_{m+k}}{(q ; q)_m(q ; q)_{m+k}} \beta^{2 m+k} \\
& =\frac{1}{2} \beta^k \frac{\left(\beta^{-1} ; q\right)_k}{(q ; q)_k}{ }_2 \varphi_1\left(\begin{array}{c}
\beta^{-1}, q^k \beta^{-1} \\
q^{k+1}
\end{array} ; q, \beta^2\right),
\end{aligned}
\ee 
where the basic hypergeometric function ${}_2 \varphi_1$ is defined as
\be
_{2}\vp_{1}\left(\begin{array}{c}
a,b\\
c
\end{array};q,z\right)=\sum_{n=0}^{\infty}\frac{\left(a,b;q\right)_{n}}{\left(c,q;q\right)_{n}}z^{n},
\ee
which truncates when either $a$ or $b$ is of the form $q^{-n},n\in\mathbb{N}$. We then apply the following conformal property:
\be \label{eq:conformal-1}
\thinspace_{2}\varphi_{1}\left(\begin{array}{c}
\beta^{-1},q^{k}\beta^{-1}\\
q^{k+1}
\end{array};q,\beta^{2}\right)=\frac{\left(q^{k}\beta;q\right)_{\infty}\left(q\beta;q\right)_{\infty}}{\left(\beta^{2};q\right)_{\infty}\left(q^{k+1};q\right)_{\infty}}\thinspace_{2}\vp_{1}\left(\begin{array}{c}
q^{-1},\beta^{-1}\\
\beta
\end{array};q,q^{k+1}\beta\right),
\ee
where the ${}_2 \vp_1$ function in the right hand side is simply:
\be \label{eq:truncate}
\begin{split}
 \thinspace_{2}\vp_{1}&\left(\begin{array}{c}
q^{-1},\beta^{-1}\\
\beta
\end{array};q,q^{k+1}\beta\right)	=\sum_{m=0}^{1}\frac{\left(q^{-1},\beta^{-1};q\right)_{m}}{\left(\beta,q;q\right)_{m}}\left(q^{k+1}\beta\right)^{m} \\
	&=1+\frac{\left(1-q^{-1}\right)\left(1-\beta^{-1}\right)}{\left(1-q\right)\left(1-\beta\right)}q^{k+1}\beta
	=1+q^{k}   .
\end{split}
\ee 
Plugging both \eqref{eq:truncate} and \eqref{eq:conformal-1} into \eqref{eq:intermidiate-1}, we conclude that 
\be
\int_{0}^{\pi}\dd\mu_{\beta}\left(\te\right)e^{2ik\te}=\frac{\beta^{k}\left(\beta^{-1};q\right)_{k}\left(1+q^{k}\right)}{2\left(q;q\right)_{k}}\frac{\left(q^{k}\beta;q\right)_{\infty}\left(q\beta;q\right)_{\infty}}{\left(\beta^{2};q\right)_{\infty}\left(q^{k+1};q\right)_{\infty}}.
\ee
Now by taking $\beta\to0$ limit:
\be
\begin{aligned}
\lim _{\beta \rightarrow 0} \beta^k\left(\beta^{-1} ; q\right)_k & =(-1)^k q^{\binom{k}{2}} \\
\lim _{\beta \rightarrow 0}(x \beta ; q)_{\infty} & =1.
\end{aligned}
\ee 
We conclude that 
\be
\int_{0}^{\pi}\dd\mu\left(\te\right)e^{2ik\te}=\frac{\left(-1\right)^{k}}{2}q^{\binom{k}{2}}\left(1+q^{k}\right).
\ee 
We move on to evaluate \eqref{eq:integral_Hn}:
\be
I_{n,k}=\int_{0}^{\pi}\dd\mu\left(\te\right)H_{n}\left(\cos\te|q\right)e^{i\left(n+2k\right)\te}.
\ee
Note that $H_n(\cos\theta|q)$ is a polynomial in $e^{2i\te}$, we can evaluate the integral term by term as:
\be
\begin{aligned}
    I_{n,k}	&=\sum_{j=0}^{n}\frac{\left(q;q\right)_{n}}{\left(q;q\right)_{j}\left(q;q\right)_{n-j}}\int_{0}^{\pi}e^{i\left(n-2j\right)\te}e^{i\left(n+2k\right)\te}\dd\mu\left(\te\right) \\
	&=\sum_{j=0}^{n}\frac{\left(q;q\right)_{n}}{\left(q;q\right)_{j}\left(q;q\right)_{n-j}}\frac{\left(-1\right)^{n+k-j}}{2}q^{\binom{n+k-j}{2}}\left(1+q^{n+k-j}\right)\\
	&=\sum_{j=0}^{n}\frac{\left(q;q\right)_{n}}{\left(q;q\right)_{j}\left(q;q\right)_{n-j}}\frac{\left(-1\right)^{k+j}}{2}q^{\binom{k+j}{2}}\left(1+q^{k+j}\right),
\end{aligned}
\ee
where in the second equality we used \eqref{eq:integral-H0}, and in the last line we changed variable $j$ to $n-j$. We want to sum over $j$ to get a compact formula for the integral. The trick of doing so is to convert $(q;q)_{n-j}$ to $(q;q)_j$ and make use of the $q$-binomial theorem. Note that:
\be
\frac{1}{\left(b;q\right)_{n-j}}=\frac{b^{j}\left(q^{1-n}b^{-1};q\right)_{j}}{\left(b;q\right)_{n}\left(-1\right)^{j}q^{\binom{j}{2}+\left(1-n\right)j}}.
\ee 
Therefore, we find
\be
\begin{aligned}
I_{n, k} & =\frac{1}{2}(-1)^k q^{\binom{k}{2}} \sum_{j=0}^n \frac{\left(q^{-n} ; q\right)_j}{(q ; q)_j} q^{(n+k) j}\left(1+q^{k+j}\right) \\
& =\frac{1}{2}(-1)^k q^{\binom{k}{2}} \frac{(q ; q)_{n+k}}{(q ; q)_k} \frac{1-q^{n+2 k}}{1-q^{n+k}}.
\end{aligned}
\ee
Similarly, for negative value of $k$, we conclude that 
\be
I_{n,-k}=\left(-1\right)^{k}q^{\binom{-k}{2}}\frac{\left(1-q^{n-2k}\right)\left(q^{1-k};q\right)_{n-1}}{2},
\ee 
which identically vanishes for $k=1,2,\cdots |n|-1$. Therefore, we obtain: \footnote{One can further show \be 
\int_{0}^{\pi}\dd\mu\left(\te\right)e^{i k\te}H_{n}\left(\cos\te|q\right)=0,\qq k\in\mathbb{N},|k|<n. 
\ee }
\be \label{eq:integral-vanish}
\int_{0}^{\pi}\dd\mu\left(\te\right)e^{i(n-2k)\te}H_{n}\left(\cos\te|q\right)=0,\qq k\in\mathbb{N},|k|<n. 
\ee 
\section{Detailed Derivations}
In this section we derive various results that appear in the main text. 
\subsection{The Overlap $\bra n | \Omega(t)\ket$}  \label{app:Overlap}
In this section we evaluate the overlap between $|\Omega(t)\ket = \exp(-i H t)|\Omega\ket $ and a state $|n\ket$ with chord number $n$. By inserting the energy eigenstate $|\te\ket$, we introduce the overlaping function $\phi_n (t)$ as:
\be  \label{eq:overlap}
\phi_n (t)\equiv \sqrt{\left(q;q\right)_{n}}\bra n|\Omega\left(t\right)\ket=\int_{0}^{\pi}\dd\mu\left(\te\right)e^{-2i\tilde{t}\cos\te}H_{n}\left(\cos\te|q\right),
\ee
where $\tilde{t}= t / \sqrt{1-q}$, and we have used the fact that $\bra\Omega|\te\ket =1$. Now we evaluate the integral by expanding the exponent in terms of partial waves as:
\be \label{eq:partial-wave}
\begin{aligned}
e^{-2 i \tilde{t} \cos \theta} & =\exp \left(-i \tilde{t}\left(e^{i \theta}+e^{-i \theta}\right)\right)=\sum_{k_1, k_2=0}^{\infty} \frac{(-i \tilde{t})^{k_1+k_2}}{k_{1}!k_{2}!} e^{i\left(k_1-k_2\right) \theta} \\
& =\sum_{k=0}^{\infty} \frac{(-i \tilde{t})^{2 k}}{k!^2}+\sum_{k=0}^{\infty} \sum_{l=1}^{\infty} \frac{(-i \tilde{t})^{2 k+l}}{k!(k+l)!}\left(e^{i l \theta}+e^{-i l \theta}\right) \\
& =\sum_{k=0}^{\infty} \frac{(-i \tilde{t})^{2 k}}{k!^2}+2 \sum_{k=0}^{\infty} \sum_{l=1}^{\infty} \frac{(-i \tilde{t})^{2 k+l}}{k!(k+l)!} \operatorname{Re}\left(e^{i l \theta}\right)\\
& = J_0 (2\tilde{t})+ \mathcal{I}(\theta) . 
\end{aligned}
\ee 
where we have separated the integrand into a $\theta$ independent part, which sums up to the Bessel function as shown in the last equality above.  We then organize the $\te$-dependent part $\mathcal{I}(\te)$ where the integrand is expressed in terms of the real part of $e^{il\theta}$ explicitly. Below we shall make use of the fact that both the measure function $\dd \mu (\te)$ and the wavefunction $H_n (\cos\te|q)$ is real. Therefore, we conclude that
\be \label{eq:integral-real}
\int_{0}^{\pi}\dd\mu\left(\te\right)H_{n}\left(\cos\te|q\right){\rm Re}\left(e^{il\te}\right)={\rm Re}\int_{0}^{\pi}\dd\mu\left(\te\right)H_{n}\left(\cos\te|q\right)e^{il\te}.
\ee
As a result, we don't need to calculate the integral for integer $l$s which lead to pure imaginary results. We have:
\be
\begin{aligned}
& \int_0^\pi \mathrm{d} \mu(\theta) e^{i k \theta} H_n(\cos \theta | q)=\int_{-\pi}^0 \mathrm{~d} \mu(\theta) e^{-i k \theta} H_n(\cos \theta | q) \\
& =\int_0^\pi \mathrm{d} \mu(\theta-\pi) e^{-i k(\theta-\pi)} H_n(\cos (\theta-\pi) | q) \\
& =(-1)^{n+k} \int_0^\pi \mathrm{d} \mu(\theta) e^{-i k \theta} H_n(\cos \theta | q) \\
& =(-1)^{n+k} \overline{\int_0^\pi \mathrm{d} \mu(\theta) e^{i k \theta} H_n(\cos \theta | q)}, \\
&
\end{aligned}
\ee
where we first change the variable $\theta$ to $-\theta$, and then shift $\theta$ to $\theta-\pi$. In the third equality we have used the fact that the measure is invariant under the shift, together with the parity condition: $H_n(-x|q)=(-1)^n H_n(x|q)$. Therefore, we conclude that for $k=n+l$, 
\be
\int_{0}^{\pi}\dd\mu\left(\te\right)e^{i\left(n+l\right)\te}H_{n}\left(\cos\te|q\right)\in\begin{cases}
\mathbb{R} & l=2,4\dots\\
i\mbr & l=1,3,5,\dots
\end{cases}
\ee
The integral \eqref{eq:integral-real} vanishes identically for $|k|<n$. Therefore, only terms satisfying $l=n+2k,k=0,1,\dots$ in \eqref{eq:partial-wave} are able to contribute to the overlapping function.  Combined with \eqref{eq:integral-H0} and \eqref{eq:integral_Hn}, we conclude that 
\be
\begin{split}
{\rm Re}\int_{0}^{\pi}\dd\mu\left(\te\right)H_{n}& \left(\cos\te|q\right)e^{il\te}\\
=&\begin{cases}
\frac{1}{2}\left(-1\right)^{m}q^{\binom{m}{2}}\left(1+q^{m}\right) & n=0,l=2m,m=1,2\cdots\\
\frac{1}{2}\left(-1\right)^{m}q^{\binom{m}{2}}\frac{\left(q;q\right)_{n+m}}{\left(q;q\right)_{m}}\frac{\left(1-q^{n+2m}\right)}{\left(1-q^{n+m}\right)} & n>0,l=n+2m,m=0,1,\cdots
\end{cases}    .
\end{split}
\ee
We then move on to calculate \eqref{eq:overlap}. Integrating against the $\te$-independent part ends up with 
\be
\int\dd\mu\left(\te\right)H_{n}\left(\cos\te|q\right)J_{0}\left(2\tilde{t}\right)=\del_{n0}J_{0}\left(2\tilde{t}\right).
\ee
We then look at the rest of integral: 
\be
\int_{0}^{\pi}\dd\mu\left(\te\right)H_{n}\left(\cos\te\right)\mathcal{I}\left(\te\right),
\ee
with $\mathcal{I}(\te)$ specified in \eqref{eq:partial-wave}. For $n=0$, only even $l$ can contribute, this leads to
\be \label{eq:n=0}
\begin{aligned}
\int_0^\pi \mathrm{d} \mu(\theta)\left[2 \sum_{k=0}^{\infty} \sum_{l=1}^{\infty} \frac{(-i \tilde{t})^{2 k+l}}{k!(k+l)!} \operatorname{Re}\left(e^{i l \theta}\right)\right] & =\sum_{k=0}^{\infty} \sum_{m=1}^{\infty} \frac{(-i \tilde{t})^{2 k+2 m}}{k!(k+2 m)!}(-1)^m q^{\binom{m}{2}}\left(1+q^m\right) \\
& =\sum_{m=1}^{\infty} q^{\binom{m}{2}}\left(1+q^m\right) \sum_{k=0}^{\infty} \frac{(-1)^k \tilde{t}^{2 k+2 m}}{k!(k+2 m)!} \\
& =\sum_{m=1}^{\infty} q^{\binom{m}{2}}\left(1+q^m\right) J_{2 m}(2 \tilde{t}).
\end{aligned}
\ee 
For $n>0$, only $l=n+2m,m=0,1,\dots$ contribute, we have:
\be\label{eq:expansion}
\begin{aligned}
\int_0^\pi \mathrm{d} \mu(\theta) H_n(\cos \theta | q)&\left[2 \sum_{k=0}^{\infty} \sum_{l=1}^{\infty} \frac{(-i \tilde{t})^{2 k+l}}{k!(k+l)!} \operatorname{Re}\left(e^{i l \theta}\right)\right] \\
&=\sum_{k=0}^{\infty} \sum_{m=0}^{\infty} \frac{(-i \tilde{t})^{2 k+n+2 m}}{k!(k+n+2 m)!}(-1)^m q^{\binom{m}{2}} \frac{(q ; q)_{n+m}}{(q ; q)_m} \frac{\left(1-q^{n+2 m}\right)}{\left(1-q^{n+m}\right)} \\
& =\sum_{m=0}^{\infty}(-i)^n q^{\binom{m}{2}} \frac{(q ; q)_{n+m}}{(q ; q)_m} \frac{\left(1-q^{n+2 m}\right)}{\left(1-q^{n+m}\right)} \sum_{k=0}^{\infty} \frac{(-1)^k \tilde{t}^{2 k+n+2 m}}{k!(k+n+2 m)!} \\
& =(-i)^n \sum_{m=0}^{\infty} q^{\binom{m}{2}} \frac{(q ; q)_{n+m}}{(q ; q)_m} \frac{\left(1-q^{n+2 m}\right)}{\left(1-q^{n+m}\right)} J_{n+2 m}(2 \tilde{t}).
\end{aligned}
\ee 
In particular, the results in \eqref{eq:n=0} can be organized in a more compact form, with the help of the identity:
\be
\sum_{m=1}^{\infty}q^{\binom{m}{2}+m}J_{2m}\left(2\tilde{t}\right)=\sum_{m=0}^{\infty}q^{\binom{m+1}{2}}J_{2m}\left(2\tilde{t}\right)=\sum_{m=0}^{-\infty}q^{\binom{m}{2}}J_{-2m}\left(2\tilde{t}\right).
\ee
Plugging back into \eqref{eq:n=0}, we conclude that 
\be
\phi_{n}\left(t\right)=\begin{cases}
\sum_{k\in\mbz}q^{\binom{k}{2}}J_{2k}\left(2\tilde{t}\right) & n=0\\
\left(-i\right)^{n}\sum_{k=0}^{\infty}q^{\binom{k}{2}}\frac{\left(q;q\right)_{n+k}}{\left(q;q\right)_{k}}\frac{1-q^{n+2k}}{1-q^{n+k}}J_{n+2k}\left(2\tilde{t}\right) & n=1,2,\cdots
\end{cases}.
\ee

\subsection{The Overlap $\bra\Delta;m_L,m_R|\Delta(t_L,t_R)\ket$}\label{app:Overlap-2}
In this section, we evaluate the projection of \( |\Delta(t_L, t_R)\rangle \) onto a state with a specified chord number. We begin with:
\begin{equation}
\langle \Delta; m_{L}, m_{R} | \Delta(t_{L}, t_{R}) \rangle = \int d\mu_{\Delta}(\theta_{L}, \theta_{R}) \, e^{iE_{L}t_{L} + iE_{R}t_{R}} (\theta_{L}, \theta_{R} | \mathcal{E}_{\Delta}^{-1} | m_{L}, m_{R}).
\end{equation}
We can expand the matrix component of \( \mathcal{E}_{\Delta}^{-1} \) in terms of \( q \)-Hermite polynomials as:
\begin{equation}
\begin{aligned}
(\theta_{L}, \theta_{R} | \mathcal{E}_{\Delta}^{-1} | m_{L}, m_{R}) &= \sum_{k=0}^{{\rm min}} \frac{(-1)^k q^{\binom{k}{2}} q^{\Delta k}}{(q;q)_k} \sqrt{\frac{(q;q)_{m}(q;q)_{n}}{(q;q)_{m-k}(q;q)_{n-k}}} \frac{H_{m-k}^{L} H_{n-k}^{R}}{\sqrt{(q;q)_{m-k}(q;q)_{n-k}}} \\
&= \sqrt{(q;q)_{m} (q;q)_{n}} \sum_{k=0}^{{\rm min}(m, n)} \frac{(-1)^k q^{\binom{k}{2}} q^{\Delta k}}{(q;q)_k} \frac{H_{m_{L}-k}^{L} H_{m_{R}-k}^{R}}{(q;q)_{m_{L}-k} (q;q)_{m_{R}-k}},
\end{aligned}
\end{equation}
where we have used the shorthand notation \( H^{L}_m = H_m(\cos \theta_L | q) \). The energy measure can be expanded as:
\begin{equation}
d\mu_{\Delta}(\theta_{L}, \theta_{R}) = d\mu(\theta_{L}) \, d\mu(\theta_{R}) \, \sum_{n=0}^{\infty} \frac{q^{\Delta n}}{(q;q)_n} H_{n}^{L} H_{n}^{R}.
\end{equation}
We then evaluate the integral over \( \theta_L \) and \( \theta_R \):
\begin{equation}
\int d\mu(\theta_{L}) \, e^{iE_{L}t_{L}} H_{n}^{L} H_{m_{L}-k}^{L} = \sum_{l_{L}=0}^{{\rm min}(n, m_{L}-k)} \binom{n, m_{L}-k}{l_{L}}_q \, \phi_{n + m_{L} - k - 2l_{L}}^*(t_{L}),
\end{equation}
where we have applied the linearization formula \eqref{eq:linearization} and use the notation
\begin{equation}
\binom{n, m}{k}_q \equiv \frac{(q;q)_n (q;q)_m}{(q;q)_k (q;q)_{m-k} (q;q)_{n-k}}.
\end{equation}
Thus, we conclude that:
\begin{equation}
\begin{split}
 \frac{\bra\Del;m_{L},m_{R}|\Del(t_{L},t_{R})\ket}{\sqrt{\left(q;q\right)_{m_{L}}\left(q;q\right)_{m_{R}}}}	&=\sum_{n=0}^{\infty}\sum_{k=0}^{{\rm min}(m_{L},m_{R})}\sum_{l_{L/R}=0}^{{\rm min}(n,m_{L/R}-k)}\frac{\left(-1\right)^{k}q^{\binom{k}{2}}q^{\Del(n+k)}}{\left(q;q\right)_{n}\left(q;q\right)_{k}\left(q;q\right)_{m_{L}-k}\left(q;q\right)_{m_{R}-k}}\\
&\times	\binom{n,m_{L}-k}{l_{L}}_{q}\binom{n,m_{R}-k}{l_{R}}_{q}\phi_{n+m_{L}-k-2l_{L}}^{*}\left(t_{L}\right)\phi_{n+m_{R}-k-2l_{R}}^{*}\left(t_{R}\right).
\end{split}
\end{equation}

\subsection{The Chord Number Generating Function $I^{\Delta}_\mu$} \label{app:gen-I}
In this section we evaluate the generating function $I^{\Delta}_\mu$  as a function of the left and right time $t_{L/R}$.  Concretely, we are going to evaluate the following integral:
\be \label{eq:I-integral-def}
\begin{aligned}
& I_{\Delta}^{\mu}\left(t_L, t_R\right)= \\
& \int_0^\pi \prod_{i=1}^4 \mathrm{~d} \mu\left(\theta_i\right)\bra\theta_1|q^{\Delta \hat{n}}| \theta_2\ket  (\theta_1, \theta_2|\mathcal{E}_{\Delta}^{-1} e^{-\mu\left(\hat{n}_L +  \hat{n}_R\right)} \mathcal{E}_{\Delta}| \theta_3, \theta_4) e^{-i\left(E_1-E_3\right) t_L-i\left(E_2-E_4\right) t_R}.
\end{aligned}
\ee
We can evaluate $\mathcal{E}^{-1}_\Delta$ to the right of with the help of the following commutation relation:
\be
\me_{\Del}^{-1}e^{-\mu(\hat{n}_{L}+\hat{n}_{R})}=\left(q^{\Del}a_{L}a_{R};q\right)_{\infty}e^{-\mu(\hat{n}_{L}+\hat{n}_{R})}=e^{-\mu(\hat{n}_{L}+\hat{n}_{R})}\left(q^{\Del}e^{-2\mu}a_{L}a_{R};q\right)_{\infty}.
\ee
The combination of the two operators exhibit the following expansion in $q^{\Delta}$:
\be
\frac{\left(q^{\Del}e^{-2\mu}a_{L}a_{R};q\right)_{\infty}}{\left(q^{\Del}a_{L}a_{R};q\right)_{\infty}}=\sum_{l=0}^{\infty}\frac{\left(e^{-2\mu};q\right)_{l}}{\left(q;q\right)_{l}}q^{\Del l}a_{L}^{l}a_{R}^{l}.
\ee 
Therefore, we can evaluate \eqref{eq:I-integral-def} by inserting the complete chord number basis $|m_1,m_2\ket $ in $\mathcal{H}_0 \otimes \mh_0$ as:
\be
\begin{split}
    (\te_{1},\te_{2}|\me_{\Del}^{-1}e^{-\mu(\hat{n}_{L}+\hat{n}_{R})}&\me_{\Del}|\te_{3},\te_{4}) =\sum_{l=0}^{\infty}(\te_{1},\te_{2}|e^{-\mu(\hat{n}_{L}+\hat{n}_{R})}\sum_{l=0}^{\infty}\frac{\left(e^{-2\mu};q\right)_{l}}{\left(q;q\right)_{l}}q^{\Del l}a_{L}^{l}a_{R}^{l}|\te_{3},\te_{4}) \\
&=\sum_{l=0}^{\infty}(\te_{1},\te_{2}|	e^{-\mu(\hat{n}_{L}+\hat{n}_{R})}|m_{1},m_{2})(m_{1},m_{2}|\frac{\left(e^{-2\mu};q\right)_{l}}{\left(q;q\right)_{l}}q^{\Del l}a_{L}^{l}a_{R}^{l}|\te_{3},\te_{4}).
\end{split}
\ee
Following our convention, the action of $a^{l}$ on a bra state $\bra m|$ the overlap $\bra\te_1,\te_2 |m_1,m_2)$ can be computed respectively as: 
\be
\bra m|a^{l}=\sqrt{\frac{\left(q;q\right)_{l+m}}{\left(q;q\right)_{m}}}\bra l+m|,\quad\bra\te_{1},\te_{2}|m_{1},m_{2}\ket=\frac{H_{m_{1}}\left(\cos\te_{1}|q\right)H_{m_{2}}\left(\cos\te_{2}|q\right)}{\sqrt{\left(q;q\right)_{m_{1}}\left(q;q\right)_{m_{2}}}}.
\ee
Therefore, we find that the integral kernel involving $\mathcal{E}_\Delta$ can be expressed as:
\be
\begin{split}
(\te_{1},\te_{2}|\me_{\Del} e^{-\mu(\hat{n}_{L}+\hat{n}_{R})}\me_{\Del}^{-1}|\te_{3},\te_{4})	&=\sum_{l,m_{1},m_{2}=0}^{\infty}\frac{\left(e^{-2\mu};q\right)_{l}e^{-\mu(m_{1}+m_{2})}}{\left(q;q\right)_{m_{1}}\left(q;q\right)_{m_{2}}\left(q;q\right)_{l}}q^{\Del l} \\
	&\times H_{m_{1}}^{1}H_{m_{2}}^{2}H_{m_{1}+l}^{3}H_{m_{2}+l}^{4},
\end{split}
\ee
where we have used the short hand notation $H^{i}_{m} \equiv H_m (\cos\te_i |q)$.  We can further expand the matter density $\bra\te_1 |q^{\Delta \hat{n}}|\te_2\ket$ as:
\be
\bra\te_{1}|q^{\Del\hat{n}}|\te_{2}\ket=\sum_{n_{1}=0}^{\infty}q^{\Del n_{1}}\bra\te_{1}|n_{1}\ket\bra n_{1}|\te_{2}\ket=\sum_{n_{1}=0}^{\infty}\frac{q^{\Del n_{1}}}{\left(q;q\right)_{n_{1}}}H_{n_{1}}^{1}H_{n_{1}}^{2}.
\ee
Therefore, the generating function $I^{\Delta}_\mu$ exhibits the following expansion:
\be
\begin{aligned}
I_{\Delta}^{\mu}\left(t_L, t_R\right) & =\int_0^\pi \prod_{i=1}^4 \mathrm{~d} \mu\left(\theta_i\right) \sum_{n_1, m_2, m_3=0}^{\infty} \frac{\left(e^{-2 \mu} ; q\right)_l q^{\Delta\left(n_1+l\right)} e^{-\mu\left(m_1+m_2\right)}}{(q ; q)_{n_1}(q ; q)_l(q ; q)_{m_1}(q ; q)_{m_2}} \\
& \times\left(H_{n_1}^1 H_{m_1}^1\right)\left(H_{n_1}^2 H_{m_2}^2\right)\left(H_{m_1+l}^3\right)\left(H_{m_2+l}^4\right) e^{-i\left(E_1-E_3\right) t_L-i\left(E_2-E_4\right) t_R}.
\end{aligned}
\ee
The integration of $\te_3$ and $\te_4$ are straightforward:
\be
\begin{split}
&\int\dd\mu\left(\te_{3}\right)H_{m_{1}+l}^{3}e^{iE_{3}t_{L}}=\phi_{m_{1}+l}^{*}\left(t_{L}\right)\\ 
&\int\dd\mu\left(\te_{4}\right)H_{m_{2}+l}^{4}e^{iE_{4}t_{L}}=\phi_{m_{2}+l}^{*}\left(t_{R}\right).
\end{split}
\ee 
To conduct the integration over $\te_1$ and $\te_2$, we can apply the linearization formula that expands the product of $q$-Hermite polynomials of its linear superpostion:
\be \label{eq:linearization}
H_{n_{1}}\left(\cos\te|q\right)H_{m_{1}}\left(\cos\te|q\right)=\sum_{k=0}^{{\rm min}\left(n_{1},m_{1}\right)}\binom{n_{1},n_{2}}{k}_{q}H_{n_{1}+m_{1}-2k}\left(\cos\te|q\right),
\ee
where the $q$-combinatoric coefficient is defined as:
\be
\binom{n_{1},n_{2}}{k}_{q}=\frac{\left(q;q\right)_{n_{1}}\left(q;q\right)_{n_{2}}}{\left(q;q\right)_{n_{1}-k}\left(q;q\right)_{n_{2}-k}\left(q;q\right)_{k}}.
\ee 
We conclude that 
\be
\begin{aligned}
I_{\Delta}^{\mu}\left(t_L, t_R\right) & =\sum_{n_1, m_1, m_2, l=0}^{\infty} \frac{\left(e^{-2 \mu} ; q\right)_l e^{-\mu\left(m_1+m_2\right)} q^{\Delta\left(n_1+l\right)}}{(q ; q)_{m_1}(q ; q)_{m_2}(q ; q)_{n_1}(q ; q)_l} \\
& \times \sum_{k_1=0}^{\min \left(n_1, m_1\right)} \sum_{k_2=0}^{\min \left(n_1, m_2\right)}\binom{n_1, m_1}{k_1}_q\binom{n_1, m_2}{k_2}_q \\
& \times \phi_{n_1+m_1-2 k_1}\left(t_L\right) \phi_{m_1+l}^*\left(t_L\right) \phi_{n_1+m_2-2 k_2}\left(t_R\right) \phi_{m_2+l}^*\left(t_R\right). 
\end{aligned}
\ee

\subsection{A Bulk Derivation of the Uncrossed 4-Point Function}
\label{app:bulk-dev}
We present a bulk derivation of \eqref{eq:integral-4pt} for the uncrossed four-point correlation function with a general temporal configuration:
\begin{equation}
I_{4} = \langle \Omega | e^{-iHt_{4}} W e^{-iHt_{3}} V e^{-iHt_{2}} V e^{-iHt_{1}} W | \Omega \rangle.
\end{equation}
The state \( |\Omega \rangle \) can be viewed as the \( \Delta \to 0 \) limit of \( |\Delta;0,0\rangle \). In the one-particle state, the contracted matter chord operator \( \wick{\c V \c V} \) acts as \( q^{\Delta_V \hat{N}} \) on \( \mathcal{H}_1 \), as established in \cite{Berkooz:2018jqr, Okuyama:2024yya, Lin_2023}. Therefore, we can write
\begin{equation}
I_{4} = \lim_{\Delta \to 0} \langle \Delta;0,0 | e^{-iH_{L}t_{4}} q^{\Delta_{W}\hat{N}} e^{-iH_{L}t_{3} - iH_{R}t_{1}} q^{\Delta_{V}\hat{N}} e^{-iH_{L}t_{2}} | \Delta;0,0 \rangle.
\end{equation}
Using isometric factorization, this becomes:
\begin{equation} \label{eq:I4-def-3}
\begin{aligned}
I_4 &= \lim _{\Delta \to 0} \int d\mu(\theta_1) \, d\mu(\theta_2) \, \langle \theta_1 | q^{\Delta \hat{n}} | \theta_2 \rangle \\
&\quad \times (0,0 | e^{-i h_L t_4} \mathcal{E}_{\Delta}^{-1} q^{\Delta_W(\hat{n}_L + \hat{n}_R)} \mathcal{E}_{\Delta} e^{-i h_L t_3 - i h_R t_1} \mathcal{E}_{\Delta}^{-1} q^{\Delta_V \hat{N}} \mathcal{E}_{\Delta} e^{-i H_L t_2} | \theta_1, \theta_2).
\end{aligned}
\end{equation}
In the \( \Delta \to 0 \) limit, the matrix component simplifies in the energy eigenbasis as follows:
\begin{equation}
\lim_{\Delta \to 0} \mathcal{E}_{\Delta}^{-1} q^{\Delta_{V}(\hat{n}_{L} + \hat{n}_{R})} \mathcal{E}_{\Delta} | \theta_{2}, \theta_{2} ) = \int d\mu(\theta_1) \, \langle \theta_1 | q^{\Delta_{V}} | \theta_2 \rangle \, | \theta_1, \theta_1 ),
\end{equation}
which follows directly from \eqref{eq:simplification-2}. Substituting back into \eqref{eq:I4-def-3}, we conclude that:
\begin{equation}
I_4 = \int \prod_{i=1}^{3} d\mu(\theta_i) \, e^{-iE_{1}(t_{1} + t_{3}) - iE_{2}t_{2} - iE_{3}t_{4}} \, \langle \theta_1 | q^{\Delta_{V}\hat{n}} | \theta_2 \rangle \, \langle \theta_3 | q^{\Delta_{W}\hat{n}} | \theta_1 \rangle.
\end{equation}

\subsection{Details on Initial Growth of $\tgvw$}\label{app:4pt-growth}
In this section, we analyze the initial growth of the normalized, time-ordered uncrossed four-point function, defined as
\be
\tilde{G}_4^{\Delta_V \Delta_W} = \frac{\langle \Omega | W(T_f) V(T_2) V(T_1) W(T_0) | \Omega \rangle}{\langle \Omega | W(T_f) W(0) | \Omega \rangle \langle \Omega | V(T_2) V(T_1) | \Omega \rangle} = \frac{G_4^{\Delta_V \Delta_W}(T_f, T_2)}{G_2^{\Delta_V}(T_{21}) G_2^{\Delta_W}(T_f)}.
\ee
Our goal is to assess the sensitivity of \(\tilde{G}_4^{\Delta_V \Delta_W}\) to the insertion of the particle \(V\), which we probe through \(\partial_{T_{21}} \tilde{G}_4^{\Delta_V \Delta_W} \big|_{T_{21} = 0}\). We find that
\be
\partial_{T_{21}} \tilde{G}_4^{\Delta_V \Delta_W} \big|_{T_{21} = 0} = \frac{\partial_{T_{21}} G_4^{\Delta_V \Delta_W}(T_f, T_{21}) \big|_{T_{21} = 0}}{G_2^{\Delta_W}(T_f)} - \frac{G_2^{\prime \Delta_V}}{(G_2^{\Delta_V})^2} \big|_{T_{21} = 0}.
\ee
The second term vanishes as \(G_2\) is an even function of \(T_{21}\), which has vanishing first order derivative at $T_{21}=0$. The derivative \(\partial_{T_{21}} G_4\) yields two terms:
\be
\partial_{T_{21}} G_4 \big|_{T_f = 0} = Y_1 + Y_2,
\ee
where 
\be
\begin{aligned}
& Y_1 = \sum_{n_1, n_2, n_3} c_{n_1, n_2, n_3} \, \phi_{n_1 + n_2}(T_f) \, \phi_{n_1 + n_3}^{\prime}(0) \, \phi_{n_2 + n_3}(-T_f), \\
& Y_2 = -\sum_{n_1, n_2, n_3} c_{n_1, n_2, n_3} \, \phi_{n_1 + n_2}^{\prime}(T_f) \, \phi_{n_1 + n_3}(0) \, \phi_{n_2 + n_3}(-T_f),
\end{aligned}
\ee
where the coefficients are given by \eqref{eq:4pt-un}
\be
c_{n_1, n_2, n_3} = \frac{q^{\Delta_V (n_1 + n_3) + \Delta_W (n_2 + n_3)}}{(q; q)_{n_1} (q; q)_{n_2} (q; q)_{n_3}}.
\ee
Note that \(\phi_n^{\prime}(0) = -i(1 - q)\delta_{n1}\) is nonzero only when \(n = 1\), leading to:
\be
Y_1 = \frac{i}{\sqrt{1 - q}} q^{\Delta_V} (1 - q^{\Delta_W}) \sum_{n = 0}^{\infty} \frac{q^{\Delta_W n}}{(q; q)_n} \, \phi_{n+1}(-T_f) \, \phi_n(T_f).
\ee
For \(Y_2\), given that \(\phi_n(0) = \delta_{n0}\), the expression simplifies to:
\be
Y_2 = -\sum_{n = 0}^{\infty} \frac{q^{\Delta_W n}}{(q; q)_n} \, \phi_n^{\prime}(T_f) \, \phi_n(-T_f).
\ee
The derivative \(\phi_n^{\prime}(T)\) decomposes as
\be
\begin{aligned}
\phi_n^{\prime}(T) & =\frac{-i}{\sqrt{1-q}} \int e^{-i E T}(2 \cos \theta) H_n \mathrm{~d} \mu(\theta) \\
& =\frac{-i}{\sqrt{1-q}}\left(\phi_{n+1}+\left(1-q^n\right) \phi_{n-1}\right)
\end{aligned},
\ee
where we have used the recursion relation for \(q\)-Hermite polynomials. Consequently,
\be
\begin{aligned}
Y_2 &= \frac{i}{\sqrt{1 - q}} \sum_{n = 0}^{\infty} \frac{q^{\Delta_W n}}{(q; q)_n} \phi_n(-T_f) \left( \phi_{n+1}(T_f) + (1 - q^n) \phi_{n-1}(T_f) \right) \\
&= \frac{i}{\sqrt{1 - q}} \sum_{n = 0}^{\infty} \frac{\phi_n(-T_f) \, \phi_{n+1}(T_f) \, q^{\Delta_W n} + q^{\Delta_W (n+1)} \, \phi_n(T_f) \, \phi_{n+1}(-T_f)}{(q; q)_n} \\
&= \frac{-i}{\sqrt{1 - q}} (1 - q^{\Delta_W}) \sum_{n = 0}^{\infty} \frac{q^{\Delta_W n}}{(q; q)_n} \phi_n(T_f) \, \phi_{n+1}(-T_f).
\end{aligned}
\ee
Thus, we obtain
\be
Y_1 + Y_2 = \frac{-i}{\sqrt{1 - q}} (1 - q^{\Delta_V})(1 - q^{\Delta_W}) \sum_{n = 0}^{\infty} \frac{q^{\Delta_W n}}{(q; q)_n} \, \phi_{n+1}(-T_f) \, \phi_n(T_f).
\ee
If we assume \(T_f \ll \sqrt{\lambda}\), we expand in terms of \(\tilde{T}_f = T_f / \sqrt{\lambda}\) and find
\be
\partial_{T_{21}} \tilde{G}_4^{\Delta_V \Delta_W} \big|_{T_{21} = 0} = (1 - q^{\Delta_V})(1 - q^{\Delta_W}) T_f + O(\tilde{T}_f^2).
\ee
If in addition \(T_{21} \ll T_f\), integrating the above expression yields
\be
\tilde{G}_4^{\Delta_V \Delta_W} = 1 + (1 - q^{\Delta_V})(1 - q^{\Delta_W}) T_f T_{21} + O(\tilde{T}_f^2).
\ee

\section{More on Energy Eigenbasis}
\label{app:energy-basis}
In this section we present more results regarding the matrix components of frequently used operators in energy eigenbasis $|\te\ket$. In particular, we show how the $6j$ symbol for the quantum group $U_{\sqrt{q}}(\mathfrak{su}(1,1))$ appears naturally as the matrix component of the dressed chord number operator $\me^{-1}_\Delta q^{\hat{n}_L +\hat{n}_R}\me_{\Delta}$. This derivation was first presented in \cite{Berkooz:2018jqr}, where it was obtained by summing over chord configurations in the presence of crossed matter chords, and later summarized in \cite{Okuyama:2024yya} as an application of the doubled-Hilbert space formalism. In this work, we begin with the matrix components of the $q$-ladder operators in the energy eigenbasis and subsequently derive the $R$-matrix by adopting the definition of the aforementioned dressed chord number operator.

We start with the following identity \cite{Berkooz:2018jqr} 
 regarding the energy eigenfunctions:
\be \label{eq:Q-func}
\sum_{p=0}^{\infty}\frac{t^{p}}{\left(q;q\right)_{p}}H_{p+m}\left(x|q\right)H_{p+n}\left(y|q\right)=\frac{\left(t^{2};q\right)_{\infty}}{\left(te^{i\left(\pm\te_{1}\pm\te_{2}\right)};q\right)_{\infty}}Q_{m,n}\left(x,y|t,q\right),
\ee 
With the normalization of state in the main text, we can reorganize the above equation as
\be
\bra\te_{1}|a^{\da m}t^{\hat{n}}a^{n}|\te_{2}\ket=\bra\te_{1}|t^{\hat{n}}|\te_{2}\ket Q_{m,n}\left(x,y|t,q\right),
\ee 
where $Q_{m,n}$ denotes the bivariate Al Salam-Chihara polynomial, and we have rewrite the prefactor in the RHS of \eqref{eq:Q-func} as the matter density with $q^{\Delta}=t$. In particular, when $n=0$, it simplifies to:
\be
Q_{m,0}\left(x,y|t,q\right)=\frac{1}{\left(t^{2};q\right)_{m}}Q_{m}\left(x|te^{\pm i\te_{2}},q\right),
\ee 
where the single variate polynomial can be expressed as a finite sum:
\be
Q_{n}\left(\cos\te_{1}|e^{\pm i\te_{2}};q\right)=e^{-in\te_{2}}\sum_{l=0}^{n}q^{l}\frac{\left(q^{l};q\right)_{n-l}\left(q^{-n};q\right)_{l}\left(e^{i\te_{2}\pm i\te_{1}};q\right)_{l}}{\left(q;q\right)_{l}}.
\ee 
Combining the previous two equations, we obtain:
\be \label{eq:creator}
\bra\te_{1}|a^{\da m}t^{\hat{n}}|\te_{2}\ket=\bra\te_{1}|t^{\hat{n}}|\te_{2}\ket\frac{Q_{m}\left(\cos\te_{1}|te^{\pm i\te_{2}};q\right)}{\left(t^{2};q\right)_{m}}.
\ee 
By taking the limit \( t \to 1 \), we deduce the matrix components of \( a^{\dagger m} \) as:
\be
\bra\te_{1}|a^{\da m}|\te_{2}\ket=\frac{\left(q^{m};q\right)_{\infty}}{\left(e^{\pm i\te_{1}\pm i\te_{2}};q\right)_{\infty}}Q_{m}\left(\cos\te_{1}|e^{\pm i\te_{2}};q\right).
\ee 
Note that when $m=0$, the above equation becomes zero for $\theta_1 \not= \theta_2$. For $\theta_1 =\theta_2$, it becomes a normalized delta function.

We can then move on to evaluate the matrix component of $\me_{\Delta} e^{-\mu (\hat{n}_L + \hat{n}_R)} \me^{-1}_\Delta$ in the doubled Hilbert space $\mh_0\otimes\mh_0$. To apply \eqref{eq:creator}, we use the reality of the inner product to convert $\me_\Delta$ to its Hermitian conjugate:
\be \label{eq:me-conjugate}
(\te_{3},\te_{4}|\me_{\Del}^{-1}e^{-\mu(\hat{n}_{L}+\hat{n}_{R})}\me_{\Del}|\te_{1},\te_{2})=(\te_{1},\te_{2}|\me_{\Del}^{\da}e^{-\mu(\hat{n}_{L}+\hat{n}_{R})}\me_{\Del}^{-1\da}|\te_{3},\te_{4}),
\ee 
where we can commute $\me_{\Delta}^{-1}$ with $e^{-\mu (\hat{n}_L + \hat{n}_R)}$as:
\be
\me_{\Del}^{\da}e^{-\mu(\hat{n}_{L}+\hat{n}_{R})}\left(q^{\Del}a_{L}^{\da}a_{R}^{\da};q\right)_{\infty}=\frac{\left(q^{\Del}e^{-2\mu}a_{L}^{\da}a_{R}^{\da};q\right)_{\infty}}{\left(q^{\Del}a_{L}^{\da}a_{R}^{\da};q\right)_{\infty}}e^{-\mu(\hat{n}_{L}+\hat{n}_{R})}.
\ee 
The prefactor then exhibits the following expansion in terms of $q^{\Delta}$:
\be
\frac{\left(q^{\Del}e^{-2\mu}a_{L}^{\da}a_{R}^{\da};q\right)_{\infty}}{\left(q^{\Del}a_{L}^{\da}a_{R}^{\da};q\right)_{\infty}}=\sum_{k=0}^{\infty}\frac{\left(e^{-2\mu};q\right)_{k}}{\left(q;q\right)_{k}}q^{\Del k}a_{L}^{\da k}a_{R}^{\da k}.
\ee 
Therefore, we evaluate \eqref{eq:me-conjugate} as:
\be
\begin{aligned}(\te_{1},\te_{2}|\me_{\Del}^{\da}e^{-\mu(\hat{n}_{L}+\hat{n}_{R})}\me_{\Del}^{-1\da}|\te_{3},\te_{4}) & =\sum_{k=0}^{\infty}\frac{\left(r^{2};q\right)_{k}}{(q;q)_{k}}q^{\Del k}(\te_{1},\te_{2}|a_{L}^{\da k}a_{R}^{\da k}e^{-\mu(\hat{n}_{L}+\hat{n}_{R})}|\te_{3},\te_{4}).
\end{aligned}
\ee 
Plugging in \eqref{eq:creator}, we find that 
\be \label{eq:intermidiate-me}
\begin{split}
(\te_{1},\te_{2}|\me_{\Del}^{\da}e^{-\mu(\hat{n}_{L}+\hat{n}_{R})}\me_{\Del}^{-1\da}|\te_{3},\te_{4})	&=(\te_{1},\te_{2}|e^{-\mu(\hat{n}_{L}+\hat{n}_{R})}|\te_{3},\te_{4})\\
&\times\sum_{k=0}^{\infty}\frac{q^{\Del k}Q_{k}\left(\cos\theta_{1}| e^{-\mu\pm i\theta_{3}},q\right)Q_{k}\left(\cos\theta_{2}| e^{-\mu\pm i\theta_{4}},q\right)}{\left(e^{-2\mu},q;q\right)_{k}}.
\end{split}
\ee 
We can use the orthogonal relation of $Q_n$s to express the above result in terms of the hypergeometric series ${}_8W_7$, as achieved in \cite{Berkooz:2018jqr}. Here we adopt a more symmetric expression which applies the definition of the $6j$-symbol of the $U_{\sqrt{q}} (\mathfrak{su}(1,1))$. If we send $\Delta \to \Delta_1$, $e^{-\mu}\to q^{\Delta}_2$, we shall find the above result is equivalent to:
\be
\begin{aligned}
\sum_{n=0}^n \frac{q^{\Delta_1 n} Q_n\left(\cos \theta_1 | q^{\Delta_2} e^{ \pm i \theta_3}, q\right) Q_n\left(\cos \theta_2 | q^{\Delta_2} e^{-\mu \pm i \theta_4}, q\right)}{\left(q^{\Delta_2}, q ; q\right)_n} \\
=\frac{\gamma_{\Del_{1}}^{\left(q\right)}\left(\te_{1},\te_{2}\right)}{\gamma_{\Del_{1}}^{\left(q\right)}\left(\te_{3},\te_{4}\right)}\gamma_{\Del_{2}}^{\left(q\right)}\left(\te_{1},\te_{3}\right)\gamma_{\Del_{2}}^{\left(q\right)}\left(\te_{2},\te_{4}\right) R_{\theta_4 \theta_2}^{(q)}\left[\begin{array}{cc}
\theta_3 & \Delta_2 \\
\theta_1 & \Delta_1
\end{array}\right].
\end{aligned}
\ee
We have used the notation for matter density introduced in \eqref{eq:R-matrix-0}. Plugging back in \eqref{eq:intermidiate-me}, and multiplying by an extra matter density $(\te_3 | q^{\Delta_1}|\te_4)$, we get an symmetric result:
\be \label{eq:R-matrix}
\begin{aligned}
\left(\theta_3\left|q^{\Delta_1 \hat{n}}\right| \theta_4\right)(\te_{3},\te_{4}|\me_{\Del}^{-1}q^{\Delta_2 (\hat{n}_{L}+\hat{n}_{R})}\me_{\Del}|\te_{1},\te_{2}) & =\gamma_{\Del_{1}}^{\left(q\right)}\left(\te_{1},\te_{2}\right)\gamma_{\Del_{1}}^{\left(q\right)}\left(\te_{3},\te_{4}\right)\gamma_{\Del_{2}}^{\left(q\right)}\left(\te_{1},\te_{3}\right)\gamma_{\Del_{2}}^{\left(q\right)}\left(\te_{2},\te_{4}\right) \\
& \times R_{\theta_4 \theta_2}^{(q)}\left[\begin{array}{ll}
\theta_3 & \Delta_2 \\
\theta_1 & \Delta_1
\end{array}\right].
\end{aligned}
\ee
It has been shown in \cite{Berkooz:2018jqr} that the $R$-matrix is closely related to the $6j$ symbol of $U_{\sqrt{q}}(\mathfrak{su}(1,1))$. Here we demonstrate that it naturally emerges as a matrix element of dressed chord number operators in the energy eigenbasis. The dressing of $q^{\Delta_2 (\hat{n}_L + \hat{n}_R)}$ by \(\me_\Delta\) accounts for the crossing between the two matter chord operators. We intend to explore the underlying group-theoretic framework for its appearance in future investigations.

In the no particle limit where we send $\Delta\to0$ in \eqref{eq:intermidiate-me}, and identify $\te_3$ with $\te_4$, by application of orthogonal relations of the Al-Salam-Chihara polynomials, one can show the matrix component becomes diagonal in  its bra entries as well: \eqref{eq:intermidiate-me} yields a $\delta$ function in the first two entries $\theta_{1,2}$, and the above formula simplifies to:
\be \label{eq:simplification-2}
(\te_{1},\te_{2}|\me^{-1}_0 e^{-\mu(\hat{n}_L+\hat{n}_R)}\me_0|\te_{3},\te_{3})=\mu^{-1}\left(\te_{2}\right)\del\left(\te_{1}-\te_{2}\right)\bra\te_{1}|e^{-\mu\hat{n}}|\te_{3}\ket.
\ee

\section{Moments of Liouvillian Operator} 
\label{app:momentsL}
In this section, we present our results for the moments of the Liouvillian operator in the single-particle state \(|\Delta;0,0\rangle\):
\begin{equation}
\mu_{n} \equiv \langle \Delta;0,0 | \mathcal{L}^n | \Delta;0,0 \rangle.
\end{equation}
It can be shown that \(\mu_{2m-1} = 0\) for \(m= 1, 2, \dots\), so the non-trivial moments are \(\mu_{2m}\), where:
\begin{equation}
\langle \Delta;0,0 | (H_L - H_R)^{2m} | \Delta;0,0 \rangle.
\end{equation}

Using the isometric factorization from Sec.~\ref{sec:factorize}, the above expression reduces to the following integral:
\begin{equation}
\int\dd\mu_{\Del}\left(\te_{1},\te_{2}\right)\bra\te_{1},\te_{2}|(H_{L}-H_{R})^{2m}|\te_{1},\te_{2}\ket
\end{equation}
Substituting the eigenvalues of the left and right Hamiltonians, \( H_{L/R} |\theta_L, \theta_R\rangle = \frac{2\cos\theta_{L/R}}{\sqrt{1-q}} |\theta_L, \theta_R\rangle \), we obtain:
\begin{equation}
\mu_{2n} = \frac{1}{(1-q)^n} \int d\mu_{\Delta}(\theta_1, \theta_2) \sum_{k=0}^{2n} \binom{2n}{k} (2\cos\theta_1)^k (2\cos\theta_2)^{2n-k}.
\end{equation}
Expanding the powers of \(2\cos\theta\) in terms of \(q\)-Hermite polynomials, as in Eq.~\eqref{eq:cos-expr}, we have:
\begin{equation}
\begin{split}
(2\cos\theta_1)^k &= \sum_{l_1=0}^{\lfloor k/2 \rfloor} c_{l_1, k} H_{k-2l_1}(\cos\theta_1|q), \\
(2\cos\theta_2)^{2m-k} &= \sum_{l_2=0}^{\lfloor m-k/2 \rfloor} c_{l_2, 2m-k} H_{2m-k-2l_2}(\cos\theta_2|q).
\end{split}
\end{equation}
Plugging this into the expression for \(\mu_{2n}\), we obtain:
\begin{equation}
\begin{aligned}
\mu_{2n} &= \sum_{k=0}^{2n} \binom{2n}{k} \frac{(-1)^k}{(1-q)^n} \sum_{l_1=0}^{\lfloor k/2 \rfloor} \sum_{l_2=0}^{\lfloor (n-k)/2 \rfloor} c_{l_1, k} c_{l_2, 2m-k} \\
& \quad \times \left[\int d\mu_{\Delta}(\theta_L, \theta_R) H_{k-2l_1}^L H_{2n-k-2l_2}^R\right].
\end{aligned}
\end{equation}
where $H^{L}_n$ denotes $H_n(\cos\te_L|q)$. The integral in the square brackets is straightforward:
\begin{equation}
\int d\mu_{\Delta}(\theta_L, \theta_R) H_{n_L}^L H_{n_R}^R = \int d\mu_L d\mu_R \sum_{k=0}^{\infty} \frac{q^{\Delta k} H_k^1 H_k^2}{(q;q)_k} H_{n_L}^L H_{n_R}^R = q^{\Delta n_L} (q;q)_{n_L} \delta_{n_L n_R}.
\end{equation}
Thus, the final result is:
\begin{equation}
\mu_{2n} = \sum_{k=0}^{2n} \sum_{l=0}^{\lfloor k/2 \rfloor} \frac{(-1)^k}{(1-q)^n} (q;q)_{k-2l} q^{\Delta(k-2l)} \binom{2n}{k} c_{l, k} c_{n+l-k, 2n-k},
\end{equation}
where the coefficients \( c_{l,k} \) are defined as:
\begin{equation} \label{eq:clk-def}
c_{l, k} = \sum_{m=0}^{l} (-1)^m q^{\binom{m+1}{2}} \frac{k-2l+2m+1}{k+1} \binom{k+1}{l-m} \binom{k-2l+m}{m}_q.
\end{equation}

\bibliography{ref}

\providecommand{\href}[2]{#2}\begingroup\raggedright\begin{thebibliography}{10}

\bibitem{Roberts:2014isa}
D.~A. Roberts, D.~Stanford and L.~Susskind, \emph{{Localized shocks}}, \href{https://doi.org/10.1007/JHEP03(2015)051}{\emph{JHEP} {\bfseries 03} (2015) 051} [\href{https://arxiv.org/abs/1409.8180}{{\ttfamily 1409.8180}}].

\bibitem{Roberts:2018mnp}
D.~A. Roberts, D.~Stanford and A.~Streicher, \emph{{Operator growth in the SYK model}}, \href{https://doi.org/10.1007/JHEP06(2018)122}{\emph{JHEP} {\bfseries 06} (2018) 122} [\href{https://arxiv.org/abs/1802.02633}{{\ttfamily 1802.02633}}].

\bibitem{Qi_2019}
X.-L. Qi and A.~Streicher, \emph{Quantum epidemiology: operator growth, thermal effects, and syk}, \href{https://doi.org/10.1007/jhep08(2019)012}{\emph{Journal of High Energy Physics} {\bfseries 2019} (2019) }.

\bibitem{Lensky:2020ubw}
Y.~D. Lensky, X.-L. Qi and P.~Zhang, \emph{{Size of bulk fermions in the SYK model}}, \href{https://doi.org/10.1007/JHEP10(2020)053}{\emph{JHEP} {\bfseries 10} (2020) 053} [\href{https://arxiv.org/abs/2002.01961}{{\ttfamily 2002.01961}}].

\bibitem{Jian_2021}
S.-K. Jian, B.~Swingle and Z.-Y. Xian, \emph{Complexity growth of operators in the syk model and in jt gravity}, \href{https://doi.org/10.1007/jhep03(2021)014}{\emph{Journal of High Energy Physics} {\bfseries 2021} (2021) }.

\bibitem{Schuster_2022}
T.~Schuster, B.~Kobrin, P.~Gao, I.~Cong, E.~T. Khabiboulline, N.~M. Linke et~al., \emph{Many-body quantum teleportation via operator spreading in the traversable wormhole protocol}, \href{https://doi.org/10.1103/physrevx.12.031013}{\emph{Physical Review X} {\bfseries 12} (2022) }.

\bibitem{Roberts_2015}
D.~A. Roberts, D.~Stanford and L.~Susskind, \emph{Localized shocks}, \href{https://doi.org/10.1007/jhep03(2015)051}{\emph{Journal of High Energy Physics} {\bfseries 2015} (2015) }.

\bibitem{shenker2015stringyeffectsscrambling}
S.~H. Shenker and D.~Stanford, \emph{Stringy effects in scrambling},  2015.

\bibitem{Roberts_2015_butterfly}
D.~A. Roberts and D.~Stanford, \emph{Diagnosing chaos using four-point functions in two-dimensional conformal field theory}, \href{https://doi.org/10.1103/physrevlett.115.131603}{\emph{Physical Review Letters} {\bfseries 115} (2015) }.

\bibitem{Hashimoto_2017}
K.~Hashimoto, K.~Murata and R.~Yoshii, \emph{Out-of-time-order correlators in quantum mechanics}, \href{https://doi.org/10.1007/jhep10(2017)138}{\emph{Journal of High Energy Physics} {\bfseries 2017} (2017) }.

\bibitem{Gu_2022}
Y.~Gu, A.~Kitaev and P.~Zhang, \emph{A two-way approach to out-of-time-order correlators}, \href{https://doi.org/10.1007/jhep03(2022)133}{\emph{Journal of High Energy Physics} {\bfseries 2022} (2022) }.

\bibitem{Parker_2019}
D.~E. Parker, X.~Cao, A.~Avdoshkin, T.~Scaffidi and E.~Altman, \emph{A universal operator growth hypothesis}, \href{https://doi.org/10.1103/physrevx.9.041017}{\emph{Physical Review X} {\bfseries 9} (2019) }.

\bibitem{Balasubramanian_2022}
V.~Balasubramanian, P.~Caputa, J.~M. Magan and Q.~Wu, \emph{Quantum chaos and the complexity of spread of states}, \href{https://doi.org/10.1103/physrevd.106.046007}{\emph{Physical Review D} {\bfseries 106} (2022) }.

\bibitem{Nielsen2005AGA}
M.~A. Nielsen, \emph{A geometric approach to quantum circuit lower bounds}, {\emph{Quantum Inf. Comput.} {\bfseries 6} (2005) 213}.

\bibitem{PhysRevD.103.026015.QueryC}
B.~Chen, B.~Czech and Z.-z. Wang, \emph{Query complexity and cutoff dependence of the ${\mathrm{cft}}_{2}$ ground state}, \href{https://doi.org/10.1103/PhysRevD.103.026015}{\emph{Phys. Rev. D} {\bfseries 103} (2021) 026015}.

\bibitem{Susskind:1994vu}
L.~Susskind, \emph{{The World as a hologram}}, \href{https://doi.org/10.1063/1.531249}{\emph{J. Math. Phys.} {\bfseries 36} (1995) 6377} [\href{https://arxiv.org/abs/hep-th/9409089}{{\ttfamily hep-th/9409089}}].

\bibitem{Maldacena:1997re}
J.~M. Maldacena, \emph{{The Large N limit of superconformal field theories and supergravity}}, \href{https://doi.org/10.4310/ATMP.1998.v2.n2.a1}{\emph{Adv. Theor. Math. Phys.} {\bfseries 2} (1998) 231} [\href{https://arxiv.org/abs/hep-th/9711200}{{\ttfamily hep-th/9711200}}].

\bibitem{Witten:1998zw}
E.~Witten, \emph{{Anti-de Sitter space, thermal phase transition, and confinement in gauge theories}}, \href{https://doi.org/10.4310/ATMP.1998.v2.n3.a3}{\emph{Adv. Theor. Math. Phys.} {\bfseries 2} (1998) 505} [\href{https://arxiv.org/abs/hep-th/9803131}{{\ttfamily hep-th/9803131}}].

\bibitem{Witten:1998qj}
E.~Witten, \emph{{Anti-de Sitter space and holography}}, \href{https://doi.org/10.4310/ATMP.1998.v2.n2.a2}{\emph{Adv. Theor. Math. Phys.} {\bfseries 2} (1998) 253} [\href{https://arxiv.org/abs/hep-th/9802150}{{\ttfamily hep-th/9802150}}].

\bibitem{Susskind:1998dq}
L.~Susskind and E.~Witten, \emph{{The Holographic bound in anti-de Sitter space}},  \href{https://arxiv.org/abs/hep-th/9805114}{{\ttfamily hep-th/9805114}}.

\bibitem{Haferkamp_2022}
J.~Haferkamp, P.~Faist, N.~B.~T. Kothakonda, J.~Eisert and N.~Yunger~Halpern, \emph{Linear growth of quantum circuit complexity}, \href{https://doi.org/10.1038/s41567-022-01539-6}{\emph{Nature Physics} {\bfseries 18} (2022) 528–532}.

\bibitem{Rabinovici:2023yex}
E.~Rabinovici, A.~S\'anchez-Garrido, R.~Shir and J.~Sonner, \emph{{A bulk manifestation of Krylov complexity}}, \href{https://doi.org/10.1007/JHEP08(2023)213}{\emph{JHEP} {\bfseries 08} (2023) 213} [\href{https://arxiv.org/abs/2305.04355}{{\ttfamily 2305.04355}}].

\bibitem{Aguilar-Gutierrez:2024nau}
S.~E. Aguilar-Gutierrez, \emph{{Towards complexity in de Sitter space from the doubled-scaled Sachdev-Ye-Kitaev model}}, \href{https://doi.org/10.1007/JHEP10(2024)107}{\emph{JHEP} {\bfseries 10} (2024) 107} [\href{https://arxiv.org/abs/2403.13186}{{\ttfamily 2403.13186}}].

\bibitem{Sergio-C:2024rka}
S.~E. Aguilar-Gutierrez, S.~Baiguera and N.~Zenoni, \emph{{Holographic complexity of the extended Schwarzschild-de Sitter space}}, \href{https://doi.org/10.1007/JHEP05(2024)201}{\emph{JHEP} {\bfseries 05} (2024) 201} [\href{https://arxiv.org/abs/2402.01357}{{\ttfamily 2402.01357}}].

\bibitem{Li:2024kfm}
T.~Li and L.-H. Liu, \emph{{Inflationary Krylov complexity}}, \href{https://doi.org/10.1007/JHEP04(2024)123}{\emph{JHEP} {\bfseries 04} (2024) 123} [\href{https://arxiv.org/abs/2401.09307}{{\ttfamily 2401.09307}}].

\bibitem{Sekino_2008}
Y.~Sekino and L.~Susskind, \emph{Fast scramblers}, \href{https://doi.org/10.1088/1126-6708/2008/10/065}{\emph{Journal of High Energy Physics} {\bfseries 2008} (2008) 065–065}.

\bibitem{Lashkari_2013}
N.~Lashkari, D.~Stanford, M.~Hastings, T.~Osborne and P.~Hayden, \emph{Towards the fast scrambling conjecture}, \href{https://doi.org/10.1007/jhep04(2013)022}{\emph{Journal of High Energy Physics} {\bfseries 2013} (2013) }.

\bibitem{Susskind:2014rva}
L.~Susskind, \emph{{Computational Complexity and Black Hole Horizons}}, \href{https://doi.org/10.1002/prop.201500092}{\emph{Fortsch. Phys.} {\bfseries 64} (2016) 24} [\href{https://arxiv.org/abs/1403.5695}{{\ttfamily 1403.5695}}].

\bibitem{Stanford:2014jda}
D.~Stanford and L.~Susskind, \emph{{Complexity and Shock Wave Geometries}}, \href{https://doi.org/10.1103/PhysRevD.90.126007}{\emph{Phys. Rev. D} {\bfseries 90} (2014) 126007} [\href{https://arxiv.org/abs/1406.2678}{{\ttfamily 1406.2678}}].

\bibitem{Brown:2015bva}
A.~R. Brown, D.~A. Roberts, L.~Susskind, B.~Swingle and Y.~Zhao, \emph{{Holographic Complexity Equals Bulk Action?}}, \href{https://doi.org/10.1103/PhysRevLett.116.191301}{\emph{Phys. Rev. Lett.} {\bfseries 116} (2016) 191301} [\href{https://arxiv.org/abs/1509.07876}{{\ttfamily 1509.07876}}].

\bibitem{Couch_2017}
J.~Couch, W.~Fischler and P.~H. Nguyen, \emph{Noether charge, black hole volume, and complexity}, \href{https://doi.org/10.1007/jhep03(2017)119}{\emph{Journal of High Energy Physics} {\bfseries 2017} (2017) }.

\bibitem{Barb_n_2019}
J.~Barbón, E.~Rabinovici, R.~Shir and R.~Sinha, \emph{On the evolution of operator complexity beyond scrambling}, \href{https://doi.org/10.1007/jhep10(2019)264}{\emph{Journal of High Energy Physics} {\bfseries 2019} (2019) }.

\bibitem{Rabinovici_2021}
E.~Rabinovici, A.~Sánchez-Garrido, R.~Shir and J.~Sonner, \emph{Operator complexity: a journey to the edge of krylov space}, \href{https://doi.org/10.1007/jhep06(2021)062}{\emph{Journal of High Energy Physics} {\bfseries 2021} (2021) }.

\bibitem{Cotler:2016fpe}
J.~S. Cotler, G.~Gur-Ari, M.~Hanada, J.~Polchinski, P.~Saad, S.~H. Shenker et~al., \emph{{Black Holes and Random Matrices}}, \href{https://doi.org/10.1007/JHEP05(2017)118}{\emph{JHEP} {\bfseries 05} (2017) 118} [\href{https://arxiv.org/abs/1611.04650}{{\ttfamily 1611.04650}}].

\bibitem{Berkooz:2018qkz}
M.~Berkooz, P.~Narayan and J.~Simon, \emph{{Chord diagrams, exact correlators in spin glasses and black hole bulk reconstruction}}, \href{https://doi.org/10.1007/JHEP08(2018)192}{\emph{JHEP} {\bfseries 08} (2018) 192} [\href{https://arxiv.org/abs/1806.04380}{{\ttfamily 1806.04380}}].

\bibitem{Berkooz:2018jqr}
M.~Berkooz, M.~Isachenkov, V.~Narovlansky and G.~Torrents, \emph{{Towards a full solution of the large N double-scaled SYK model}}, \href{https://doi.org/10.1007/JHEP03(2019)079}{\emph{JHEP} {\bfseries 03} (2019) 079} [\href{https://arxiv.org/abs/1811.02584}{{\ttfamily 1811.02584}}].

\bibitem{Lin_2023}
H.~W. Lin and D.~Stanford, \emph{A symmetry algebra in double-scaled syk}, \href{https://doi.org/10.21468/scipostphys.15.6.234}{\emph{SciPost Physics} {\bfseries 15} (2023) }.

\bibitem{Susskind:2021esx}
L.~Susskind, \emph{{Entanglement and Chaos in De Sitter Space Holography: An SYK Example}}, \href{https://doi.org/10.22128/jhap.2021.455.1005}{\emph{JHAP} {\bfseries 1} (2021) 1} [\href{https://arxiv.org/abs/2109.14104}{{\ttfamily 2109.14104}}].

\bibitem{Susskind:2022dfz}
L.~Susskind, \emph{{Scrambling in Double-Scaled SYK and De Sitter Space}},  \href{https://arxiv.org/abs/2205.00315}{{\ttfamily 2205.00315}}.

\bibitem{Lin:2022nss}
H.~Lin and L.~Susskind, \emph{{Infinite Temperature's Not So Hot}},  \href{https://arxiv.org/abs/2206.01083}{{\ttfamily 2206.01083}}.

\bibitem{Susskind:2022bia}
L.~Susskind, \emph{{De Sitter Space, Double-Scaled SYK, and the Separation of Scales in the Semiclassical Limit}},  \href{https://arxiv.org/abs/2209.09999}{{\ttfamily 2209.09999}}.

\bibitem{Berkooz:2022mfk}
M.~Berkooz, M.~Isachenkov, M.~Isachenkov, P.~Narayan and V.~Narovlansky, \emph{{Quantum groups, non-commutative AdS$_{2}$, and chords in the double-scaled SYK model}}, \href{https://doi.org/10.1007/JHEP08(2023)076}{\emph{JHEP} {\bfseries 08} (2023) 076} [\href{https://arxiv.org/abs/2212.13668}{{\ttfamily 2212.13668}}].

\bibitem{Goel:2023svz}
A.~Goel, V.~Narovlansky and H.~Verlinde, \emph{{Semiclassical geometry in double-scaled SYK}}, \href{https://doi.org/10.1007/JHEP11(2023)093}{\emph{JHEP} {\bfseries 11} (2023) 093} [\href{https://arxiv.org/abs/2301.05732}{{\ttfamily 2301.05732}}].

\bibitem{Narovlansky:2023lfz}
V.~Narovlansky and H.~Verlinde, \emph{{Double-scaled SYK and de Sitter Holography}},  \href{https://arxiv.org/abs/2310.16994}{{\ttfamily 2310.16994}}.

\bibitem{Susskind:2023hnj}
L.~Susskind, \emph{{De Sitter Space has no Chords. Almost Everything is Confined.}}, \href{https://doi.org/10.22128/jhap.2023.661.1043}{\emph{JHAP} {\bfseries 3} (2023) 1} [\href{https://arxiv.org/abs/2303.00792}{{\ttfamily 2303.00792}}].

\bibitem{Rahman:2024vyg}
A.~A. Rahman and L.~Susskind, \emph{{Infinite Temperature is Not So Infinite: The Many Temperatures of de Sitter Space}},  \href{https://arxiv.org/abs/2401.08555}{{\ttfamily 2401.08555}}.

\bibitem{Rahman:2024iiu}
A.~A. Rahman and L.~Susskind, \emph{{$p$-Chords, Wee-Chords, and de Sitter Space}},  \href{https://arxiv.org/abs/2407.12988}{{\ttfamily 2407.12988}}.

\bibitem{Almheiri:2024ayc}
A.~Almheiri and F.~K. Popov, \emph{{Holography on the Quantum Disk}},  \href{https://arxiv.org/abs/2401.05575}{{\ttfamily 2401.05575}}.

\bibitem{Almheiri:2024xtw}
A.~Almheiri, A.~Goel and X.-Y. Hu, \emph{{Quantum gravity of the Heisenberg algebra}}, \href{https://doi.org/10.1007/JHEP08(2024)098}{\emph{JHEP} {\bfseries 08} (2024) 098} [\href{https://arxiv.org/abs/2403.18333}{{\ttfamily 2403.18333}}].

\bibitem{Milekhin:2024vbb}
A.~Milekhin and J.~Xu, \emph{{On scrambling, tomperature and superdiffusion in de Sitter space}},  \href{https://arxiv.org/abs/2403.13915}{{\ttfamily 2403.13915}}.

\bibitem{Lin:2022rbf}
H.~W. Lin, \emph{{The bulk Hilbert space of double scaled SYK}}, \href{https://doi.org/10.1007/JHEP11(2022)060}{\emph{JHEP} {\bfseries 11} (2022) 060} [\href{https://arxiv.org/abs/2208.07032}{{\ttfamily 2208.07032}}].

\bibitem{Xu:2024hoc}
J.~Xu, \emph{{Von Neumann Algebras in Double-Scaled SYK}},  \href{https://arxiv.org/abs/2403.09021}{{\ttfamily 2403.09021}}.

\bibitem{Milekhin:2024ToAppear}
A.~Milekhin and J.~Xu, {In Preparation}.

\bibitem{mukhametzhanov2023largepsykchord}
B.~Mukhametzhanov, \emph{Large p syk from chord diagrams},  2023.

\bibitem{Okuyama:2024gsn}
K.~Okuyama, \emph{{More on doubled Hilbert space in double-scaled SYK}}, \href{https://doi.org/10.1016/j.physletb.2024.138858}{\emph{Phys. Lett. B} {\bfseries 855} (2024) 138858} [\href{https://arxiv.org/abs/2404.02833}{{\ttfamily 2404.02833}}].

\bibitem{Okuyama:2024yya}
K.~Okuyama, \emph{{Doubled Hilbert space in double-scaled SYK}},  \href{https://arxiv.org/abs/2401.07403}{{\ttfamily 2401.07403}}.

\bibitem{Choi:2019bmd}
C.~Choi, M.~Mezei and G.~S\'arosi, \emph{{Exact four point function for large $q$ SYK from Regge theory}}, \href{https://doi.org/10.1007/JHEP05(2021)166}{\emph{JHEP} {\bfseries 05} (2021) 166} [\href{https://arxiv.org/abs/1912.00004}{{\ttfamily 1912.00004}}].

\bibitem{Streicher_2020}
A.~Streicher, \emph{Syk correlators for all energies}, \href{https://doi.org/10.1007/jhep02(2020)048}{\emph{Journal of High Energy Physics} {\bfseries 2020} (2020) }.

\bibitem{Sergio2024towards}
S.~E. Aguilar-Gutierrez, \emph{Towards complexity in de sitter space from the double-scaled sachdev-ye-kitaev model},  2024.

\bibitem{Harlow:2021dfp}
D.~Harlow and J.-q. Wu, \emph{{Algebra of diffeomorphism-invariant observables in Jackiw-Teitelboim gravity}}, \href{https://doi.org/10.1007/JHEP05(2022)097}{\emph{JHEP} {\bfseries 05} (2022) 097} [\href{https://arxiv.org/abs/2108.04841}{{\ttfamily 2108.04841}}].

\bibitem{Bhattacharjee:2022ave}
B.~Bhattacharjee, P.~Nandy and T.~Pathak, \emph{{Krylov complexity in large q and double-scaled SYK model}}, \href{https://doi.org/10.1007/JHEP08(2023)099}{\emph{JHEP} {\bfseries 08} (2023) 099} [\href{https://arxiv.org/abs/2210.02474}{{\ttfamily 2210.02474}}].

\bibitem{Tang:2023ocr}
H.~Tang, \emph{{Operator Krylov complexity in random matrix theory}},  \href{https://arxiv.org/abs/2312.17416}{{\ttfamily 2312.17416}}.

\bibitem{Tang:2024xgg}
H.~Tang, \emph{{Entanglement entropy in type II$_1$ von Neumann algebra: examples in Double-Scaled SYK}},  \href{https://arxiv.org/abs/2404.02449}{{\ttfamily 2404.02449}}.

\bibitem{caputa2021geometrykrylovcomplexity}
P.~Caputa, J.~M. Magan and D.~Patramanis, \emph{Geometry of krylov complexity},  2021.

\bibitem{Mertens:2022irh}
T.~G. Mertens and G.~J. Turiaci, \emph{{Solvable models of quantum black holes: a review on Jackiw\textendash{}Teitelboim gravity}}, \href{https://doi.org/10.1007/s41114-023-00046-1}{\emph{Living Rev. Rel.} {\bfseries 26} (2023) 4} [\href{https://arxiv.org/abs/2210.10846}{{\ttfamily 2210.10846}}].

\bibitem{Saad:2019pqd}
P.~Saad, \emph{{Late Time Correlation Functions, Baby Universes, and ETH in JT Gravity}},  \href{https://arxiv.org/abs/1910.10311}{{\ttfamily 1910.10311}}.

\bibitem{Hashimoto_2023}
K.~Hashimoto, K.~Murata, N.~Tanahashi and R.~Watanabe, \emph{Krylov complexity and chaos in quantum mechanics}, \href{https://doi.org/10.1007/jhep11(2023)040}{\emph{Journal of High Energy Physics} {\bfseries 2023} (2023) }.

\bibitem{Gorsky_2020}
A.~Dymarsky and A.~Gorsky, \emph{Quantum chaos as delocalization in krylov space}, \href{https://doi.org/10.1103/PhysRevB.102.085137}{\emph{Phys. Rev. B} {\bfseries 102} (2020) 085137}.

\bibitem{2025KComplexity}
M.~Ambrosini, E.~Rabinovici, A.~Sánchez-Garrido, R.~Shir and J.~Sonner, \emph{Operator k-complexity in dssyk: Krylov complexity equals bulk length},  2025.

\bibitem{Verlinde:2024zrh}
H.~Verlinde and M.~Zhang, \emph{{SYK Correlators from 2D Liouville-de Sitter Gravity}},  \href{https://arxiv.org/abs/2402.02584}{{\ttfamily 2402.02584}}.

\bibitem{Gaiotto:2024kze}
D.~Gaiotto and H.~Verlinde, \emph{{SYK-Schur duality: Double scaled SYK correlators from $N=2$ supersymmetric gauge theory}},  \href{https://arxiv.org/abs/2409.11551}{{\ttfamily 2409.11551}}.

\bibitem{Milekhin:2023bjv}
A.~Milekhin and J.~Xu, \emph{{Revisiting Brownian SYK and its possible relations to de Sitter}}, \href{https://doi.org/10.1007/JHEP10(2024)151}{\emph{JHEP} {\bfseries 10} (2024) 151} [\href{https://arxiv.org/abs/2312.03623}{{\ttfamily 2312.03623}}].

\bibitem{Berkooz:2020xne}
M.~Berkooz, N.~Brukner, V.~Narovlansky and A.~Raz, \emph{{The double scaled limit of Super--Symmetric SYK models}}, \href{https://doi.org/10.1007/JHEP12(2020)110}{\emph{JHEP} {\bfseries 12} (2020) 110} [\href{https://arxiv.org/abs/2003.04405}{{\ttfamily 2003.04405}}].

\bibitem{Boruch:2023bte}
J.~Boruch, H.~W. Lin and C.~Yan, \emph{{Exploring supersymmetric wormholes in $ \mathcal{N} $ = 2 SYK with chords}}, \href{https://doi.org/10.1007/JHEP12(2023)151}{\emph{JHEP} {\bfseries 12} (2023) 151} [\href{https://arxiv.org/abs/2308.16283}{{\ttfamily 2308.16283}}].

\bibitem{Lam:2018pvp}
H.~T. Lam, T.~G. Mertens, G.~J. Turiaci and H.~Verlinde, \emph{{Shockwave S-matrix from Schwarzian Quantum Mechanics}}, \href{https://doi.org/10.1007/JHEP11(2018)182}{\emph{JHEP} {\bfseries 11} (2018) 182} [\href{https://arxiv.org/abs/1804.09834}{{\ttfamily 1804.09834}}].

\bibitem{qdisk1999-1}
D.~Shklyarov, S.~Sinel'shchikov and L.~Vaksman, \emph{On function theory in quantum disc: Integral representations},  1999.

\bibitem{qdisk1999-2}
D.~Shklyarov, S.~Sinel'shchikov and L.~Vaksman, \emph{On function theory in quantum disc: q-differential equations and fourier transform},  1999.

\bibitem{Blommaert:2023wad}
A.~Blommaert, T.~G. Mertens and S.~Yao, \emph{{The q-Schwarzian and Liouville gravity}},  \href{https://arxiv.org/abs/2312.00871}{{\ttfamily 2312.00871}}.

\bibitem{Gesteau:2024rpt}
E.~Gesteau and H.~Liu, \emph{{Toward stringy horizons}},  \href{https://arxiv.org/abs/2408.12642}{{\ttfamily 2408.12642}}.

\bibitem{Askey1983AGO}
R.~A. Askey and M.~E.~H. Ismail, \emph{A generalization of ultraspherical polynomials}, .

\end{thebibliography}\endgroup
\bibliographystyle{JHEP}

\end{document}